\documentclass[mathpazo]{csiam-ls}
\usepackage{graphicx}
\usepackage{cleveref}
\usepackage{xcolor}
\renewcommand\thisnumber{x}
\renewcommand\thisyear{2026}
\renewcommand\thismonth{xx}
\renewcommand\thisvolume{x}
\renewcommand\doi{10.4208/csiam-ls.SO-2025-0022}

\begin{document}
%%%%% title : short title may not be used but TITLE is required.
% \title{TITLE}
% \title[short title]{TITLE}
\title{Cell-cell Communication Inference and Analysis: \\Biological Mechanisms, Computational Approaches,\\and Future Opportunities}

%%%%% author(s) :
% single author:
% \author[name in running head]{AUTHOR\corrauth}
% [name in running head] is NOT OPTIONAL, it is a MUST.
% Use \corrauth to indicate the corresponding author.
% Use \email to provide email address of author.
% \footnote and \thanks are not used in the heading section.
% Another acknowlegments/support of grants, state in Acknowledgments section
% \section*{Acknowledgments}
  % \author[O.~Author]{Only Author\corrauth}
  % \address{School of Mathematical Sciences, Beijing Normal University,
  % Beijing 100875, P.R. China}
%\email{{\tt author@email} (O.~Author)}

% multiple authors:
% Note the use of \affil and \affilnum to link names and addresses.
% The author for correspondence is marked by \corrauth.
% use \emails to provide email addresses of authors
% e.g. below example has 3 authors, first author is also the corresponding
%      author, author 1 and 3 having the same address.
% \author[Z. Zhang et~al.]{Zhengru Zhang\affil{1}\comma\corrauth,
%       Author Chan\affil{2}~and Author Zhao\affil{1}}
% \address{\affilnum{1}\ School of Mathematical Sciences,
%          Beijing Normal University,
%          Beijing 100875, P.R. China. \\
%           \affilnum{2}\ Department of Mathematics,
%           Hong Kong Baptist University, Hong Kong SAR.}
% \emails{{\tt zhang@email} (Z.~Zhang), {\tt chan@email} (A.~Chan),
%          {\tt zhao@email} (A.~Zhao)}
% \footnote and \thanks are not used in the heading section.
% Another acknowlegments/support of grants, state in Acknowledgments section
% \section*{Acknowledgments}
\author[X. Cheng et~al.]{Xiangzheng Cheng\affil{1}, Haili Huang\affil{1}, Ye Su\affil{1}, Qing Nie\affil{2,3,4},\\ Xiufen Zou\affil{1,*} and Suoqin Jin\affil{1}\comma\corrauth}
\address{
\affilnum{1}\ School of Mathematics and Statistics, Wuhan University, Wuhan 430072, China. \\
\affilnum{2}\ NSF-Simons Center for Multiscale Cell Fate Research, University of California, Irvine, Irvine, CA 92697, USA. \\
\affilnum{3}\ Department of Mathematics, University of California, Irvine, Irvine, CA 92697, USA \\  
\affilnum{4}\ Department of Developmental and Cell Biology, University of California, Irvine, Irvine, CA 92697, USA}
\emails{{\tt sqjin@whu.edu.cn} (S. Jin), {\tt xfzou@whu.edu.cn} (X. Zou)}
%
%same address:
%\author[F. Author and A.~Co-Author,]{First Author and A.~Co-Author\corrauth}
%\address{address of First Author and His Best Friend}
%

%%%%% Begin Abstract %%%%%%%%%%%
\begin{abstract}
In multicellular organisms, cells coordinate their activities through cell-cell communication (CCC), which is crucial for development, tissue homeostasis, and disease progression. Recent advances in single-cell and spatial omics technologies provide unprecedented opportunities to systematically infer and analyze CCC from these omics data, either by integrating prior knowledge of ligand-receptor interactions (LRIs) or through \textit{de novo} approaches. A variety of computational methods have been developed, focusing on methodological innovations, accurate modeling of complex signaling mechanisms, and investigation of broader biological questions. These advances have greatly enhanced our ability to analyze CCC and generate biological hypotheses. Here, we introduce the biological mechanisms and modeling strategies of CCC, and provide a focused overview of more than 140 computational methods for inferring CCC from single-cell and spatial transcriptomic data, emphasizing the diversity in methodological frameworks and biological questions. Finally, we discuss the current challenges and future opportunities in this rapidly evolving field, and summarize available methods in an interactive online resource (https://cellchat.whu.edu.cn) to facilitate more efficient method comparison and selection.

\end{abstract}
%%%%% end %%%%%%%%%%%

%%%%% AMS/PACs/Keywords %%%%%%%%%%%
%\pac{}

%%%% maketitle %%%%%
\maketitle

%%%% Start %%%%%%

\section{Introduction}\label{sec1}

Cell-cell communication (CCC) refers to the process by which cells exchange signals to coordinate their activities through soluble signaling molecules or direct physical contact. It forms the fundamental basis for maintaining homeostasis and achieving functional integration in multicellular organisms. This process involves a series of highly specific molecular mechanisms, requiring the coordinated participation of various molecules—such as ligands, receptors, ions, and metabolites—as well as membrane structures, including tight junctions, gap junctions, and immunological synapses\cite{atakan2014molecular,heldin2016signals,alberts2022molecular,LimMayer2024,clark2018biology}. Elucidating the mechanisms underlying CCC not only deepens our understanding of development and disease, but also identifies potential targets for therapeutic interventions\cite{CCCreviewFan2020,CCCreviewSu2024}.

Traditionally, the study of CCC has relied heavily on labor-intensive experimental approaches, such as histological tissue section analysis, in vitro co-culture systems, and in vivo genetic manipulations. These methods are often time-consuming and technically demanding, typically permitting the investigation of communication between only a limited number of cell types and signals, thereby hindering a comprehensive elucidation of CCC networks. The rapid advancement of single-cell RNA sequencing (scRNA-seq) technologies has revolutionized CCC research\cite{CCCreviewFan2020,RN448}. The scRNA-seq data capture gene expression patterns relevant to CCC, providing indirect insights into the abundance of proteins involved in signal transduction. Building on this foundation, numerous computational tools have been developed to infer CCC by integrating scRNA-seq data with prior knowledge of ligand-receptor interactions (LRIs)\cite{CCCreviewFan2020,RN448,RN449,CCCreviewJin2022,CCCreviewNie2023,RN447,CCCreviewSu2024,cesaro2025advances}, thus yielding biologically interpretable insights. However, spatial information on cells is inherently lost in scRNA-seq data. Given that most CCC events occur within a short spatial distance, the emerging spatial transcriptomics brings new opportunities to improve CCC analysis\cite{STreview2021integrating,STyuan2021}. The ongoing technological progress, such as single-cell and spatial multi-omics\cite{spatialomicsScience2023,reviewMultiOmics2023}, and increasingly complex biological questions continue to drive the evolution of computational approaches, enabling more comprehensive and in-depth analyses.

This review aims to equip biologically focused researchers with practical guidance on when and how to apply existing CCC tools, while providing computational biologists with the biological foundations and methodological advances of CCC to facilitate the development of novel CCC analysis tools. While related reviews exist\cite{CCCreviewFan2020,RN448,RN449,CCCreviewJin2022,CCCreviewNie2023,RN447,CCCreviewSu2024,cesaro2025advances}, here we emphasize the biological mechanisms and modeling strategies that underpin CCC, and critically examine the most recent and innovative computational approaches. This review begins with a systematic overview of the biological foundations of CCC, including various types of intercellular and intracellular signaling mechanisms. We then present a general modeling framework for CCC inference, and review a range of computational methods developed in recent years, with an emphasis on their methodological diversification and the specific biological questions they aim to address. Finally, we discuss key challenges and emerging opportunities in this rapidly advancing field. To aid the research community, we provide an interactive, web-based summary of all reviewed computational methods (available at https://cellchat.whu.edu.cn). This resource details each method's overview, computational principles, and addressed biological questions to facilitate efficient method comparison and selection. It features interactive filtering based on methodological characteristics and includes a submission portal for researchers to contribute new tools, ensuring the resource remains current.

\section{Biological Mechanisms of Cell-Cell Communication}\label{sec2}
\subsection{Types of Cell-Cell Communication}

Precise, reliable, and reproducible mechanisms of CCC have evolved to coordinate biological activities among cells within multicellular organisms. Distinct communication systems operate across various tissues and organs to transmit information from signaling cells to their target cells\cite{alberts2022molecular,LimMayer2024,atakan2014molecular}. The chemical signals used by signaling cells for information transmission are known as ligands, while the corresponding binding molecules on target cells are called receptors. Ligands comprise diverse molecules—including proteins (e.g., growth factors, cytokines), peptides, small molecules (e.g., metabolites, neurotransmitters, hormones), and lipids (e.g., prostaglandins, eicosanoids)—that bind to cell-surface receptors. A separate class of hydrophobic ligands (e.g., steroids, gases) diffuses across the plasma membrane to bind intracellular receptors. Receptors are broadly classified by localization and function: cell-surface (transmembrane) receptors and intracellular (cytoplasmic or nuclear) receptors. Cell-surface receptors are further categorized as ion channel-linked, G-protein-coupled, or enzyme-linked\cite{heldin2016signals}. Upon ligand binding, receptors undergo conformational changes that activate intracellular signaling molecules within target cells, and further initiate signal transduction cascades.

\begin{figure}[!tbh]
  \centering
  \includegraphics[scale=0.475, trim=12 0 25 0, clip]{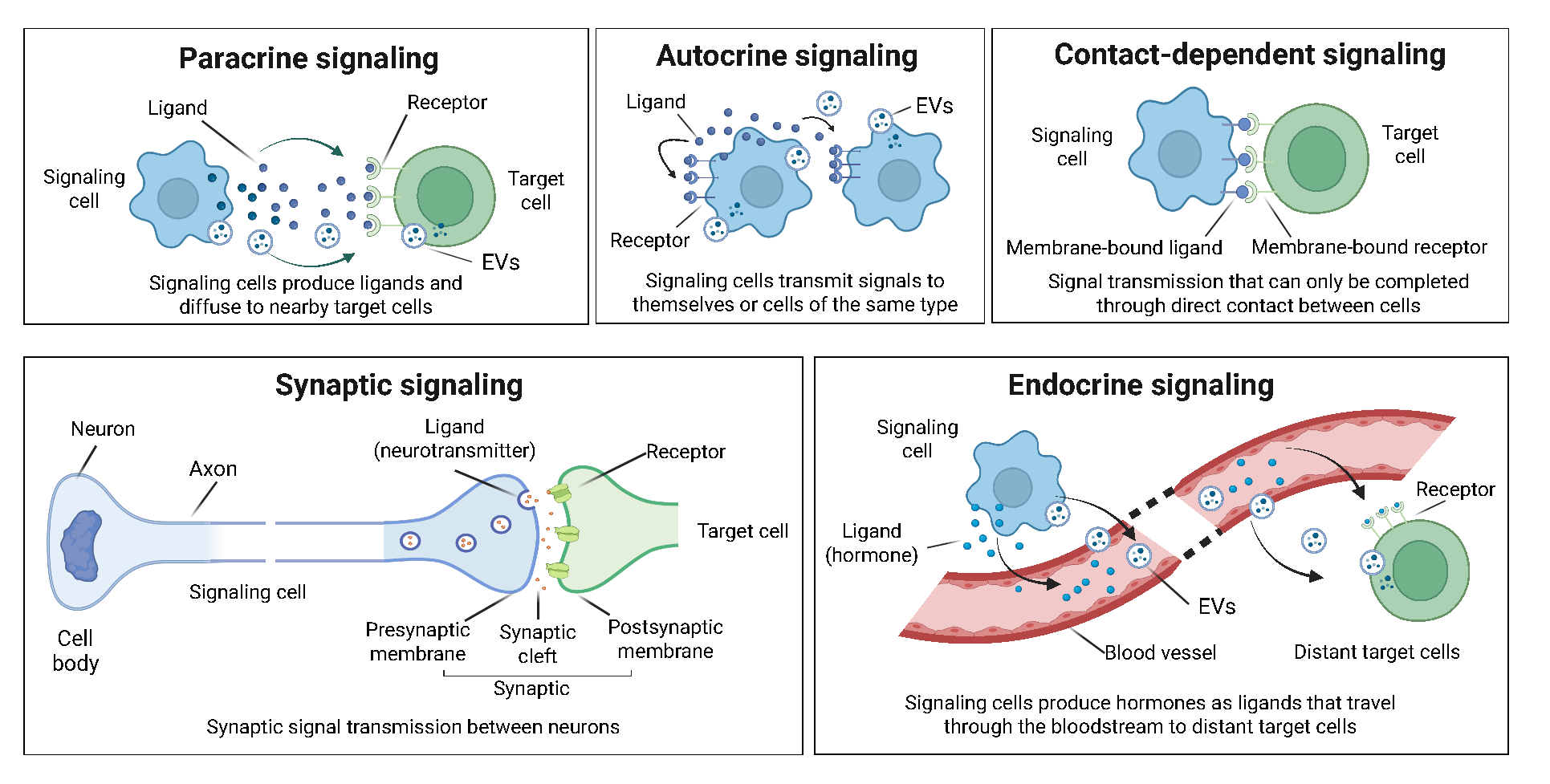}
  \caption{Diverse types of cell-cell communication. Paracrine signaling involves signaling molecules that act on neighboring cells within a localized area by diffusing through the extracellular space. Autocrine signaling refers to signals that act on the same cell that produces them or on a population of identical cell types. Contact-dependent signaling relies on direct physical contact between neighboring cells, such as direct membrane contact or gap junctions. Synaptic signaling enables long-distance communication through the specialized synaptic structures of neurons. Endocrine signaling involves signaling molecules produced by signaling cells that reach distant target cells via extracellular fluids, such as blood. Additionally, extracellular vesicles (EVs) also mediate intercellular signaling by delivering a range of bioactive molecules (e.g., proteins, lipids, nucleic acids) through paracrine, autocrine, and endocrine mechanisms. This figure is inspired by a previous work\cite{alberts2022molecular}. Created in BioRender. S. Jin, 2026, https://BioRender.com/trsm0kl.}\label{fig1}
\end{figure}

CCC can be broadly classified into five types based on underlying principles: paracrine, autocrine, contact-dependent, synaptic, and endocrine signaling (\Cref{fig1}). These categories are primarily distinguished by the distance over which ligands act to reach their target cells\cite{clark2018biology,LimMayer2024,atakan2014molecular}.

Paracrine signaling is a short-range mechanism of CCC in which ligand-producing cells secrete signaling molecules that diffuse locally and act exclusively on neighboring target cells, without requiring direct physical contact, as illustrated in \Cref{fig1}. In this mode of signaling, ligands released by signaling cells traverse the extracellular matrix (ECM) and bind specifically to receptors on adjacent target cells. The spatial distribution of ligands within the ECM forms a concentration gradient that is essential for effective signal transmission, enabling target cells to generate differential responses depending on ligand concentration. After signal transduction, ligands are normally quickly degraded by enzymes or removed by neighboring cells. These regulatory processes help maintain ligand concentration homeostasis within the ECM, thereby minimizing interference with subsequent paracrine signaling. Prompt removal of ligands following signal activation also limits their spatial spread, reducing the likelihood of erroneous interactions with non-cognate target cells outside the intended signaling domain.

Autocrine signaling is a specialized form of paracrine signaling in which cells secrete signaling molecules that act on themselves or on other cells of the same type (\Cref{fig1}). In a population of genetically identical cells, the collective signal intensity produced by the group inherently exceeds that of a single cell, thereby enhancing coordinated population-level behavior. During early embryonic development, autocrine signaling plays a critical role in promoting cellular proliferation and differentiation, as well as maintaining a supportive pregnancy-associated microenvironment. In contrast, dysregulated autocrine signaling is strongly associated with the uncontrolled proliferation of cancer cells. Excessive autocrine production of growth factors can drive pathological hyperproliferation, thereby contributing to oncogenesis.

Contact-dependent signaling, also known as juxtacrine signaling, occurs through direct physical contact between the signaling and target cells\cite{fagotto1996cell} (\Cref{fig1}). This form of communication can be classified into three major types. The first involves membrane-bound signaling molecules on the surface of the signaling cell that specifically bind to receptors on the plasma membrane of an adjacent target cell. The second type entails the direct delivery of ligands from the signaling cell into the intracellular space of the target cell, where they subsequently interact with receptors. The third mechanism involves the formation of intercellular channels that facilitate the direct transfer of small signaling molecules between the cytoplasms of neighboring cells. Well-characterized examples include gap junctions in animal cells and plasmodesmata in plant cells. Notably, large macromolecules such as proteins and nucleic acids are generally unable to pass through these channels.

The signal transduction mechanisms described above primarily involve short-range communication. However, long-distance CCC is essential for coordinating organism-wide biological activities, as illustrated in \Cref{fig1}. The human nervous system represents the most sophisticated form of long-distance signaling, comprising vast networks of neurons that communicate via specialized synaptic structures. Synaptic signaling enables rapid, long-distance, yet contact-dependent communication between neurons\cite{kornberg2014communicating}. Although the cell bodies of communicating neurons may be distant, their elongated axons and dendrites transport signals directly to specialized sites of cell-cell contact. This architecture ensures efficient, spatiotemporally precise information transfer regardless of the distances between the cell bodies of contacting cells. A synapse is a highly specialized structure for information transfer, comprising a presynaptic membrane, a synaptic cleft, and a postsynaptic membrane (\Cref{fig1}). Upon arrival of an electrical impulse at the synaptic terminal, neurotransmitters are released from synaptic vesicles into the cleft. These diffusing neurotransmitters then bind to receptors on the postsynaptic membrane, triggering a new electrical signal in the target cell. This process mediates a critical bidirectional conversion between electrical and chemical signaling.

Another major form of long-distance signaling is endocrine signaling (\Cref{fig1}). In this process, endocrine cells secrete hormones that are transported through the bloodstream to distant target cells to convey regulatory information. Unlike synaptic signaling, which transmits biological information within milliseconds via direct neuron-to-neuron communication, endocrine signaling depends on circulatory distribution, resulting in significantly slower transmission speeds\cite{desjardins1981endocrine}. The signaling molecules involved in this pathway—collectively known as hormones—are typically characterized by low water solubility, a property that facilitates their long-distance transport in the blood. In contrast, water-soluble signaling molecules are generally involved in paracrine signaling. Despite its slower onset, endocrine signaling exerts more prolonged effects on target cells. Even a small quantity of hormone molecules can elicit sustained regulatory responses, playing a vital role in the modulation of diverse physiological processes.

Notably, beyond classical ligand-receptor interactions, extracellular vesicle (EV)-mediated CCC has garnered significant interest for its potential applications as novel disease biomarkers and drug delivery systems. EVs are membrane-bound vesicles released by cells via plasma membrane budding and fission or multivesicular body formation, encapsulating diverse bioactive molecules such as proteins, lipids, and nucleic acids. Once released into the extracellular space, EVs interact with the extracellular and pericellular matrix, facilitating their transport to recipient cells through autocrine, paracrine, or endocrine routes\cite{van2022challenges}. Upon reaching a target cell, EVs exert their influence through three primary mechanisms: direct binding to cell-surface receptors, fusion with plasma membrane to deliver cargo, or internalization into the cytoplasm \cite{gurung2021exosome}. Unlike conventional signaling, EV-mediated communication enables the coordinated delivery of multiple signaling molecules, potentially eliciting more complex, pleiotropic responses in recipient cells\cite{liu2023review}. This multifaceted interaction repertoire introduces an additional layer of complexity to CCC networks.

\subsection{Signal Transduction in Cell-Cell Communication}

\begin{figure}[!tbh]
  \centering
  \includegraphics[scale=0.72]{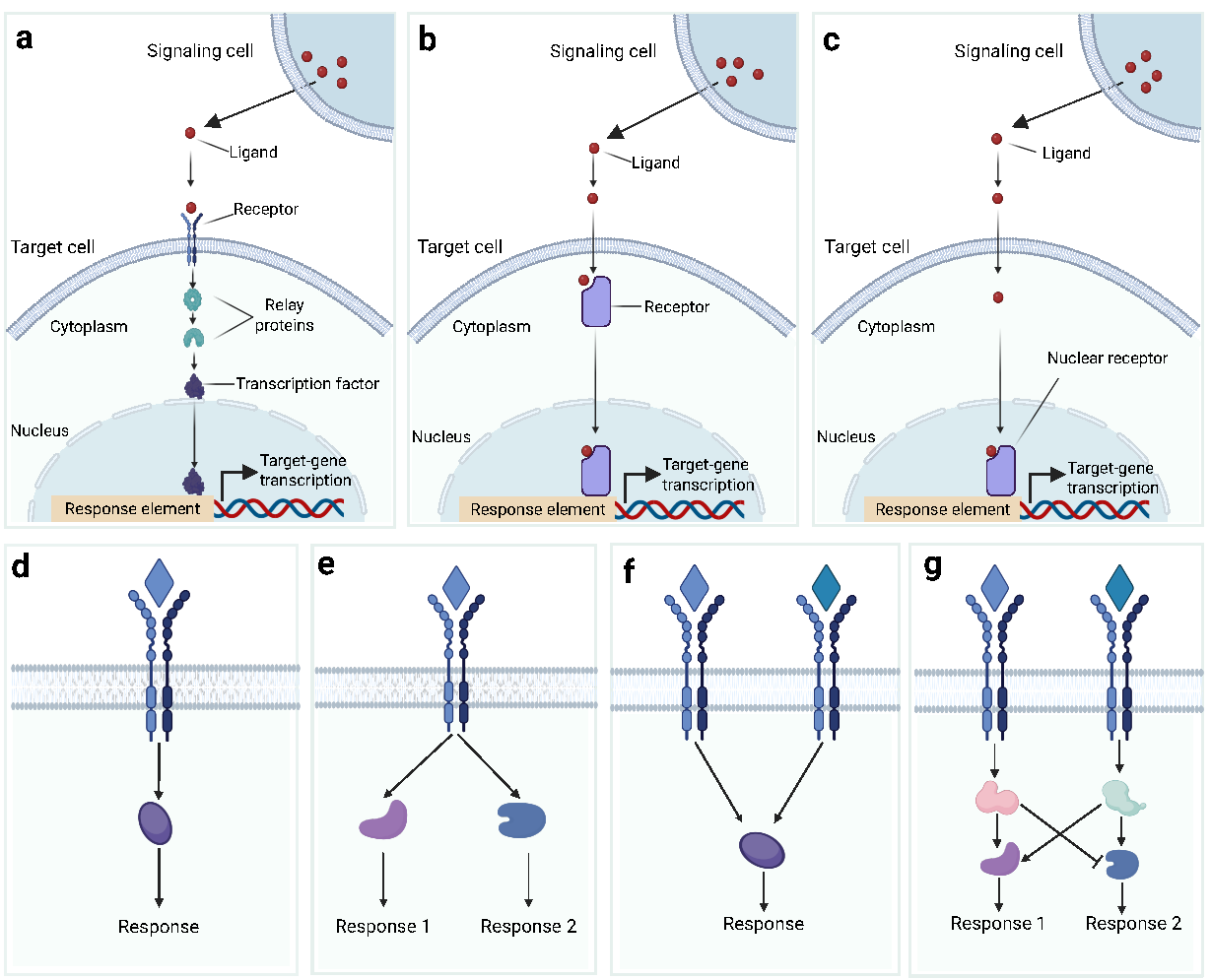}
  \caption{Cellular signal transduction mediated by different receptor types and major modes of intracellular signaling crosstalk. (a) Cell-surface receptor signaling: Ligand binding activates a cascade of cytoplasmic relay proteins and transcription factors, resulting in the nuclear translocation of transcription factors to regulate target gene expression. (b) Cytoplasmic receptor signaling: The ligand diffuses across the plasma membrane and binds its receptor in the cytoplasm. The resulting ligand-receptor complex then translocates to the nucleus, where it binds specific DNA regulatory sequences to direct transcriptional regulation. (c) Nuclear receptor signaling: The ligand diffuses across the plasma membrane and into the nucleus, where it binds a nuclear receptor. The ligand-receptor complex then directly modulates gene transcription. (d) Single-pathway activation: A lone intracellular signaling pathway is initiated by ligand-receptor binding. (e) Multiple-pathway activation: A single ligand-receptor pair triggers several distinct signaling pathways. (f) Signal convergence: Multiple, distinct ligand-receptor pairs activate a common downstream signaling pathway. (g) Pathway crosstalk: Interdependent signaling pathways, triggered by different ligand-receptor pairs, mutually influence one another. This figure is inspired by a previous work\cite{Avior2013} and Pearson Education, Inc. Created in BioRender. S. Jin, 2026, https://BioRender.com/xa3ut41.}
  \label{fig2}
\end{figure}

Upon ligand binding, a receptor transduces information from the signaling cell by initiating an intracellular signaling cascade that ultimately regulates gene expression in the target cell. This regulatory process, known as signal transduction, operates through distinct mechanisms depending on the receptor type. For signaling molecules that bind cell-surface receptors, the receptor typically orchestrates a cytoplasmic cascade of relay proteins. This cascade modulates target gene expression by activating specific transcription factors (TFs) within the nucleus (\Cref{fig2}a). In contrast, intracellular receptors—often TFs themselves—are located in the cytoplasm or nucleus and are activated by hydrophobic ligands that are able to diffuse across the plasma membrane (\Cref{fig2}b and c). Following ligand binding, the ligand-receptor complex directly interacts with specific DNA regulatory sequences to activate or repress transcription.

The signaling cascade involves intracellular signal propagation. Rather than proceeding as a simple linear chain, intracellular signaling operates as a complex regulatory network composed of various signaling molecules. These molecules can be broadly classified into second messengers and signaling proteins. Second messengers are small intracellular molecules that mediate rapid and broad signal dissemination. In contrast, signaling proteins are large macromolecules primarily responsible for highly specific signal transduction events. Upon receptor activation, second messengers transmit signals by rapidly altering their intracellular concentrations, thereby modulating the activity of downstream effectors. Signaling proteins, due to their large molecular size, undergo minimal concentration changes but exert their effects by activating downstream signaling components or modulating second messenger levels\cite{newton2016second}.

Furthermore, during intracellular signal transduction, the addition or removal of phosphate groups plays a critical role in regulating most signaling processes. As a result, signaling proteins often function as molecular switches. Two major classes of these switches are proteins modified through phosphorylation of specific amino acid residues by upstream kinases, and G proteins, which function by cycling between an active GTP-bound state and an inactive GDP-bound state. In addition, specific interactions among signaling proteins further facilitate the transmission and modulation of intracellular signals. Ultimately, the signaling cascade regulates the expression of target genes via activated TFs, thereby inducing a range of biological responses, including cell proliferation, differentiation, and metabolic reprogramming\cite{nishi2015crosstalk}.

\subsection{Signal Crosstalk in Cell-Cell Communication}

Signaling pathways do not operate in isolation but function within an integrated network of protein–protein interactions. The signaling mode in which intracellular pathways influence one another is referred to as signaling crosstalk, which helps to integrate signals from multiple inputs in different ways, giving rise to the vast range of cellular responses. Signaling crosstalk typically manifests in several distinct patterns\cite{RN442,RN444,RN445} (\Cref{fig2}d-g). The first pattern is single signal response, wherein ligand-receptor binding activates only one intracellular signaling pathway (\Cref{fig2}d). The second pattern is signal bifurcation, in which a single ligand-receptor pair simultaneously activates multiple signaling pathways (\Cref{fig2}e). For instance, fibroblast growth factors (FGFs) can activate the RTK pathway, the STAT pathway, and an additional pathway involving lipid metabolism remodeling and increased intracellular calcium levels\cite{RN439}. The third pattern is convergent signaling, where distinct ligand-receptor pairs converge to activate a shared intracellular pathway, such as during T lymphocyte differentiation\cite{RN439} (\Cref{fig2}f). The fourth pattern is modulatory crosstalk, where separate pathways initiated by different ligand-receptor pairs interact such that one pathway modulates—either enhances or inhibits—the transmission of another (\Cref{fig2}g). For example, Hh signaling can potentiate Wnt pathway activity, while Wnt signaling, in turn, modulates Hh effectors—a dynamic interplay essential in tissue regeneration and cancer progression\cite{orzechowska2023sonic}.

\section{Modeling Strategies for Cell-Cell Communication }\label{sec3}

\begin{figure}[p]
  \centering
  \includegraphics[scale=0.41, trim=20 0 15 0, clip]{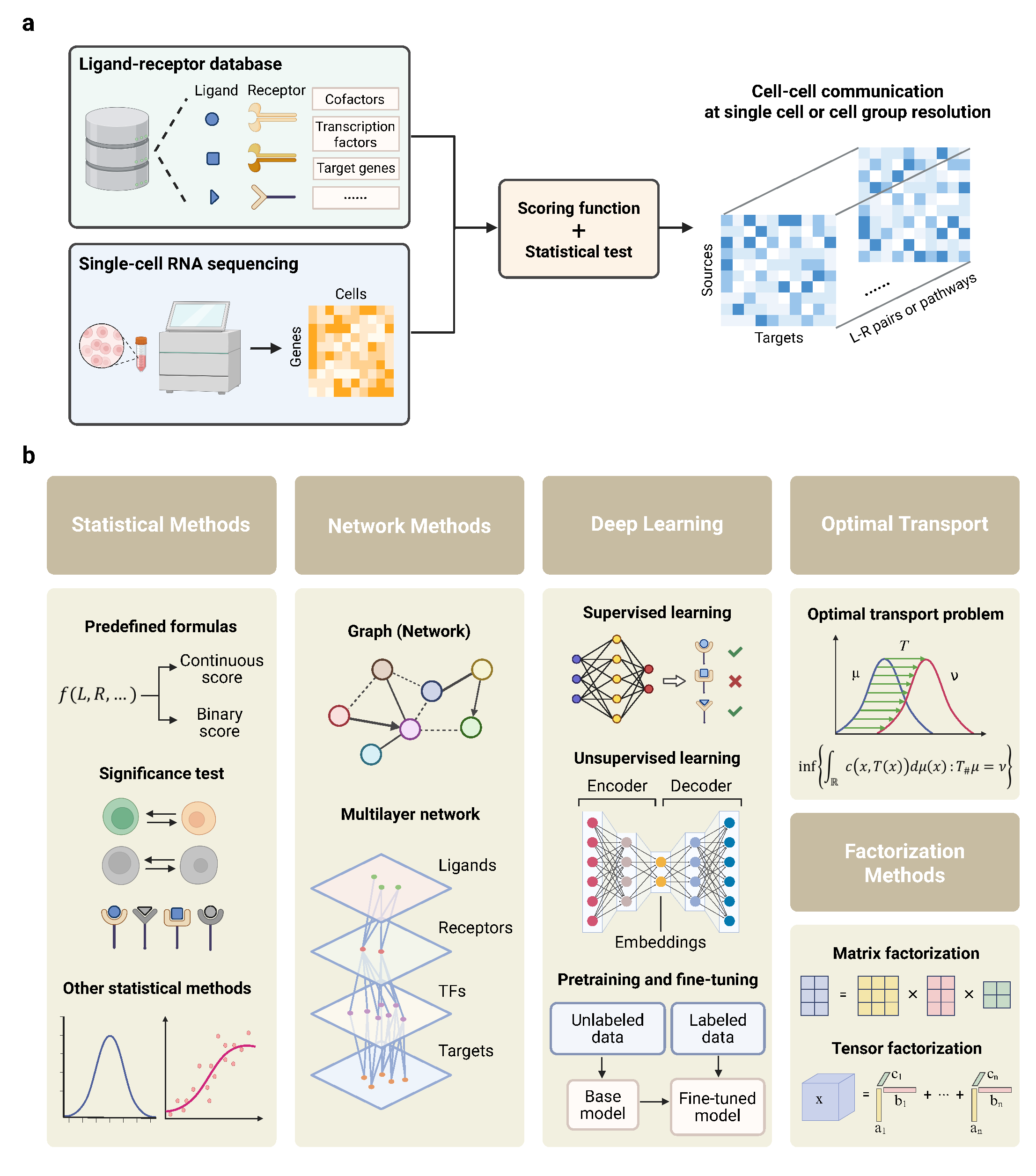}
  \caption{General strategies and computational methodologies for inferring CCC from gene expression data. (a) A typical workflow for CCC inference. Gene expression profiles from single-cell or spatial transcriptomics serve as proxies for protein abundance. These data are integrated with prior knowledge, including genes encoding interacting ligands and receptors, co-factors, transcription factors and target genes, to infer CCC. A scoring function quantifies the likelihood or strength of potential CCC events, followed by statistical testing to assess significance. The final output predicts CCC at single-cell or cell-group resolution. (b) Common computational methodologies. The five primary strategies for CCC inference include statistical methods, network methods, deep learning, optimal transport, and factorization methods. Created in BioRender. S. Jin, 2026, https://BioRender.com/llyvm5g.}
  \label{fig3}
\end{figure}

Based on the biological principles of cellular signal transduction, computational strategies have been developed to systematically investigate and characterize CCC from single-cell and spatial omics data. These computational approaches typically require at least two types of input data, as illustrated in \Cref{fig3}a. The first is the gene expression profile obtained from single-cell transcriptomics or spatial transcriptomics, and the second is prior knowledge of LRIs curated from experimental literature and public databases. Gene expression levels serve as indirect proxies for protein abundance, while the prior knowledge is utilized to identify and extract genes encoding interacting proteins from the expression data, thereby narrowing the analysis to protein activities that are potentially involved in mediating CCC.

To infer the CCC between cells, a commonly adopted assumption is that the likelihood of communication between a sender and a receiver cell is positively correlated with the expression levels of the ligand and its receptor in the corresponding cell. A scoring function is thus constructed to quantify the communication probability or strength mediated by each ligand-receptor pair between different cells or cell types (\Cref{fig3}a). To enhance the robustness of the inference and reduce the impact of background noise, statistical methods such as permutation testing are subsequently employed to identify statistically significant CCC.

Moreover, sophisticated computational methods can be developed by considering diverse biological mechanisms and experimental designs. First, many ligand-receptor interactions operate in a multi-subunit architecture, and therefore their heteromeric complexes should be accurately represented. Second, cofactors, including soluble agonists and antagonists, and co-stimulatory and co-inhibitory membrane-bound receptors, introduce additional effects on the core interaction between ligands and receptors, which can be incorporated into CCC modeling. Indeed, many signaling pathways, such as BMP, WNT and TGF-$\beta$, are prominently modulated by their cofactors, both positively and negatively\cite{heldin2016signals}. Third, downstream signaling response is often a key indicator of whether a cell has genuinely received a signal, greatly helping to mitigate false positives and establish causal relationships in signal transduction. Fourth, since secreted signaling and contact-dependent signaling operate over different spatial ranges, spatial distance should also be taken into account. Finally, comparative analysis across complex experimental designs—including multiple biological replicates and conditions—is essential. For example, analyzing temporal data can reveal the dynamic evolution of CCC, while cross-condition comparisons enable the systematic identification of altered signaling mechanisms and potential therapeutic targets.

\section{Computational Methods for Inferring Cell-Cell Communication}\label{sec4}

\begin{figure}[p]
  \centering
  \includegraphics[scale=0.855]{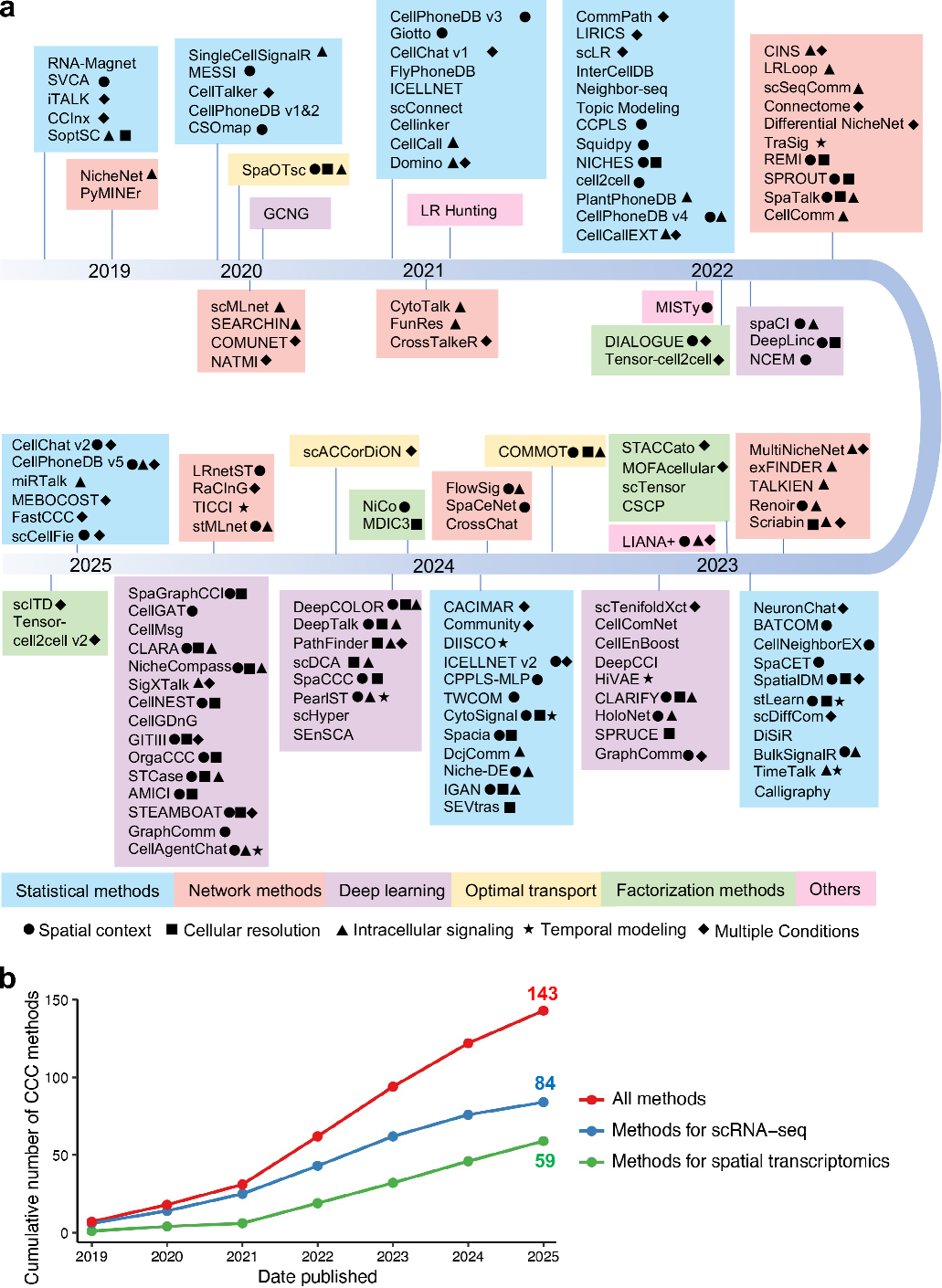}
  \caption{Timeline and growth of CCC inference methods. (a) The timeline of 143 CCC inference methods organized by publication or preprint year. Tools based on different methodological principles are distinguished using colored boxes, including statistical methods, network methods, deep learning methods, optimal transport methods, and factorization methods. Symbols denote the specific biological questions addressed. Methods without any symbols are general tools for inferring CCC between cell groups from scRNA-seq data. (b) Cumulative growth of CCC methods. The plot shows the cumulative number of CCC tools developed for scRNA-seq and spatial transcriptomics data over time. The number of methods published up to October 2025 is indicated.}
  \label{fig4}
\end{figure}

\begin{figure}[p]
  \centering
  \includegraphics[scale=0.839]{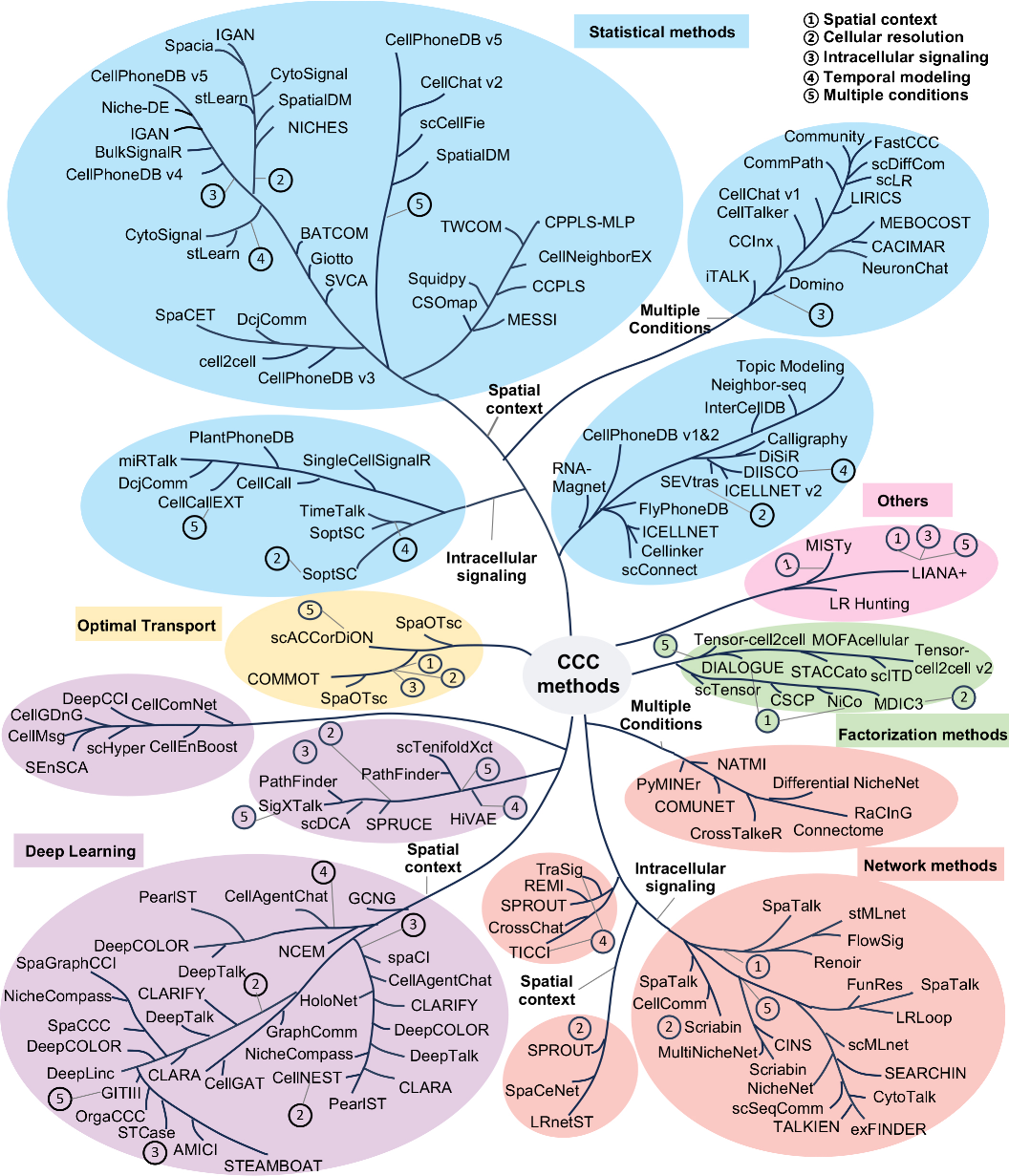}
  \caption{Methodological phylogeny of CCC tools. The evolutionary tree classifies 143 computational methods by their core computational strategies (main branches, colored) and the specific biological questions they address (sub-branches). This visualization, inspired by a previous review\cite{RN447}, illustrates the evolutionary relationships and functional diversity of CCC tools.}
  \label{fig5}
\end{figure}

The advent of single-cell and spatially resolved omics technologies, especially transcriptomics, has driven an abundance of computational tools for modeling CCC (\Cref{fig4}). These tools exhibit growing methodological diversity and vary in the biological features they are designed to investigate (\Cref{fig5}). This diversity reflects ongoing efforts and innovations by researchers to explore CCC mechanisms from multiple perspectives and to develop analytical approaches tailored to specific biological questions. Existing methods for CCC analysis can be broadly classified into five categories based on their core inference strategies (\Cref{fig3}b): (1) statistical methods, (2) network methods, (3) deep learning methods, (4) optimal transport methods, and (5) factorization methods. For each category, we first outline the core modeling principles and then evaluate how current tools address key biological questions. This evaluation is structured around five analytical aspects: spatial constraints, cellular resolution (e.g., single-cell level or spot-level resolution), intracellular signaling, temporal dynamics, and cross-condition comparison.

\subsection{Statistical Methods}

Statistical methods can be broadly classified into two major strategies(\Cref{fig3}b). The first and most prevalent strategy employs customized scoring functions to quantify the interaction strength between cell pairs, followed by statistical hypothesis testing or thresholding to evaluate significance. The second strategy utilizes more sophisticated statistical models to characterize CCC patterns.

Tools implementing the first strategy primarily diverge in their design of CCC scores and statistical tests. Some apply explicit rules to constrain signaling gene expression, producing binary interaction scores\cite{RN332,RN334,RN370}. More commonly, tools generate continuous scores by calculating the mean, product, or correlation of ligand and receptor expression\cite{RN263,RN338,RN277,RN343,RN266,RN313,RN344,RN289,zheng2022mebocost,jakobsson2021scconnect,RN380,RN414,shao2025extracellular,burdziak2023epigenetic}. The significance of these scores is typically assessed against a null distribution by randomly permuting cell labels. Specific scoring functions reflect distinct biological considerations. CellPhoneDB\cite{RN343}, for instance, defines a communication score as the mean of the average ligand expression in one cluster and the average receptor expression in another. To account for protein complexes, it uses the minimum average expression among subunits of ligands or receptors. Other tools incorporate more refined biological models. Given the additional modulation of co-factors (e.g., agonists, antagonists, co-receptors) on the core interaction between ligands and receptors, CellChat\cite{RN342} estimates the level of multimeric ligands and receptors by the geometric mean of their subunits, and designs its scoring function based on the law of mass action to integrate all known molecular interactions. Specifically, the communication probability $P_{i,j}$ from cell groups $i$ to $j$ for a ligand-receptor pair $k$ is modeled as:

\begin{equation}\label{eq1}
  \begin{aligned}
    P_{i,j}^k &= \frac{L_i R_j}{K_h+L_i R_j}\times\left(1+\frac{AG_i}{K_h+AG_i}\right)\cdot\left(1+\frac{AG_j}{K_h+AG_j}\right)\\
              &\quad \times\frac{K_h}{K_h+AN_i}\cdot\frac{K_h}{K_h+AN_j}\times\frac{n_i n_j}{n^2},\\
    L_i &= \sqrt[m_1]{L_{i,1}\dots L_{i,m_1}},\ R_j=\sqrt[m_2]{R_{j,1}\dots R_{j,m_2}}\cdot\frac{1+RA_j}{1+RI_j}.
  \end{aligned}
\end{equation}
Here, $L_i$ represents the expression level of ligand $L$ in cell group $i$, and $L_{i,1},\dots ,L_{i,m_1}$ denote the expression levels of the $m_1$ subunits of ligand $L$ in cell group $i$. Receptor $R$ and its $m_2$ subunits in cell group $j$ are defined analogously. $RA_j$ and $RI_j$ represent the average expression levels of co-stimulatory and co-inhibitory receptors in cell group $j$, respectively, while $AG$ and $AN$ denote the average expression levels of soluble agonists and antagonists. $n$ is the total number of cells in the dataset, while $n_i$ and $n_j$ denote the numbers of cells in cell groups $i$ and $j$, respectively. $K_h$ denotes the parameter in the Hill function.

The second strategy employs more advanced statistical methods to model CCC\cite{RN354,RN355,RN414,RN275BulkSignalR,RN356cell2cell,RN367MESSI,RN386}. Several tools within this category estimate communication strength through formal statistical frameworks. For instance, CCPLS\cite{RN355} applies partial least squares (PLS) regression, interpreting regression coefficients as the influence of neighboring cell types—mediated by CCC—on highly variable gene expression. FastCCC\cite{RN414} offers a highly scalable, permutation-free approach that uses fast Fourier transformation (FFT)-based convolution for rapid p-value calculation and a modular scoring system. Other tools utilize statistical methods primarily for the interpretation of inferred interactions. For example, after evaluating the significance of LRIs using null distributions of Spearman correlation coefficients between ligands and receptors, as well as between receptors and their downstream target genes, BulkSignalR\cite{RN275BulkSignalR} associates LRIs with specific cell types by constructing a LASSO regression model. Building upon communication score computation, cell2cell\cite{RN356cell2cell} employs a genetic algorithm to prioritize ligand-receptor pairs that best correlate with physical distance. Additionally, MESSI\cite{RN367MESSI} identifies both intracellular and intercellular signaling genes using a Mixture of Experts (MoE) model with multi-task learning. Topic Modeling\cite{RN386} infers genes affected by intercellular interactions through latent Dirichlet allocation (LDA). Collectively, these tools highlight the diversity of statistical approaches used in CCC analysis.

Spatial omics provides an unprecedented opportunity to systematically investigate CCC on native tissues. Incorporating spatial context into CCC analysis allows the calculation of communication scores to be constrained within a short distance range, aligning with the intuitive assumption that cells in close proximity are more likely to interact. Giotto\cite{RN364Giotto} identifies significantly interacting cell type pairs by randomly permuting cell type labels within a predefined spatial neighborhood network. Communication scores are then calculated only for cell pairs deemed spatially proximal and likely to interact based on cell type. CellNeighborEX\cite{RN358CellNeighborEX} leverages spatial transcriptomics data to identify neighbor-dependent gene expression by comparing transcriptomes in heterotypic versus homotypic cellular contexts, providing an unbiased approach to uncover genes regulated by direct cell contact and the immediate microenvironment, thereby revealing a layer of intercellular communication that complements and extends beyond canonical ligand-receptor analysis. SpaCET\cite{RN374SpaCET} deconvolves immune and stromal cell lineage fractions in tumors by integrating cancer-specific genomic patterns and hierarchical regression, and it infers CCC by requiring both spatial colocalization and significant ligand-receptor co-expression within spots, providing a spatially grounded framework to analyze interactions in the tumor microenvironment. Niche-DE\cite{RN423Niche-DE} is a regression-based statistical framework that identifies cell-type-specific genes whose expression is modulated by the local cellular context, and reveals LRIs that drive niche-differential expression patterns from spatial transcriptomics data.

Most tools compute communication scores using gene expression profiles aggregated at the cell group level. Although this strategy helps mitigate the sparsity of scRNA-seq data, it fails to fully exploit the advantages of single-cell resolution. Consequently, it cannot capture cell-specific interactions within cell groups and may introduce bias due to predefined cell clusters. To address these limitations, several recent methods have been developed to infer interactions directly between individual cell (or spot) pairs\cite{RN303, RN410CytoSignal, RN432Spacia}. stLearn\cite{RN382stLearn} integrates spatial information to compute communication scores in both `within-spot' and `between-spot' modes, while also accounting for cell type diversity to facilitate analyses across diverse biological contexts. Spacia\cite{RN432Spacia} directly models CCC at single-cell resolution using a multi-instance learning framework and explicitly captures complex many-to-one relationships between sender and receiver cells. Conceptually different from other CCC inference methods, SpatialDM\cite{RN378SpatialDM} explicitly incorporates spatial distance through a spatial weight matrix. It identifies significant ligand-receptor pairs and their local interaction hotspots using a bivariate Moran's I statistic, bypassing the need for permutation tests via an analytical null distribution. This makes SpatialDM both statistically rigorous and highly scalable to datasets comprising millions of cells. CytoSignal\cite{RN410CytoSignal} predicts signaling locations and activity at single-cell resolution by computing a weighted average of ligand expression across a cell’s neighborhood and then multiplying by the receptor expression in the target cell.

Based on the established importance of analyzing CCC, understanding the subsequent intracellular signaling cascades and transcriptional regulation is crucial for deciphering the complete cellular response mechanism. To bridge this gap, computational methods increasingly leverage curated pathway databases, such as KEGG\cite{kanehisa2000kegg} and Reactome\cite{croft2010reactome}, to connect LRIs to potential downstream genes through hypothesis testing or correlation analysis\cite{RN282,RN312,RN286,RN345}. Among these tools, SingleCellSignalR\cite{RN286} infers CCC from scRNA-seq data using a curated ligand-receptor database and a regularized score that provides a stable threshold to control false positives, facilitating the identification of high-confidence interactions for downstream validation. A more integrated approach is exemplified by CellPhoneDB v5\cite{RN316CellPhoneDBv5}. It extends beyond LRI identification with its CellSign module\cite{RN345}, which is designed to connect interactions to intracellular signaling and transcriptional outcomes. This module utilizes curated knowledge of receptor-TF relationships and downstream TF activity data to model signal propagation. DcjComm\cite{RN411} is a versatile computational tool that employs an NMF-based joint learning model to simultaneously identify functional gene modules, perform dimension reduction, and cluster cells, followed by a statistical model to infer CCC by integrating ligand-receptor pairs with downstream TF-target gene activities. Other tools incorporate the expression of downstream genes directly into the calculation of communication scores to provide a more holistic inference of CCC, such as SoptSC\cite{wang2019cell}, CellCall\cite{RN268}, BulkSignalR\cite{RN275BulkSignalR} and IGAN\cite{RN417IGAN}. The recent method BulkSignalR\cite{RN275BulkSignalR} addresses the challenge of inferring CCC from bulk and spatial multi-cellular data by statistically integrating LRIs with downstream pathway activity, providing a robust and functionally contextualized approach to decipher cellular networks from these prevalent data types. IGAN\cite{RN417IGAN} leverages spatial transcriptomics to construct intercellular gene association networks between adjacent cells by performing a non-parametric statistical model. It further elucidates the full upstream and downstream pathway context and thereby reveals communication heterogeneity and pathway-level mechanisms.

Incorporating temporal dynamics into CCC analysis enables a more nuanced understanding of cell state transitions and extends the analytical scope of computational tools, which is particularly valuable in contexts such as embryonic development, cancer progression, and other temporally evolving biological processes. stLearn\cite{RN382stLearn} introduces a framework that integrates gene expression with spatial information to reconstruct spatiotemporal trajectories, thereby uncovering spatial branching events that occur over time within tissues. Beyond trajectory-focused tools\cite{RN382stLearn,wang2019cell}, TimeTalk\cite{RN282} combines trajectory inference with causal testing to identify regulatory relationships between ligand-receptor pairs and temporally relevant TFs during early embryogenesis. DIISCO\cite{RN413}, on the other hand, utilizes Gaussian process regression models to characterize dynamic changes in cell type proportions and to predict time-resolved intercellular communication patterns.

Advances in experimental technologies have increased the availability of multi-condition datasets, spurring demand for computational methods that compare CCC across biological contexts. Such comparative analysis is crucial for identifying altered signaling mechanisms that drive cell fate decisions and phenotypic transitions. Common strategies for this task involve coupling differential CCC network analysis with differential gene expression profiling\cite{RN332,RN301,zheng2022mebocost,zhao2023inferring}, or applying statistical tests (e.g., Fisher’s exact test, permutation tests) to communication scores\cite{RN334,RN263,RN338,RN407}. Beyond these approaches, CellChat\cite{RN342,jin2025cellchat} employs network topology analysis, using centrality measures to identify key signaling sources, targets, mediators, and influencers within a network, and further groups pathways based on functional and topological similarity. Several tools are specifically designed for differential CCC analysis. scDiffCom\cite{RN263} provides a statistically rigorous framework to identify dysregulated LRIs and cell-type pairs between conditions (e.g., aging or disease), powering the scAgeCom atlas of age-related communication changes in mouse tissues. Community\cite{RN407} uniquely integrates cell type abundance, active fraction, and expression level to pinpoint altered interactions and disentangle compensatory mechanisms in case-control studies. CellCallEXT\cite{gao2022cellcallext} extends its precursor CellCall\cite{RN268} by incorporating transcription factor activity alterations between conditions to infer disease-associated changes in both intercellular communication and intracellular signaling. To address scalability in large-scale studies, FastCCC\cite{RN414} introduces a permutation-free framework for analytical p-value computation. Its key innovation is a reference-based paradigm that leverages a large human atlas, providing a healthy baseline to robustly identify dysregulated communications in disease. CACIMAR\cite{RN403} proposes a conservation score to identify CCC events that are conserved across different species. For spatial transcriptomics, SpatialDM\cite{RN378SpatialDM} uses generalized linear models to test if experimental conditions significantly influence interaction density across multiple samples.

\subsection{Network Methods}

Network methods exploit the connectivity structure of CCC by representing cells (or cell types) and genes as nodes and depicting interactions as directed edges, with edge weights corresponding to interaction strengths (\Cref{fig3}b). These methods elucidate patterns of intercellular communication by analyzing the topological and statistical properties of nodes and edges using graph theory, network analysis, and probabilistic graphical models.

The properties of graphs and networks enable intuitive and precise modeling of gene regulatory processes, which is particularly beneficial for tools that integrate downstream signaling and represent genes as nodes. Some tools\cite{RN306,RN302,RN283,RN377SpaTalk} focus on analyzing intracellular signaling while other tools build comprehensive signaling networks that integrate both intercellular and intracellular communication\cite{RN340,RN281,RN401,RN349,sang2025unraveling,RN350,RN454stMLnet}. These tools infer directed regulatory networks from ligands to downstream genes by integrating prior biological knowledge with gene expression data to identify specifically expressed genes or significant signaling pathways, thereby enabling network pruning. For example, to predict ligand-target links between interacting cell types, NicheNet\cite{RN349,sang2025unraveling} employs the Personalized PageRank (PPR) algorithm to compute a regulatory potential score for all ligand-target gene pairs from prior knowledge sources. This approach prioritizes ligands and their potential target genes by using the ligand of interest as the seed node and calculating a signaling importance score for each gene within the ligand-signaling network. The core approach PPR is formulated as:

\begin{equation}\label{eq2}
  v=\left(1-d\right)\times u + d\times W\times v.
\end{equation}
Here, $v$ denotes the vector of importance scores for all nodes in the network, derived from the steady-state distribution of the random walker; $u$ represents the personal preference for each node, with a value of 1 assigned to the ligand of interest and 0 to all other genes; $W$ is the normalized adjacency matrix of the weighted ligand-signaling network. By applying PPR to each ligand and performing a cutoff, the $n\times m$ ligand-gene signaling importance score matrix is obtained. This matrix is then multiplied by the $m\times m$ weighted adjacency matrix of the gene regulatory network, yielding a $n\times m$ ligand-target regulatory potential score matrix $L$, where $n$ is the number of ligands considered and $m$ is the number of possible target genes. Let $l_{ij}$ denote the regulatory potential score of ligand $i$ for target gene $j$, which can be computed as follows:

\begin{equation}\label{eq25}
  l_{ij} = \sum_{k=1}^m\left(PPR_{ik}\times GRN_{kj}\right),
\end{equation}
with $PPR_{ik}$ the importance of gene $k$ in the signaling of ligand $i$ and $GRN_{kj}$ the edge weight from gene $k$ to target gene $j$.

To account for spatial constraints of CCC, methods have been developed to emphasize CCC analysis within proximal cells or spots\cite{RN306,RN371Renoir,RN373SpaCeNet,RN377SpaTalk,RN379,RN454stMLnet,RN415FlowSig,RN418,RN409}. For example, SpaTalk\cite{RN377SpaTalk} scores the ligand-receptor-target signaling network between spatially proximal cells by constructing a KNN cell graph network and a prior ligand-receptor-TF knowledge graph and estimating the probability of downstream TF activation via a random walk algorithm. Different from SVCA\cite{RN385SVCA} that models the expression of genes independently of each other, SpaCeNet\cite{RN373SpaCeNet} employs a Gaussian graphical model to decompose the observed cellular profiles into contributions arising from cellular variability and cellular interactions, which captures complex multivariate relationships between genes, and disentangles the intracellular from intercellular gene correlations using spatial conditional independence. Renoir\cite{RN371Renoir} maps spatially resolved ligand-target activities and identifies communication niches by integrating spatial transcriptomics with scRNA-seq data, using a comprehensive neighborhood scoring system that accounts for cellular composition, receptor expression, and spatial proximity to reveal context-specific signaling microenvironments. Building on the earlier method scMLnet\cite{RN350}, which constructs a multilayer signaling network comprising three subnetworks including ligand-receptor, receptor-TF and TF-target gene, stMLnet\cite{RN454stMLnet} models the spatial-temporal distribution of ligand concentration $u(x, y, z)$ during the diffusion process using a partial differential equation (PDE), and then computes the signaling strength of ligand-receptor pairs between specific cell pairs. The PDE used to model the ligand diffusion process is formulated as follows:

\begin{equation}\label{eq26}
\frac{\partial u(x,y,z)}{\partial t} = D\Delta u(x,y,z),\quad (x,y,z)\in\mathbb{R}^3\backslash B_1,
\end{equation}
where $\Delta$ is the Laplace operator; $D$ is the diffusion coefficient; $B_1$ represents a unit ball indicating the sender cell. Incorporating the law of mass action, the signaling strength $LRS_j^k$ of the ligand-receptor pair $k$ at the receiver cell $j$ is defined as follows:

\begin{equation}\label{eq27}
LRS_j^k = \sum_{i=1}^n\left(\frac{1}{d_{ij}}L_i^kR_j^k\right).
\end{equation}
Here $L_i^k$ and $R_j^k$ are the corresponding ligand expression in the sender cell $i$ and receptor expression in the receiver cell $j$, respectively. $d_{ij}$ represents the relative distance between the cell $i$ and the cell $j$. Subsequently, the nonlinear regulatory relationship between ligand-receptor signaling activity and target gene expression is modeled using random forest regression. Suppose the inferred multilayer network contains $m$ target genes, and each target gene $TG_t$ is linked to $n_t$ ligand-receptor pairs. Accordingly, $m$ random forest regression models are constructed—one for each of the $m$ target genes—as follows:

\begin{equation}\label{eq28}
TG_t = f_t\left(LRS^{1_t}, LRS^{2_t}, \dots, LRS^{n_t}\right),\quad t = 1, 2, \dots,m.
\end{equation}
The signaling activities $LRS^{1_t}, LRS^{2_t}, \dots, LRS^{n_t}$ across the receiver cells are used as input to random forest regression model $f_t$ to predict the expression of $TG_t$.

Unlike methods that rely on prior pathway annotations, CytoTalk\cite{RN318} performs \textit{de novo} prediction of complete signal transduction pathways mediated by LRIs between two cell types from single-cell transcriptomics. It constructs two intracellular networks based on the mutual information of gene pairs—one for the sender and one for the receiver cells—and connects them via known LRIs. An optimal signaling network that considers genes with high cell type specificity and close connection to high-scoring ligand-receptor pairs is then identified by solving a network propagation-based prize-collecting Steiner forest problem. In addition, LRLoop\cite{RN288} detects feedback signaling loops formed by pairs of LRIs through a network propagation algorithm. Collectively, these methods highlight the increasing diversity and sophistication of modeling strategies tailored to specific biological questions.

Leveraging network-based frameworks, TraSig\cite{RN292} and TICCI\cite{RN434} address intercellular interactions among dynamically transitioning cells during differentiation and development. TraSig first employs a probabilistic graphical model to infer the pseudotemporal ordering of cells, and subsequently applies a sliding window approach to reconstruct gene expression profiles along each trajectory edge, thereby enabling the inference of LRIs between cell clusters that overlap in pseudotime. In contrast, TICCI integrates intercellular association probabilities and communication probabilities to construct a class k-nearest neighbor (KNN) graph with weighted edges, which is then used to assess cellular differentiation states and identify branching trajectories.

For datasets derived from different conditions, network topological metrics can be applied to compare differences between CCC networks\cite{RN280,RN319,RN273,RN397}.CINS\cite{RN285} is a multi-step framework that uniquely leverages cell-type proportion changes across conditions in scRNA-seq data to first infer a Bayesian network of differential cell-type interactions and then uses a constrained regression model to identify the key ligand-receptor pairs mechanistically responsible for these interactions, providing a powerful tool for hypothesis generation in case-control studies. Among these, COMUNET\cite{RN280} introduces a dissimilarity metric that accounts not only for the presence or absence of edges but also their weights and directions when comparing edge sets between two layers representing ligand-receptor pairs in a multilayer CCC network. When multiple samples are available, Scriabin\cite{RN352} is a computational framework that performs comparative CCC analysis at true single-cell resolution by constructing cell-cell interaction matrices, employing a binning strategy for large-scale comparisons, and identifying co-expressed interaction programs that are inferred using the Weighted Gene Correlation Network Analysis (WGCNA) framework. Scriabin powerfully reveals heterogeneous and rare communication events obscured by agglomerative methods. MultiNicheNet\cite{RN336} prioritizes differentially expressed and active ligand-receptor pairs across multi-condition and multi-sample datasets by aggregating multiple criteria—such as differential expression, cell type specificity, and sample-level expression—into a final weighted score.

\subsection{Deep Learning Methods}

Deep learning has demonstrated remarkable effectiveness across a wide range of applications in the field of single-cell and spatial omics, with recent increasing efforts of developing computational methods to characterize complex nonlinear relationships in CCC (\Cref{fig3}b). These methods excel at learning latent representations of cells and genes, potentially uncovering CCC patterns overlooked by traditional approaches. Particularly notable is the growing application of graph neural networks (GNNs), which naturally model structural relationships and dependencies in graphs representing cellular or gene interactions.

Spatial transcriptomics enables construction of cell-cell spatial graphs that provide contextual information for training neural networks\cite{RN375spaCI,RN363,RN365GraphComm,RN366HoloNet,RN369,RN404CellGAT,zohora2025cellnest,RN422NicheCompass,qi2025interpretable,Xiao20240821608964GITIII}. For instance, CellGAT\cite{RN404CellGAT} employs graph attention networks to integrate scRNA-seq data with protein interaction knowledge, predicting context-specific LRIs with multi-omics evidence. In contrast, DeepCOLOR\cite{RN328DeepCOLOR} uses a variational autoencoder to recover cell-cell colocalization networks at single-cell resolution by integrating single-cell and spatial transcriptomes.

Training neural networks to predict gene expression in receptor cells enables the inference of CCC and the understanding of how LRIs affect downstream targets.
HoloNet\cite{RN366HoloNet} and scDCA\cite{RN428scDCA} identify CCC events influencing transcriptional responses through multi-view networks that predict target gene expression and interpret the relative contribution of each view via attention scores. In HoloNet, each view corresponds to a CCC network defined by a specific ligand-receptor pair, whereas in scDCA, each view represents the interaction network between a specific pair of cell types. HoloNet represents the first deep learning method to incorporate detailed signaling mechanisms—including molecular diffusion, cofactors, and LRI-specific interaction ranges—building upon traditional approaches like CellChat. CellAgentChat\cite{RN324CellAgentChat} predicts gene expression profiles from ligand-receptor gene expression and infers regulatory relationships through receptor perturbation analysis. Other tools focus on reconstructing intracellular gene regulatory networks\cite{RN357CLARIFY,RN375spaCI,RN422NicheCompass,RN425PathFinder,RN426}. For example, spaCI\cite{RN375spaCI} maps genes into a latent space using adaptive graph models with attention mechanisms, and constrains interacting gene pairs to be positioned closer together, thereby facilitating the prediction of LRIs and their upstream regulators from the learned representations. CLARIFY\cite{RN357CLARIFY} reconstructs intracellular interaction networks using pre-inferred gene regulatory networks and provides downstream regulatory information for the inference of intercellular communication networks. CLARA\cite{RN406CLARA} is a spatially-aware, transformer-inspired method that infers CCC at the resolution of individual cell pairs by modeling contextual relationships between ligands and receptors within local cellular neighborhoods, providing a highly granular view of the cellular interactome. By integrating gene expression profiles with spatial locations, CCC networks can be directly reconstructed without prior knowledge of ligand-receptor pairs or restriction to immediate neighborhoods\cite{RN362DeepLinc,RN357CLARIFY,RN412DeepTalk,RN433,feng2025orgaccc}. For example, DeepLinc\cite{RN362DeepLinc} can reveal potential long-range interactions by using variational graph autoencoder with adversarial learning. NicheCompass\cite{RN422NicheCompass} interprets cellular interactions through decoder weights, and GITIII\cite{Xiao20240821608964GITIII} characterizes the influence of neighboring cells on the central cell using a single-layer graph transformer model.

To investigate the temporal dynamics of CCC, HiVAE\cite{RN394HiVAE}  and CellAgentChat\cite{RN324CellAgentChat} incorporate pseudotime trajectories inferred by existing tools to characterize cellular interactions. 
HiVAE quantifies information flow between different cell types by computing transfer entropy along the pseudotemporal ordering of cells from scRNA-seq data. CellAgentChat calculates communication scores only between cells within the same pseudotime bin or in adjacent bins, thereby reflecting the temporal proximity of interacting cells. It further trains a neural network to predict gene expression levels of each cell at the next time point, conditioned on the effects of cellular interactions. 

Currently, relatively few deep learning methods have been developed to identify differential CCC across biological conditions. Among existing approaches for scRNA-seq data, scTenifoldXct\cite{RN294scTenifoldXct} formulates CCC inference as a manifold alignment problem. It uses neural networks to learn low-dimensional representations of gene expression from two cell types—optionally incorporating known ligand-receptor pairs—and projects a coupled gene-pair similarity matrix from two samples into a unified latent space. Differential LRIs are then identified by computing Euclidean distances between these gene-pair representations. In a different strategy, PathFinder\cite{RN425PathFinder} employs a graph transformer, taking gene expression data and predefined gene-gene interaction paths as input, and is trained to predict the condition label of individual cells. The resulting path weights indicate the relative importance of each interaction in distinguishing between conditions.

Finally, we focus on the technical implementation of deep learning methods, highlighting architectural designs and training strategies. Current approaches employ diverse frameworks—including unsupervised and supervised learning strategies—implemented through variational autoencoders (VAEs)\cite{kingma2022autoencodingvariationalbayes}, graph neural networks (GNNs), and Transformers. When applied to spatial transcriptomics, these architectures capture nonlinear dependencies, spatial organization, and interaction ranges inherent to CCC. In unsupervised learning-based tools, most methods employ autoencoders—particularly variational autoencoders (VAEs)\cite{kingma2022autoencodingvariationalbayes}—to learn feature representations that capture potential CCC patterns\cite{RN328DeepCOLOR,RN272SPRUCE,RN394HiVAE,RN422NicheCompass,RN426,RN357CLARIFY,RN362DeepLinc,RN433,feng2025orgaccc,qi2025interpretable}. These models are typically optimized to minimize the discrepancy between the input and reconstructed data, and use the learned latent representations and reconstructed data to infer CCC. For instance, OrgaCCC\cite{feng2025orgaccc} leverages cellular gene expression profiles, cell-cell spatial graph, and known LRIs-derived gene–gene graph to reconstruct both the intercellular interaction network and the LRI network through two orthogonal graph autoencoders operating at the cell and gene levels, respectively. SPRUCE\cite{RN272SPRUCE} captures interaction topics for each cell-cell pair and identifies ligand and receptor genes enriched within each topic by employing an embedded topic model built upon a variational autoencoder framework, which achieves unbiased identification of interpretable cell states across multiple datasets by characterizing CCC patterns. Beyond autoencoders, scHyper\cite{RN429scHyper} is a hypergraph neural network-based method that models CCC as a global, high-order network, leveraging the discrepancy between nodes' intrinsic and contextual embeddings to reconstruct non-linear interaction scores and provide a systems-level view of cellular crosstalk. CellNEST\cite{zohora2025cellnest} adopts graph attention networks (GATs) along with contrastive learning to integrate ligand–receptor information with spatial position at single cell or spot resolution, where unsupervised training is performed by maximizing the Jensen-Shannon divergence between the original network and its corrupted counterpart. 
Finally, attention scores are used to quantify the probability of CCC, which is computed as follows:

\begin{equation}\label{eq3}
  \alpha_{i,j}=Tanh\left(a^{\mathsf{T}}\left[W_vh_i+W_vh_j+W_ee_{i,j}\right]\right).
\end{equation}
Here, $h_i$, $h_j$ are vertex feature vectors for vertices $i$ and $j$ of the input graph, where a vertex corresponds to a cell or spot; $e_{i,j}$ represents the edge feature vector from $j$ to $i$; $W_v$ is a learnable weight matrix, while $W_e$ is the equivalent matrix for edge features; the attention $a$ is a learnable parameter.

In addition to CellNEST, attention mechanisms have been effectively leveraged by several recent methods such as GITIII\cite{Xiao20240821608964GITIII}, AMICI\cite{RN456AMICI} and Steamboat\cite{LiangSteamboat}. GITIII\cite{Xiao20240821608964GITIII} infers spatially-resolved CCC at single-cell resolution from imaging-based spatial transcriptomics data by leveraging a single-layer graph transformer model. It models how a cell's transcriptional state is shaped by the cell type compositions and expression profiles of its spatial neighbors, providing a powerful solution for datasets with sparse ligand-receptor coverage. Building upon this concept, AMICI\cite{RN456AMICI} utilizes a sparsely regularized, multi-headed attention module to adaptively identify interactions across multiple spatial scales, delineate spatially dependent cellular sub-populations, and connect these interactions to downstream functional consequences in receiver cells. Furthermore, Steamboat\cite{LiangSteamboat}  applies a multi-head attention model to dissect cellular interactions across scales by decomposing a cell's gene expression into distinct components: intrinsic programs, local communication from neighbors, and long-range interactions. A distinctive feature of Steamboat is its capacity for in silico spatial perturbation, enabling the prediction of cellular responses to dynamic microenvironmental changes.

A fundamental challenge in applying supervised learning to CCC inference is the scarcity of gold-standard datasets with experimentally validated interactions, making reliable label definition difficult. To address this, several methods\cite{RN363,RN365GraphComm,RN391,RN284,RN404CellGAT,RN405CellMsg,RN430SEnSCA,peng2025predicting} curate positive labels from experimentally documented LRIs, treating all other pairs as negatives. The trained models are subsequently used to infer previously unannotated LRIs. For example, CellMsg\cite{RN405CellMsg} leverages graph convolutional networks on multimodal protein features to \textit{de novo} predict a high-confidence ligand-receptor interactome, which it then integrates with scRNA-seq data via a three-point estimation method to quantify communication strength, offering an enhanced, feature-driven approach to decipher CCC networks. Furthermore, SEnSCA\cite{RN430SEnSCA} constructs negative samples of LRIs by replacing the common practice of randomly selecting negative samples from unlabeled ligand-receptor pairs with K-means clustering, thereby reducing label noise and sample distribution bias that could weaken model predictive performance. Other tools adopt alternative strategies, such as thresholding communication scores or aggregating outputs from established statistical methods, to predefine interaction labels as ground truth\cite{RN375spaCI,RN325,RN429scHyper}. Another strategy shifts the prediction target from interactions themselves to downstream effects. In this approach, some models are trained to predict gene expression\cite{RN366HoloNet,RN369,RN324CellAgentChat} or cell type classifications\cite{RN425PathFinder,RN428scDCA}, with signaling activity subsequently assessed post-hoc through analysis of model weights or sensitivity analyses.

Recent advances have produced sophisticated hybrid models that integrate multiple deep learning architectures to address the complexity of CCC. Unlike single-architecture approaches, these hybrid frameworks offer modular and flexible designs that can be tailored to specific biological questions. For instance, DeepTalk\cite{RN412DeepTalk} employs a hybrid training strategy combining self-supervised pretraining with supervised fine-tuning to infer CCC at single-cell resolution. The model first learns global connectivity patterns through masked node prediction on large-scale cell graphs, then fine-tunes on specific datasets for binary edge classification, significantly enhancing generalization capability. GraphComm\cite{RN365GraphComm} represents another hybrid approach, constructing directed graphs from scRNA-seq data and applying GATs to integrate protein complex and pathway information. The method leverages a comprehensive knowledge base of over 30,000 protein interaction pairs to capture detailed cellular localization and intracellular signaling patterns, substantially improving CCC inference accuracy. The integration of large language models (LLMs) represents a particularly promising direction. SpaCCC\cite{RN431SpaCCC} fine-tunes pretrained LLMs on gene expression prediction tasks, enabling the embedding of ligands and receptors into a shared latent space. This unified embedding facilitates the identification of potential LRIs and subsequent CCC inference.

\subsection{Optimal Transport Methods}

Methods based on optimal transport conceptualize CCC as a resource allocation problem (\Cref{fig3}b). These methods typically infer the most probable communication paths between cells by minimizing the transportation cost of signaling molecules across spatial locations. SpaOTsc\cite{ RN376SpaOTsc} is a pioneering approach that formulates two sequential optimal transport problems to decode spatial cellular organization and communication. The method first aligns scRNA-seq data (source distribution) with spatial transcriptomics data (target distribution) by computing an optimal transport plan that minimizes the cost based on gene expression dissimilarity. The resulting optimal transport distance quantifies spatial proximity between cell pairs. Subsequently, SpaOTsc infers CCC networks by solving the second optimal transport problem where sender and receiver cells are modeled as source and target distributions, with intercellular distances defining the transport cost. The optimal transport plan directly yields the likelihood of communication for each cell pair, integrating both spatial constraints and expression levels of ligands, receptors, and downstream genes. In addition, SpaOTsc trains a random forest model to predict downstream gene expression, thereby estimating the spatial range over which specific signaling pathways operate. To account for the competition between different ligand and receptor species as well as spatial distances between cells, COMMOT\cite{cang2023screeningCOMMOT} develops a collective optimal transport method to handle complex molecular interactions and spatial constraints. Mathematically, the collective optimal transport problem is formulated as follows:

\begin{equation}\label{eq4}
  \begin{array}{c}
    \displaystyle
    \min_{P \in T} \quad
    \sum_{(i,j)\in I} \left\langle P_{i,j,\cdot,\cdot},\, C_{(i,j)} \right\rangle_F + \sum_i F(\mu_i) + \sum_j F(\nu_j), \\

    \displaystyle
    T = \left\{ P \in \mathbb{R}_+^{n_l \times n_r \times n_s \times n_s}:\ P_{i,j,\cdot,\cdot} = 0 \text{ for } (i,j) \notin I,\ \sum_{j,l} P_{i,j,k,l} \leq X^{L}_{i,k},\ \sum_{i,k} P_{i,j,k,l} \leq X^{R}_{j,l} \right\}, \\

    \displaystyle
    \mu_i(k) = X^{L}_{i,k} - \sum_{j,l} P_{i,j,k,l}, \quad
    \nu_j(l) = X^{R}_{j,l} - \sum_{i,k} P_{i,j,k,l},
  \end{array}
\end{equation}
where $X_{i,k}^L$ and $X_{j,l}^R$ represent the expression level of ligand $i$ on spot $k$ and the expression level of receptor $j$ on spot $l$. $F$ penalizes the untransported mass $\mu_i$ and $\nu_j$. $P_{i,j,k,l}$ is the coupling matrix scoring the signaling strength from spot $k$ to spot $l$ through the pair consisting of the ligand $i$ and receptor $j$, while $I$ denotes the index set of bindable ligand-receptor pairs. $C_{(i,j)}$ is the cost matrix based on the thresholded distance matrix, designed to impose spatial distance constraints. The collective optimal transport problem determines a collection of optimal transport plans for all pairs of ligand and receptor species that can be coupled simultaneously, thereby enabling the consideration of competition among different species.

Apart from modeling CCC via optimal transport, ScACCorDION\cite{RN427} focuses on characterizing changes in intercellular communication across multiple biological conditions. It represents the communication network between cell types within each sample as a directed weighted graph, and applies optimal transport to compute the Wasserstein distance between these graphs. Additionally, ScACCorDION identifies communication events that differ between sample groups by estimating the barycenters of a collection of CCC networks.

\subsection{Factorization Methods}

These methods infer CCC and their latent patterns using matrix or tensor factorization, capable of simultaneously analyzing multiple factors influencing CCC (\Cref{fig3}b). Among matrix factorization-based tools\cite{RN419,zhang2023defining,RN424Nico}, CSCP\cite{zhang2023defining} aims to reveal communication patterns between cell types at a fine-grained level. It identifies pairs of cell subgroups with strong and similar intercellular crosstalk signals by solving a coupled non-negative matrix factorization problem, which minimizes differences in ligand or receptor expression between cells while maximizing the activity scores of crosstalk signals mediated by ligand–receptor pairs. By integrating scRNA-seq data with spatial transcriptomics data, NiCo\cite{RN424Nico} captures intra-cell type variability through a non-negative matrix factorization framework based on the concept of niche composition, leading to the identification of latent factors for each cell type. These latent factors are then used to infer relevant signaling mediators and cell type pairs with covarying factors, thereby enabling the construction of intercellular interaction networks between cell types. 

Methods like DIALOGUE\cite{jerby2022dialogueDIALOGUE}, MOFAcellular\cite{flores2023multicellular} and scITD\cite{RN323} uncover latent multicellular programs from single-cell gene expression data by applying penalized matrix decomposition\cite{witten2009penalized}, multi-omics factor analysis (MOFA)\cite{argelaguet2018multi,argelaguet2020mofa} and Tucker tensor decomposition\cite{tucker1966some}, respectively. Potential signaling is then identified based on the ligands and receptors enriched in the multicellular programs. Notably, MOFAcellular is a flexible multi-view integration framework, which enables the inclusion of additional tissue-level descriptions in the model, such as cell type compositions, spatial relationships, and CCC scores.

By organizing the inferred CCC between cell types into a tensor, several methods, including scTensor\cite{RN399}, Tensor-cell2cell\cite{RN339} and STACCato\cite{Dai20231215571918}, have been proposed to uncover CCC patterns as well as their unique combinations of cell types and ligand-receptor pairs by using non-negative Tucker decomposition\cite{kim2007nonnegative}, CANDECOMP/PARAFAC (CP) decomposition\cite{carroll1970analysis,harshman1970foundations} and Tucker decomposition, respectively. Tensor-cell2cell is the pioneering method for uncovering context-driven CCC patterns by constructing a 4D-communication tensor. Mathematically, Tensor-cell2cell constructs a $C\times P\times S\times T$ fourth-order tensor $\chi$, where $C$, $P$, $S$, $T$ correspond to the number of contexts, ligand-receptor pairs, sender cells and receiver cells respectively. Tensor factorization is then performed by iteratively optimizing the following objective function until convergence:

\begin{equation}\label{eq5}
  \min_{c,p,s,t} \lVert\chi - \sum_{r=1}^{R} c^r \otimes p^r \otimes s^r \otimes t^r\rVert_F^2,
\end{equation}
where $\otimes$ denotes the outer product and $c^r$, $p^r$, $s^r$ and $t^r$ are vectors of the factor $r$, each containing the loadings of the respective elements along a specific dimension of the tensor, enabling the further identification and interpretation of context-dependent communication patterns. Following this idea, STACCato identifies condition-related CCC events while also explicitly adjusting for sample-level confounding variables (e.g., batch, age and gender). More recently, Tensor-cell2cell v2\cite{Armingol20221102514917} performs integrative analysis of protein- and metabo-litemediated CCC by using coupled tensor component analysis, thereby facilitating the joint interpretation of CCC patterns across multiple modalities.

\section{Intuitive Visualization and Systems Analysis of Cell-Cell Communication}\label{sec45}

\begin{figure}[p]
  \centering
  \includegraphics[scale=0.73]{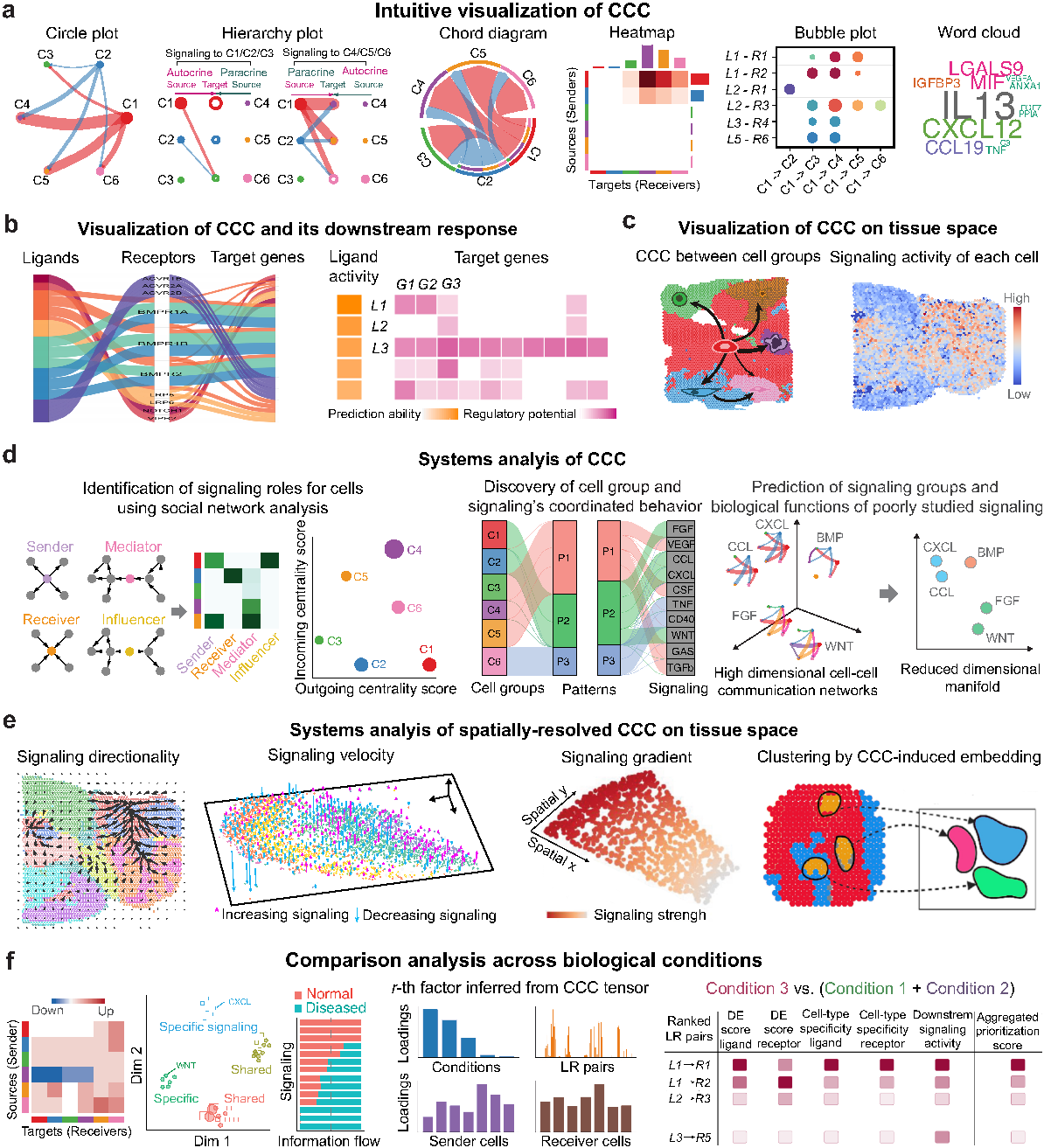}
  \caption{Visualization and analysis of CCC. (a) Common intuitive visualizations of inferred CCC networks. (b) The alluvial plot showing ligand–receptor–target gene relationships. (c) Spatial communication patterns depicting signaling directionality and strength on spatial tissues. (d) Examples of analysis techniques of CCC for scRNA-seq data, including identification of key signaling roles using network centrality analysis, pattern recognition to uncover coordinated cellular behaviors, and manifold learning to group functionally similar signaling pathways. (e) Examples of analysis techniques of CCC for spatial transcriptomics data, including inference of signaling directionality, mapping of signaling velocity and spatial gradients, and identification of microenvironment-induced cell states. (f) Comparative CCC analysis across biological conditions, highlighting differential communication patterns and context-specific interactions.}
  \label{fig45}
\end{figure}

While most computational tools focus primarily on inferring CCC, several methods provide sophisticated visualization and systems-level analysis capabilities to help interpret complex communication networks (\Cref{fig45}).

For network visualization, tools employ diverse graphical representations: CellChat\cite{RN342,jin2025cellchat} utilizes circle plots, hierarchical diagrams, chord diagrams, heatmaps, bubble plots, and word clouds to effectively illustrate intercellular communication patterns and highlight signaling variations across cell types (\Cref{fig45}a). Similarly, CellCall\cite{RN268} employs alluvial plots to trace cascading relationships from ligands through receptors to downstream target genes, while NicheNet\cite{RN349,sang2025unraveling} provides heatmaps displaying ligand-target regulatory potential and predicted ligand activity (\Cref{fig45}b). In spatial contexts, CellChat visualizes communication links directly on tissue architecture, and COMMOT\cite{cang2023screeningCOMMOT} maps signaling activity to individual cellular locations (\Cref{fig45}c).

Systems-level analysis applies quantitative approaches to decode complexity within CCC networks and identify emergent properties. CellChat employs network centrality metrics to identify key signaling sources, targets, mediators, and influencers within communication networks, revealing critical components of cellular microenvironments (\Cref{fig45}d). Pattern recognition through matrix factorization enables prediction of primary incoming/outgoing signals for specific cell types and uncovers coordinated responses across diverse populations. Furthermore, CellChat clusters signaling pathways by defining similarity measures and applying manifold learning from both functional and topological perspectives, facilitating identification of signaling groups with shared architectures and biological interpretation of less-characterized pathways.

Spatial analysis reveals distinctive organizational patterns of CCC (\Cref{fig45}e). COMMOT interpolates communication patterns into vector fields to visualize spatial directionality of signal transmission and reception. CytoSignal\cite{RN410CytoSignal} introduces signaling velocity by leveraging RNA velocity of ligands and receptors, enabling identification and visualization of dynamic signal intensity changes across tissue space. Meanwhile, MintFlow\cite{Akbarnejad2025} and GITIII\cite{Xiao20240821608964GITIII} generate microenvironment-induced embeddings that enable fine-grained cell type separation based on communication-driven cellular states.

Comparative analysis across biological conditions detects complex CCC changes under varying contexts. CellChat identifies altered signaling pathways and ligand-receptor pairs by analyzing differences in network architecture, information flow, and through joint manifold learning of intrinsic CCC structures (\Cref{fig45}f). Beyond the pairwise comparison, Tensor-cell2cell\cite{RN339} employs tensor decomposition to uncover context-driven communication patterns with unique combinations of cell types and ligand-receptor pairs, visualizing context loadings, ligand-receptor pair loadings, and sender/receiver cell loadings to facilitate pattern interpretation across diverse conditions (\Cref{fig45}f). STACCato\cite{Dai20231215571918} identifies condition-related CCC while adjusting for sample-level variables (e.g., batch, age, gender) using supervised tensor regression. MultiNicheNet\cite{RN336} addresses differential CCC analysis in complex multifactorial designs, accommodating inter-sample heterogeneity while correcting for batch effects and covariates through ranking-based prioritization integrated into an aggregated score that highlights significantly altered interactions for downstream interpretation (\Cref{fig45}f).

\section{Challenges and Opportunities}\label{sec5}

We have systematically surveyed over 143 computational tools for CCC analysis, categorizing them according to their underlying computational principles. While the field has diversified and advanced considerably in recent years, significant challenges remain alongside emerging opportunities.

\subsection{Resources, Validation and Benchmarking}

The development of CCC analysis tools relies on diverse prior knowledge resources and methodological frameworks. Systematic comparisons reveal that choices of resources and methods substantially impact inference results, thereby influencing downstream biological interpretations\cite{LIANA}.

Accurate recapitulation of molecular interactions is essential for biologically meaningful CCC prediction. However, ligand–receptor databases vary considerably in content—including multisubunit complex annotation, functional classifications (e.g., secreted signaling or contact-dependent signaling, molecular function and subcellular location), inclusion of intracellular signaling events, and the total number of curated interactions. Detailed comparisons of existing databases are available in a previous review\cite{cesaro2025advances}. In addition to the protein-mediated ligand-receptor databases, resources are emerging for metabolite–protein interactions (e.g., MACC\cite{gao2024macc}, MetalinksDB\cite{farr2024metalinksdb}, MRCLinkdb\cite{zhang2024predicting}) and EV-mediated interactions (e.g., EV-COMM\cite{chen2024ev}, miRTalkDB\cite{shao2025extracellular}). Since most tools rely on their own or referenced databases—and results are highly sensitive to database quality—researchers may draw different conclusions using different methods. A unified, well-annotated, high-quality interaction repository is urgently needed to standardize and enhance the reliability of CCC inference.

\begin{figure}[!tbh]
  \centering
  \includegraphics[scale=0.61, trim=20 0 20 0, clip]{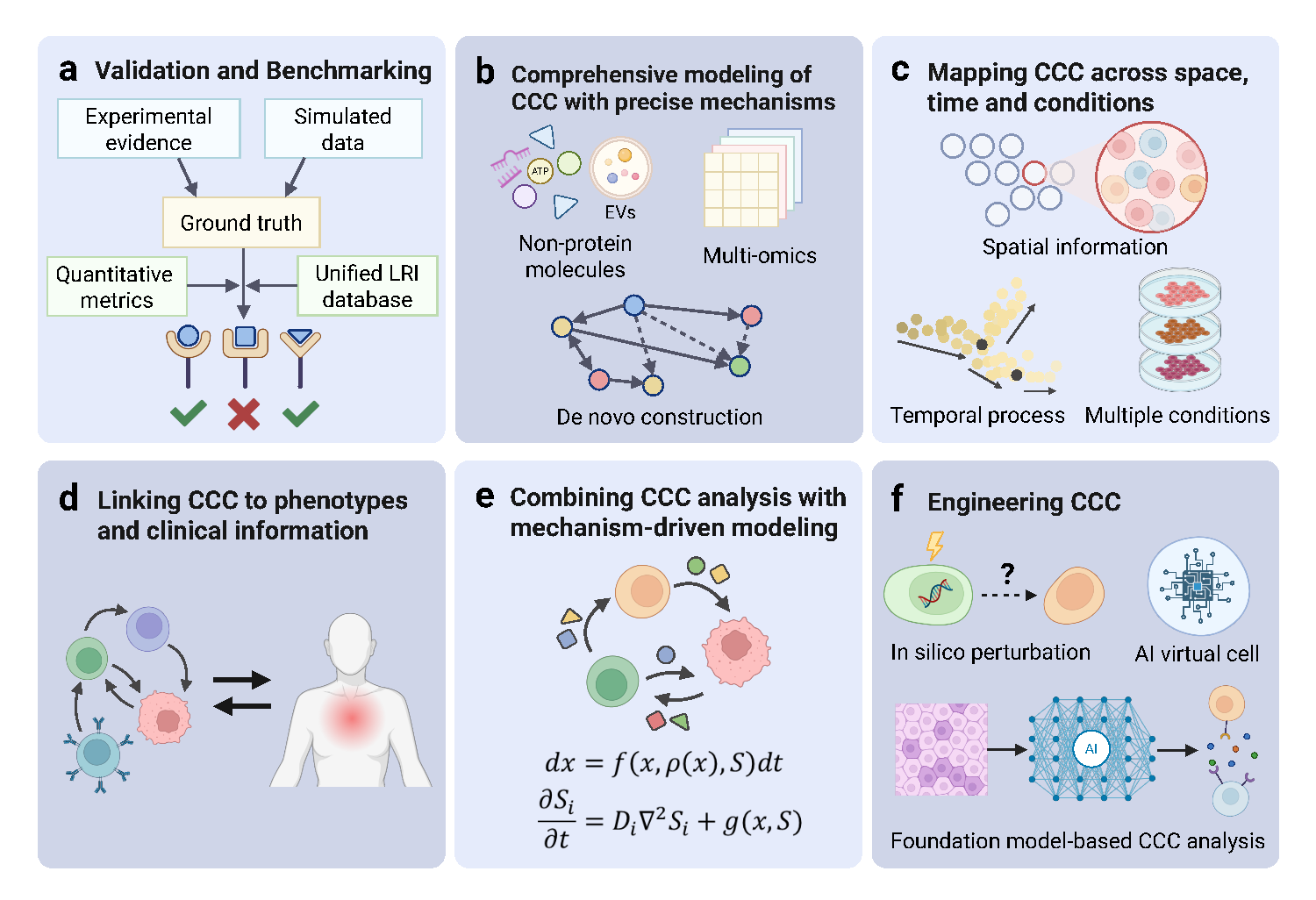}
  \caption{Challenges and opportunities in CCC analysis. (a) Validation and benchmarking challenges. Inferred CCC results are influenced by the choice of prior knowledge resources. However, the absence of experimental ground truth and standardized prior knowledge resources hinders systematic tool evaluation. (b) Modeling CCC with precise mechanisms and other modalities. Most CCC studies focus on protein-mediated ligand–receptor interactions, overlooking non-protein signaling and complex regulatory mechanisms. Advances in multi-omics technologies and \textit{de novo} network reconstruction now enable more comprehensive CCC inference and the discovery of previously uncharacterized CCC. (c) Spatiotemporal and cross-condition analysis. Integrating spatial and temporal context, along with comparative studies across conditions, can reveal deeper biological insights—yet these approaches face considerable technical and analytical hurdles. (d) Clinical and phenotypic integration. Linking phenotypic and clinical data to CCC helps identify key CCC in disease progression, supporting the development of precision medicine. (e) Mechanistic-informed modeling. Further refinement to incorporate communication more explicitly will not only improve the accuracy of the learned cellular dynamics but also offer new insights into temporal CCC dynamics and cellular responses to altered cellular communication.Mathematical models of signaling mechanisms allow researchers to probe regulatory dynamics under experimentally inaccessible conditions. (f) Engineering and intervention. Emerging computational approaches—including in silico perturbation, foundation models, and virtual cells—open new avenues for predictive and engineering-based CCC research, though model interpretability remains a critical challenge. Created in BioRender. S. Jin, 2026, https://BioRender.com/y1iwqii.}
  \label{fig6}
\end{figure}

A major challenge remains the scarcity of experimental ground truth for evaluating computational predictions (\Cref{fig6}a). Recent experimental advances—including barcode-based (BRICseq\cite{BRICseq2020}, RABID-seq\cite{RABID-seq2021}) and droplet-based (ProximID\cite{ProximID2018}, PIC-seq\cite{PIC-seq2020}) technologies—now enable tracing of sender–receiver contacts at single-cell resolution, thereby facilitating validation of predicted CCC. For instance, PIC-seq identifies context-specific cell interactions and differentially regulated genes, while SPEAC-seq\cite{wheeler2023droplet} captures soluble ligands to link sender-cell perturbations with receiver-cell responses. Although these methods improve measurement throughput and accuracy, most remain limited to direct cellular contacts and cannot simultaneously profile multiple ligand–receptor interactions or cell pairs. Further innovation in experimental techniques and curation of ground-truth datasets are essential for systematic validation of computational predictions.

Computational benchmarking using simulated data with known ground truth has been attempted in several studies\cite{li2025scmultisim,RN454stMLnet,cang2023screeningCOMMOT,tanevski2022explainable}. However, accurately simulating CCC remains challenging due to the complexity and redundancy of signaling in multicellular systems. Systematic tool assessment is further complicated by the diversity of underlying databases, biological assumptions, data compatibility, and parameter settings. Although recent benchmarking efforts have established more systematic evaluation frameworks—particularly for scRNA-seq-based methods\cite{RN305,benchmarkCCCWang}—development of objective and standardized quantitative metrics remains crucial. Accessible and unified platforms such as LIANA+\cite{dimitrov2024liana+} are emerging to address this need. With over 50 computational methods now available for spatial transcriptomics data, systematic benchmarking of these spatially aware tools is urgently required to guide future methodological development.

\subsection{Comprehensive modeling of CCC with precise mechanisms}

While most CCC inference methods focus on protein-mediated LRIs, non-protein signaling mechanisms, including those mediated by small molecules such as metabolites and neurotransmitters, also play vital roles in biological systems and require systematic investigation (\Cref{fig6}b). For instance, profiling enzymes and transporters involved in non-protein signaling enables inference of metabolite-mediated CCC\cite{zheng2022mebocost} or neuron–neuron communication\cite{zhao2023inferring,jakobsson2021scconnect} from scRNA-seq data. A central challenge in metabolite-mediated CCC analysis lies in accurately estimating metabolic activity. Recently, a scalable tool called scCellFie\cite{Armingol20250509653038} has been developed to infer metabolic activity per single cell or spatial spot from transcriptomic data. Non-protein signaling molecules exhibit greater diversity and more complex regulatory mechanisms than proteins, necessitating careful consideration in modeling.

Beyond metabolite signaling, EVs are increasingly recognized as key mediators of CCC in both prokaryotic and eukaryotic organisms, with considerable clinical potential as biomarkers and drug delivery vehicles\cite{van2022challenges}. Recent years have seen the emergence of databases, technologies, and computational tools to elucidate EV-mediated communication dynamics. For example, EV-COMM\cite{chen2024ev} compiles literature-curated EV-mediated interactions across cells and species, linking EV cargo to downstream pathways and functional outcomes. Similarly, miRTalkDB\cite{shao2025extracellular} catalogs experimentally supported EV-derived miRNAs and their target interactions in humans, mice, and rats, providing a foundation for inferring EV–miRNA-mediated CCC. Functional RNAs within EVs—such as mRNAs and miRNAs—can modulate recipient cell behavior and are critical to intercellular signaling. However, standard scRNA-seq protocols are optimized for cellular RNAs and often fail to capture EV RNAs effectively\cite{miceli2024extracellular}. Bridging this technical gap requires specialized experimental and computational workflows that account for the low abundance, heterogeneity, and fragmented nature of EV transcripts. Recent advances enable characterization of EV transcriptomic features, estimation of cellular secretion activity, and related analyses\cite{luo2022transcriptomic,he2024sevtras,Zhao20250624660984}. For instance, SEVtras\cite{he2024sevtras} identifies droplets containing small extracellular vesicles and infers a vesicle signal score to estimate single-cell secretion activity from scRNA-seq data. More recently, miRTalk\cite{shao2025extracellular} has emerged as the first method to systematically infer EV-derived miRNA-mediated CCC by integrating module scores for EV biogenesis, secretion signatures, RNA-induced silencing complex–related genes, and miRNA–target expression. Despite these advances, the mechanisms and applications of EV-mediated CCC remain incompletely understood. Further investigation—particularly into vesicle heterogeneity in spatial contexts—will deepen insights into EV-driven communication.

Most current CCC studies rely on mRNA expression as a proxy for protein abundance, an assumption often undermined by post-transcriptional regulation and complex protein trafficking processes. The advent of single-cell and spatial multi-omics technologies—including proteomics\cite{stoeckius2017simultaneous,peterson2017multiplexed}, epigenomics\cite{cusanovich2015multiplex}, and metabolomics\cite{lan2024single,mao2025single}—enables more comprehensive and granular characterization of CCC. Integrating single-cell ATAC-seq data, for example, can illuminate transcriptional regulation dynamics\cite{jin2020scai}. Moreover, intracellular post-translational modifications (PTMs) play central roles in signal integration and gene regulation. Emerging multimodal profiling technologies such as Phospho-seq\cite{blair2025phospho} and SIGNAL-seq\cite{Opzoomer20240223581433} allow measurement of PTMs, offering deep insights into cell signaling and transcriptional responses. SIGNAL-seq, in particular, enables simultaneous profiling of mRNA ligand–receptor pairs, intracellular PTMs, and transcriptional states in thousands of single cells. Expanding analytical tools to incorporate such multi-omics data will improve both inference and validation of CCC and support more effective benchmarking.

Advances in computational methods now enable more accurate modeling of complex CCC mechanisms, including signaling competition and cooperation, feedback regulation, and higher-order interactions. For example, COMMOT\cite{cang2023screeningCOMMOT} accounts for competition among ligands and receptors; SigXTalk\cite{hou2025dissecting} quantifies signaling crosstalk fidelity and specificity using scRNA-seq data; and LRLoop\cite{RN288} identifies intercellular feedback loops. A remaining challenge is the integration of diverse mechanisms to evaluate intercellular communication potential at a global scale. With continued research, dynamic and mechanistic models of CCC are expected to become feasible, promoting a systems-level understanding of cellular behavior.

Beyond conventional CCC inference, new computational approaches identify external signals not captured in the data, detect hierarchical communication structures, and reveal intercellular signaling flows. exFINDER\cite{RN281} infers external signals and their downstream networks by leveraging prior knowledge of ligand–receptor–transcription factor pathways and scRNA-seq data, addressing signals originating outside the profiled cellular population. CrossChat\cite{RN409} detects hierarchical structures in CCC through two complementary strategies: CrossChatH performs multi-resolution cell clustering to reveal global communicative hierarchies, and CrossChatT identifies local tree structures among ligands and receptors based on inclusive or disjoint expression patterns. To identify LRIs where intracellular processes mediate signal inflow and trigger outflow of other intercellular signals, FlowSig\cite{RN415FlowSig} uses graphical causal modeling with conditional independence tests to infer a completed partially directed acyclic graph (CPDAG) representing intercellular flows.

\subsection{\textit{De novo} construction of CCC}

\textit{De novo} construction of cellular signaling networks represents an important frontier to infer CCC directly from single-cell and spatial omics data without reliance on prior knowledge of ligand-receptor pairs. This approach offers potential to uncover novel signaling pathways and intercellular interactions mediated by complex mechanisms, thereby providing fresh biological insights. CytoTalk\cite{RN318} pioneered this paradigm by reconstructing signaling pathways from scRNA-seq data without using known pathway annotations. Its algorithm formulates this as a prize-collecting Steiner forest (PCSF) problem, identifying context-specific, parsimonious subnetworks that connect salient genes within and between cell types. The method operates on the rationale that cell-type-specific signals are encoded in highly expressed genes that are topologically proximate to active ligand-receptor pairs within a background protein-protein interaction network. While this approach reveals context-specific signaling cascades absent from canonical pathways, its utility is constrained by the completeness of the background network and computational demands for large datasets. For spatial transcriptomics with limited gene panels and inadequate ligand-receptor coverage, \textit{de novo} construction offers distinct advantages. Methods including DeepLinc\cite{RN362DeepLinc}, Spacia\cite{RN432Spacia}, GITIII\cite{Xiao20240821608964GITIII}, AMICI\cite{RN456AMICI} and Steamboat\cite{LiangSteamboat} infer CCC by integrating gene expression with spatial proximity without preselecting genes based on prior knowledge. However, omitting ligand-receptor information presents a fundamental challenge: establishing causal relationships for predicted interactions and distinguishing whether associated genes function as signal mediators or downstream targets.

\subsection{Mapping CCC across space, time and conditions}

Spatial context is crucial for accurate CCC inference (\Cref{fig6}c), yet current spatial transcriptomics technologies vary substantially in cellular resolution, gene coverage, and signal-to-noise ratio, creating significant challenges for single-cell resolution analysis. Low-resolution methods like Visium\cite{staahl2016visualization} capture mixed cell populations within individual spots, complicating heterogeneity studies, while high-resolution \textit{in situ} techniques such as Xenium and CosMx\cite{he2022high} offer single-cell resolution but suffer from limited transcript detection. The large scale, technical noise, and sparsity of high-resolution spatial data impose stringent requirements on CCC method design.

Temporal dynamics represent another critical dimension, enabling elucidation of communication mechanism evolution. Common approaches integrate cell lineage trajectory analysis with CCC inference to identify stage-specific LRIs along developmental trajectories\cite{RN282,RN292}, while other models leverage multi-timepoint data to capture interaction dynamics\cite{RN413}. However, transcriptomic data alone provides limited temporal resolution, as ligand and receptor expression may not coincide temporally with functional communication events, underscoring the need for multi-omics integration.

CCC events occur across diverse temporal scales, with different processes—ligand-receptor binding, signal transduction, transcription factor activation, and downstream responses—operating at distinct rates that reflect biological function and pathway specificity. Paracrine signaling, relying on molecular diffusion, proceeds slowly: a typical protein requires hours to diffuse 1 millimeter in aqueous environments. In contrast, endocrine signaling molecules transported via fluid flow in the circulatory system can cover the same distance in approximately 3 seconds\cite{muller2011extracellular}. Synaptic signaling operates even faster, with neuronal electrical impulses exceeding 100 m/s and neurotransmitter diffusion across synaptic clefts occurring within milliseconds\cite{purves2017neuroscience}. Receptor activation is typically rapid—G protein-coupled receptors activate within milliseconds to microseconds\cite{gruteser2022examination}. Cellular responses also vary considerably: protein modification-based responses (cell motility, secretion, metabolism) occur within seconds to minutes, while gene expression changes requiring new protein synthesis generally demand minutes to hours\cite{alberts2022molecular}. Integrating dynamic equations that capture these temporal behaviors would enable more precise modeling of communication dynamics and enhance biological interpretation.

Comparative CCC analysis across conditions reveals both conserved and context-specific signaling patterns, but batch effects and technical covariates from multi-individual, multi-tissue, or multi-condition datasets can obscure genuine biological variation. Consequently, developing robust batch correction strategies, modeling inter-individual variability, and accurately identifying biologically meaningful CCC changes remain substantial challenges.

\subsection{Linking CCC to phenotypes and clinical information}

Disease progression involves dynamic tissue alterations and complex clinical manifestations. Integrating phenotypic and clinical data with CCC analysis can reveal interactions that play crucial roles in disease initiation and progression, offering new opportunities for phenotype manipulation, prognostic biomarker discovery, and therapeutic targeting\cite{CCCreviewSu2024}. Such integration is essential for elucidating disease mechanisms and advancing precision medicine (\Cref{fig6}d). While some studies have identified phenotype-associated signaling pathways\cite{RN425PathFinder} or used ligand–receptor interactions for prognosis prediction\cite{yuan2019systematic}, a key challenge lies in the heterogeneity of phenotypic and clinical data, and how to effectively integrate such data with molecular-level communication information within computational models. Furthermore, since phenotypic changes often arise from coordinated activity across multiple pathways—and vice versa—methods capable of disentangling these complex regulatory relationships are needed.

\subsection{Combining CCC analysis with mechanism-driven modeling}

Mathematical modeling enables exploration of biological processes across parameter ranges often inaccessible experimentally, facilitating evaluation of system dynamics under diverse conditions\cite{daneshpour2019modeling} (\Cref{fig6}e). For instance, modeling revealed how IL-2’s dual role in T-cell proliferation and apoptosis contributes to immune homeostasis\cite{hart2014paradoxical}. Combining mechanistic models with CCC analysis can help interpret inferred signaling patterns, validate pathway functionality, and uncover how signaling dynamics influence system homeostasis or disease progression\cite{myers2021mechanistic}. This approach requires accurate representation of biological mechanisms and dynamic relationships, along with appropriate model assumptions—challenges that intensify in multi-scale systems involving diverse spatial and temporal dimensions.

Traditional mathematical modeling is typically limited to few cell types and pathways. Integrating genome-wide CCC analysis with mechanistic modeling remains challenging, though recent advances in learning cellular dynamics from single-cell or spatial transcriptomics show promise\cite{zhang2025integrating}. Current methods for inferring RNA velocity or reconstructing vector fields from snapshot data, however, often overlook intercellular communication. Emerging approaches address this gap: one study incorporated a transformer module to capture CCC while learning cellular vector fields\cite{jiang2025learning}, and another developed CytoBridge, an unbalanced mean-field Schrödinger bridge method, to model cellular dynamics and interactions from time-series single-cell data\cite{zhang2025modelingcelldynamicsinteractions,zhang2025deciphering}. These examples use neural networks to learn CCC without prior ligand–receptor information. Further refinement to incorporate communication more explicitly will not only improve the accuracy of the learned cellular dynamics but also offer new insights into temporal CCC dynamics and cellular responses to altered cellular communication.

\subsection{Engineering CCC}

Rapid advances in single-cell omics and artificial intelligence are fostering goal-directed, intervention-based approaches in CCC research, enabling deeper mechanistic insight and improved prediction. Key emerging directions include in silico perturbation, foundation model–based analysis, and virtual cell modeling (\Cref{fig6}f). In silico perturbation allows rapid assessment of genetic or chemical interventions on signaling and cell behavior, overcoming experimental limitations. For example, CellAgentChat uses an agent-based model to simulate perturbations via rule modifications, supporting novel intervention design\cite{RN324CellAgentChat}. In addition, foundation models pretrained on large-scale single-cell and spatial datasets excel at tasks like cell annotation, multi-omic integration, and perturbation prediction\cite{RN339LLM,hao2024large}, yet adapting them for CCC analysis remains difficult due to ground-truth scarcity and pathway complexity. Virtual cells—computational models simulating cellular behavior and interactions—can leverage AI to replicate molecular, cellular, and tissue dynamics across conditions\cite{bunne2024build}. However, robust evaluation frameworks and attention to interpretability are essential to ensure these methods reliably address biological questions.

\section{Conclusion}\label{sec6}

Cell–cell communication encompasses diverse, finely regulated mechanisms whose investigation holds significant promise for biomedicine and precision health. Transcriptomics-based computational methods for inferring CCC have advanced substantially, evolving from generic approaches to specialized tools addressing distinct biological questions through varied methodological principles. These developments support a more integrated and biologically meaningful understanding of communication networks. Despite this progress, important challenges and emerging opportunities remain, particularly given rapid innovations in spatial omics technologies and artificial intelligence, which are poised to offer transformative perspectives for future CCC analysis.

%%%% Acknowledgments %%%%%%%%
\section*{Acknowledgments}
The authors thank Dr. Peijie Zhou in Peking University and Dr. Axel Almet in University of California, Irvine for reading the manuscript and helpful suggestions.

The work was supported by a Major Research Plan of the National Natural Science Foundation of China (Grant No. 92374108). 

X. Cheng and H. Huang contributed equally to this work.

%%%% Bibliography  %%%%%%%%%%
% \bibliographystyle{abbrv}
\bibliographystyle{unsrt}
\bibliography{references}

@article {Akbarnejad2025,
	author = {Akbarnejad, Amir and Steele, Lloyd and Jafree, Daniyal J. and Birk, Sebastian and Sallese, Marta Rosa and Rademaker, Koen and Boxall, Adam and Rumney, Benjamin and Tudor, Catherine and Patel, Minal and Prete, Martin and Makarchuk, Stanislaw and Li, Tong and Stanley, Heather and Foster, April Rose and Roberts, Kenny and Trinh, Andrew L. and Villa, Carlo Emanuele and Testa, Giuseppe and Mahil, Satveer and Mehrjou, Arash and Smith, Catherine and Vakili, Sattar and Clatworthy, Menna R. and Mitchell, Thomas and Bayraktar, Omer Ali and Haniffa, Muzlifah and Lotfollahi, Mohammad},
	title = {Mapping and reprogramming microenvironment-induced cell states in human disease using generative {AI}},
	elocation-id = {2025.06.24.661094},
	year = {2025},
	doi = {10.1101/2025.06.24.661094},
	publisher = {Cold Spring Harbor Laboratory},
	URL = {https://www.biorxiv.org/content/early/2025/06/26/2025.06.24.661094},
	eprint = {https://www.biorxiv.org/content/early/2025/06/26/2025.06.24.661094.full.pdf},
	journal = {bioRxiv}
}

@article{STreview2021integrating,
   author = {Longo, Sophia K. and Guo, Margaret G. and Ji, Andrew L. and Khavari, Paul A.},
   title = {Integrating single-cell and spatial transcriptomics to elucidate intercellular tissue dynamics},
   journal = {Nature Reviews Genetics},
   volume = {22},
   number = {10},
   pages = {627-644},
   ISSN = {1471-0064},
   DOI = {10.1038/s41576-021-00370-8},
   url = {https://doi.org/10.1038/s41576-021-00370-8},
   year = {2021},
   type = {Journal Article}
}

@article{reviewMultiOmics2023,
   author = {Vandereyken, K. and Sifrim, A. and Thienpont, B. and Voet, T.},
   title = {Methods and applications for single-cell and spatial multi-omics},
   journal = {Nat Rev Genet},
   volume = {24},
   number = {8},
   pages = {494-515},
   ISSN = {1471-0064 (Electronic)
1471-0056 (Print)
1471-0056 (Linking)},
   DOI = {10.1038/s41576-023-00580-2},
   url = {https://www.ncbi.nlm.nih.gov/pubmed/36864178},
   year = {2023},
   type = {Journal Article}
}

@article{spatialomicsScience2023,
   author = {Bressan, D. and Battistoni, G. and Hannon, G. J.},
   title = {The dawn of spatial omics},
   journal = {Science},
   volume = {381},
   number = {6657},
   pages = {eabq4964},
   DOI = {10.1126/science.abq4964},
   url = {https://www.ncbi.nlm.nih.gov/pubmed/37535749},
   year = {2023},
   type = {Journal Article}
}

@article{STyuan2021,
   author = {Dries, R. and Chen, J. and Del Rossi, N. and Khan, M. M. and Sistig, A. and Yuan, G. C.},
   title = {Advances in spatial transcriptomic data analysis},
   journal = {Genome Res},
   volume = {31},
   number = {10},
   pages = {1706-1718},
   DOI = {10.1101/gr.275224.121},
   url = {https://www.ncbi.nlm.nih.gov/pubmed/34599004},
   year = {2021},
   type = {Journal Article}
}

@article{hou2025dissecting,
  title={Dissecting crosstalk induced by cell-cell communication using single-cell transcriptomic data},
  author={Hou, Jiawen and Zhao, Wei and Nie, Qing},
  journal={Nature Communications},
  volume={16},
  number={1},
  pages={5970},
  year={2025},
  publisher={Nature Publishing Group UK London}
}

@article{sang2025unraveling,
  title={Unraveling cell--cell communication with {NicheNet} by inferring active ligands from transcriptomics data},
  author={Sang-Aram, Chananchida and Browaeys, Robin and Seurinck, Ruth and Saeys, Yvan},
  journal={Nature Protocols},
  volume={20},
  pages={1439--1467},
  year={2025},
  publisher={Nature Publishing Group UK London}
}

@article{jiang2025learning,
  title={Learning collective multi-cellular dynamics from temporal {scRNA-seq} via a transformer-enhanced {Neural} {SDE}},
  author={Jiang, Qi and Zhang, Lei and Li, Longquan and Wan, Lin},
  journal={arXiv preprint arXiv:2505.16492},
  year={2025}
}

@article{zhang2025integrating,
  title={Integrating dynamical systems modeling with spatiotemporal {scRNA-seq} data analysis},
  author={Zhang, Zhenyi and Sun, Yuhao and Peng, Qiangwei and Li, Tiejun and Zhou, Peijie},
  journal={Entropy},
  volume={27},
  number={5},
  pages={453},
  year={2025}
}

@article{myers2021mechanistic,
  title={Mechanistic and data-driven models of cell signaling: Tools for fundamental discovery and rational design of therapy},
  author={Myers, Paul J and Lee, Sung Hyun and Lazzara, Matthew J},
  journal={Current opinion in systems biology},
  volume={28},
  pages={100349},
  year={2021},
  publisher={Elsevier}
}

@misc{kingma2022autoencodingvariationalbayes,
   title={Auto-Encoding Variational Bayes}, 
   author={Diederik P Kingma and Max Welling},
   year={2022},
   eprint={1312.6114},
   archivePrefix={arXiv},
   primaryClass={stat.ML},
   url={https://arxiv.org/abs/1312.6114}, 
}

@article{LIANA,
  title={Comparison of methods and resources for cell-cell communication inference from single-cell {RNA-Seq} data},
  author={Dimitrov, Daniel and T{\"u}rei, D{\'e}nes and Garrido-Rodriguez, Martin and Burmedi, Paul L and Nagai, James S and Boys, Charlotte and Ramirez Flores, Ricardo O and Kim, Hyojin and Szalai, Bence and Costa, Ivan G and others},
  journal={Nature communications},
  volume={13},
  number={1},
  pages={3224},
  year={2022},
  publisher={Nature Publishing Group UK London}
}

@article{li2025scmultisim,
  title={{scMultiSim}: simulation of single-cell multi-omics and spatial data guided by gene regulatory networks and cell--cell interactions},
  author={Li, Hechen and Zhang, Ziqi and Squires, Michael and Chen, Xi and Zhang, Xiuwei},
  journal={Nature Methods},
  volume={22},
  pages={982--993},
  year={2025},
  publisher={Nature Publishing Group US New York}
}

@article{cang2023screeningCOMMOT,
  title={Screening cell--cell communication in spatial transcriptomics via collective optimal transport},
  author={Cang, Zixuan and Zhao, Yanxiang and Almet, Axel A and Stabell, Adam and Ramos, Raul and Plikus, Maksim V and Atwood, Scott X and Nie, Qing},
  journal={Nature methods},
  volume={20},
  number={2},
  pages={218--228},
  year={2023},
  publisher={Nature Publishing Group US New York}
}

@article{tanevski2022explainable,
  title={Explainable multiview framework for dissecting spatial relationships from highly multiplexed data},
  author={Tanevski, Jovan and Flores, Ricardo Omar Ramirez and Gabor, Attila and Schapiro, Denis and Saez-Rodriguez, Julio},
  journal={Genome biology},
  volume={23},
  number={1},
  pages={97},
  year={2022},
  publisher={Springer}
}

@article{dimitrov2024liana+,
  title={{LIANA+} provides an all-in-one framework for cell--cell communication inference},
  author={Dimitrov, Daniel and Sch{\"a}fer, Philipp Sven Lars and Farr, Elias and Rodriguez-Mier, Pablo and Lobentanzer, Sebastian and Badia-i-Mompel, Pau and Dugourd, Aurelien and Tanevski, Jovan and Ramirez Flores, Ricardo Omar and Saez-Rodriguez, Julio},
  journal={Nature Cell Biology},
  volume={26},
  number={9},
  pages={1613--1622},
  year={2024},
  publisher={Nature Publishing Group UK London}
}

@article{zheng2022mebocost,
   author = {Zheng, Rongbin and Zhang, Yang and Tsuji, Tadataka and Gao, Xinlei and Shamsi, Farnaz and Wagner, Allon and Yosef, Nir and Cui, Kui and Chen, Hong and Kiebish, Michael A and Aristizabal-Henao, Juan J and Narain, Niven R and Zhang, Lili and Tseng, Yu-Hua and Chen, Kaifu},
   title = {{MEBOCOST} maps metabolite-mediated intercellular communications using single-cell {RNA-seq}},
   journal = {Nucleic Acids Research},
   volume = {53},
   number = {12},
   pages = {gkaf569},
   year = {2025},
   month = {06},
   issn = {1362-4962},
   doi = {10.1093/nar/gkaf569},
   url = {https://doi.org/10.1093/nar/gkaf569},
   eprint = {https://academic.oup.com/nar/article-pdf/53/12/gkaf569/63587518/gkaf569.pdf},
}

@article{zhao2023inferring,
  title={Inferring neuron-neuron communications from single-cell transcriptomics through {NeuronChat}},
  author={Zhao, Wei and Johnston, Kevin G and Ren, Honglei and Xu, Xiangmin and Nie, Qing},
  journal={Nature Communications},
  volume={14},
  number={1},
  pages={1128},
  year={2023},
  publisher={Nature Publishing Group UK London}
}

@article{jakobsson2021scconnect,
  title={{scConnect}: a method for exploratory analysis of cell--cell communication based on single-cell {RNA-sequencing} data},
  author={Jakobsson, Jon ET and Spjuth, Ola and Lagerstr{\"o}m, Malin C},
  journal={Bioinformatics},
  volume={37},
  number={20},
  pages={3501--3508},
  year={2021},
  publisher={Oxford University Press}
}

@article{shao2025extracellular,
  title={Extracellular vesicle-derived {miRNA-mediated} cell--cell communication inference for single-cell transcriptomic data with {miRTalk}},
  author={Shao, Xin and Yu, Lingqi and Li, Chengyu and Qian, Jingyang and Yang, Xinyu and Yang, Haihong and Liao, Jie and Fan, Xueru and Xu, Xiao and Fan, Xiaohui},
  journal={Genome Biology},
  volume={26},
  number={1},
  pages={95},
  year={2025},
  publisher={Springer}
}

@article{stoeckius2017simultaneous,
  title={Simultaneous epitope and transcriptome measurement in single cells},
  author={Stoeckius, Marlon and Hafemeister, Christoph and Stephenson, William and Houck-Loomis, Brian and Chattopadhyay, Pratip K and Swerdlow, Harold and Satija, Rahul and Smibert, Peter},
  journal={Nature methods},
  volume={14},
  number={9},
  pages={865--868},
  year={2017},
  publisher={Nature Publishing Group US New York}
}

@article{peterson2017multiplexed,
  title={Multiplexed quantification of proteins and transcripts in single cells},
  author={Peterson, Vanessa M and Zhang, Kelvin Xi and Kumar, Namit and Wong, Jerelyn and Li, Lixia and Wilson, Douglas C and Moore, Renee and McClanahan, Terrill K and Sadekova, Svetlana and Klappenbach, Joel A},
  journal={Nature biotechnology},
  volume={35},
  number={10},
  pages={936--939},
  year={2017},
  publisher={Nature Publishing Group US New York}
}

@article{cusanovich2015multiplex,
  title={Multiplex single-cell profiling of chromatin accessibility by combinatorial cellular indexing},
  author={Cusanovich, Darren A and Daza, Riza and Adey, Andrew and Pliner, Hannah A and Christiansen, Lena and Gunderson, Kevin L and Steemers, Frank J and Trapnell, Cole and Shendure, Jay},
  journal={Science},
  volume={348},
  number={6237},
  pages={910--914},
  year={2015},
  publisher={American Association for the Advancement of Science}
}

@article{lan2024single,
  title={Single Cell mass spectrometry: Towards quantification of small molecules in individual cells},
  author={Lan, Yunpeng and Zou, Zhu and Yang, Zhibo},
  journal={TrAC Trends in Analytical Chemistry},
  volume={174},
  pages={117657},
  year={2024},
  publisher={Elsevier}
}

@article{mao2025single,
  title={Single-Cell Simultaneous Metabolome and Transcriptome Profiling Revealing Metabolite-Gene Correlation Network},
  author={Mao, Xiying and Xia, Dandan and Xu, Miao and Gao, Yan and Tong, Le and Lu, Chen and Li, Weiqi and Xie, Runmin and Liu, Qinghuai and Jiang, Dechen and others},
  journal={Advanced Science},
  volume={12},
  number={4},
  pages={2411276},
  year={2025},
  publisher={Wiley Online Library}
}

@article{jin2020scai,
  title={{scAI}: an unsupervised approach for the integrative analysis of parallel single-cell transcriptomic and epigenomic profiles},
  author={Jin, Suoqin and Zhang, Lihua and Nie, Qing},
  journal={Genome biology},
  volume={21},
  number={1},
  pages={25},
  year={2020},
  publisher={Springer}
}

@article{staahl2016visualization,
  title={Visualization and analysis of gene expression in tissue sections by spatial transcriptomics},
  author={St{\aa}hl, Patrik L and Salm{\'e}n, Fredrik and Vickovic, Sanja and Lundmark, Anna and Navarro, Jos{\'e} Fern{\'a}ndez and Magnusson, Jens and Giacomello, Stefania and Asp, Michaela and Westholm, Jakub O and Huss, Mikael and others},
  journal={Science},
  volume={353},
  number={6294},
  pages={78--82},
  year={2016},
  publisher={American Association for the Advancement of Science}
}

@article{he2022high,
  title={High-plex imaging of {RNA} and proteins at subcellular resolution in fixed tissue by spatial molecular imaging},
  author={He, Shanshan and Bhatt, Ruchir and Brown, Carl and Brown, Emily A and Buhr, Derek L and Chantranuvatana, Kan and Danaher, Patrick and Dunaway, Dwayne and Garrison, Ryan G and Geiss, Gary and others},
  journal={Nature biotechnology},
  volume={40},
  number={12},
  pages={1794--1806},
  year={2022},
  publisher={Nature Publishing Group US New York}
}

@article{yuan2019systematic,
  title={Systematic expression analysis of ligand-receptor pairs reveals important cell-to-cell interactions inside glioma},
  author={Yuan, Dongsheng and Tao, Yiran and Chen, Geng and Shi, Tieliu},
  journal={Cell communication and signaling},
  volume={17},
  number={1},
  pages={48},
  year={2019},
  publisher={Springer}
}

@article{daneshpour2019modeling,
  title={Modeling cell--cell communication for immune systems across space and time},
  author={Daneshpour, Hirad and Youk, Hyun},
  journal={Current Opinion in Systems Biology},
  volume={18},
  pages={44--52},
  year={2019},
  publisher={Elsevier}
}

@article{hart2014paradoxical,
  title={Paradoxical signaling by a secreted molecule leads to homeostasis of cell levels},
  author={Hart, Yuval and Reich-Zeliger, Shlomit and Antebi, Yaron E and Zaretsky, Irina and Mayo, Avraham E and Alon, Uri and Friedman, Nir},
  journal={Cell},
  volume={158},
  number={5},
  pages={1022--1032},
  year={2014},
  publisher={Elsevier}
}

@article{bunne2024build,
  title={How to build the virtual cell with artificial intelligence: Priorities and opportunities},
  author={Bunne, Charlotte and Roohani, Yusuf and Rosen, Yanay and Gupta, Ankit and Zhang, Xikun and Roed, Marcel and Alexandrov, Theo and AlQuraishi, Mohammed and Brennan, Patricia and Burkhardt, Daniel B and others},
  journal={Cell},
  volume={187},
  number={25},
  pages={7045--7063},
  year={2024},
  publisher={Elsevier}
}

@article{RN424NiCo,
   author = {Agrawal, Ankit and Thomann, Stefan and Basu, Sukanya and Gr{\"u}n, Dominic},
   title = {{NiCo} identifies extrinsic drivers of cell state modulation by niche covariation analysis},
   journal = {Nature Communications},
   volume = {15},
   number = {1},
   pages = {10628},
   ISSN = {2041-1723},
   year = {2024},
   type = {Journal Article}
}

@article{RN415FlowSig,
   author = {Almet, Axel A and Tsai, Yuan-Chen and Watanabe, Momoko and Nie, Qing},
   title = {Inferring pattern-driving intercellular flows from single-cell and spatial transcriptomics},
   journal = {Nature Methods},
   volume = {21},
   number = {10},
   pages = {1806-1817},
   ISSN = {1548-7091},
   year = {2024},
   type = {Journal Article}
}

@article{RN339,
   author = {Armingol, Erick and Baghdassarian, Hratch M. and Martino, Cameron and Perez-Lopez, Araceli and Aamodt, Caitlin and Knight, Rob and Lewis, Nathan E.},
   title = {Context-aware deconvolution of cell–cell communication with {Tensor-cell2cell}},
   journal = {Nature Communications},
   volume = {13},
   number = {1},
   pages = {3665},
   ISSN = {2041-1723},
   DOI = {10.1038/s41467-022-31369-2},
   url = {https://doi.org/10.1038/s41467-022-31369-2},
   year = {2022},
   type = {Journal Article}
}

@article{RN356cell2cell,
   author = {Armingol, Erick and Ghaddar, Abbas and Joshi, Chintan J and Baghdassarian, Hratch and Shamie, Isaac and Chan, Jason and Her, Hsuan-Lin and Berhanu, Samuel and Dar, Anushka and Rodriguez-Armstrong, Fabiola},
   title = {Inferring a spatial code of cell-cell interactions across a whole animal body},
   journal = {PLoS computational biology},
   volume = {18},
   number = {11},
   pages = {e1010715},
   ISSN = {1553-734X},
   year = {2022},
   type = {Journal Article}
}

@article{RN385SVCA,
   author = {Arnol, Damien and Schapiro, Denis and Bodenmiller, Bernd and Saez-Rodriguez, Julio and Stegle, Oliver},
   title = {Modeling Cell-Cell Interactions from Spatial Molecular Data with Spatial Variance Component Analysis},
   journal = {Cell Reports},
   volume = {29},
   number = {1},
   pages = {202-211},
   year = {2019},
   type = {Journal Article}
}

@article{RN357CLARIFY,
   author = {Bafna, Mihir and Li, Hechen and Zhang, Xiuwei},
   title = {{CLARIFY}: cell--cell interaction and gene regulatory network refinement from spatially resolved transcriptomics},
   journal = {Bioinformatics},
   volume = {39},
   number = {Supplement\_1},
   pages = {i484-i493},
   ISSN = {1367-4811},
   DOI = {10.1093/bioinformatics/btad269},
   url = {https://doi.org/10.1093/bioinformatics/btad269},
   year = {2023},
   type = {Journal Article}
}

@article{RN283,
   author = {Baruzzo, G. and Cesaro, G. and Di Camillo, B.},
   title = {Identify, quantify and characterize cellular communication from single-cell {RNA} sequencing data with {scSeqComm}},
   journal = {Bioinformatics},
   volume = {38},
   number = {7},
   pages = {1920-1929},
   ISSN = {1367-4811 (Electronic)
1367-4803 (Linking)},
   DOI = {10.1093/bioinformatics/btac036},
   url = {https://www.ncbi.nlm.nih.gov/pubmed/35043939},
   year = {2022},
   type = {Journal Article}
}

@article{RN422NicheCompass,
   author = {Birk, Sebastian and Bonafonte-Pard{\`a}s, Irene and Feriz, Adib Miraki and Boxall, Adam and Agirre, Eneritz and Memi, Fani and Maguza, Anna and Yadav, Anamika and Armingol, Erick and Fan, Rong},
   title = {Quantitative characterization of cell niches in spatially resolved omics data},
   journal = {Nature Genetics},
   volume = {57},
   pages = {897-909},
   ISSN = {1061-4036},
   year = {2025},
   type = {Journal Article}
}

@article{RN336,
	author = {Browaeys, Robin and Gilis, Jeroen and Sang-Aram, Chananchida and De Bleser, Pieter and Hoste, Levi and Tavernier, Simon and Lambrechts, Diether and Seurinck, Ruth and Saeys, Yvan},
	title = {{MultiNicheNet}: a flexible framework for differential cell-cell communication analysis from multi-sample multi-condition single-cell transcriptomics data},
	elocation-id = {2023.06.13.544751},
	year = {2023},
	doi = {10.1101/2023.06.13.544751},
	publisher = {Cold Spring Harbor Laboratory},
	URL = {https://www.biorxiv.org/content/early/2023/06/14/2023.06.13.544751},
	eprint = {https://www.biorxiv.org/content/early/2023/06/14/2023.06.13.544751.full.pdf},
	journal = {bioRxiv}
}

@article{RN349,
   author = {Browaeys, Robin and Saelens, Wouter and Saeys, Yvan},
   title = {{NicheNet}: modeling intercellular communication by linking ligands to target genes},
   journal = {Nature Methods},
   volume = {17},
   number = {2},
   pages = {159-162},
   ISSN = {1548-7105},
   DOI = {10.1038/s41592-019-0667-5},
   url = {https://doi.org/10.1038/s41592-019-0667-5},
   year = {2020},
   type = {Journal Article}
}

@article{RN286,
   author = {Cabello-Aguilar, S. and Alame, M. and Kon-Sun-Tack, F. and Fau, C. and Lacroix, M. and Colinge, J.},
   title = {{SingleCellSignalR}: inference of intercellular networks from single-cell transcriptomics},
   journal = {Nucleic Acids Res},
   volume = {48},
   number = {10},
   pages = {e55},
   ISSN = {1362-4962 (Electronic)
0305-1048 (Print)
0305-1048 (Linking)},
   DOI = {10.1093/nar/gkaa183},
   url = {https://www.ncbi.nlm.nih.gov/pubmed/32196115},
   year = {2020},
   type = {Journal Article}
}

@article{RN376SpaOTsc,
   author = {Cang, Zixuan and Nie, Qing},
   title = {Inferring spatial and signaling relationships between cells from single cell transcriptomic data},
   journal = {Nature Communications},
   volume = {11},
   number = {1},
   pages = {2084},
   ISSN = {2041-1723},
   DOI = {10.1038/s41467-020-15968-5},
   url = {https://doi.org/10.1038/s41467-020-15968-5},
   year = {2020},
   type = {Journal Article}
}

@article{RN350,
   author = {Cheng, Jinyu and Zhang, Ji and Wu, Zhongdao and Sun, Xiaoqiang},
   title = {Inferring microenvironmental regulation of gene expression from single-cell {RNA} sequencing data using {scMLnet} with an application to {COVID-19}},
   journal = {Briefings in Bioinformatics},
   volume = {22},
   number = {2},
   pages = {988-1005},
   ISSN = {1477-4054},
   DOI = {10.1093/bib/bbaa327},
   url = {https://doi.org/10.1093/bib/bbaa327},
   year = {2020},
   type = {Journal Article}
}

@article{RN301,
   author = {Cherry, C. and Maestas, D. R. and Han, J. and Andorko, J. I. and Cahan, P. and Fertig, E. J. and Garmire, L. X. and Elisseeff, J. H.},
   title = {Computational reconstruction of the signalling networks surrounding implanted biomaterials from single-cell transcriptomics},
   journal = {Nat Biomed Eng},
   volume = {5},
   number = {10},
   pages = {1228-1238},
   ISSN = {2157-846X (Electronic)
2157-846X (Linking)},
   DOI = {10.1038/s41551-021-00770-5},
   url = {https://www.ncbi.nlm.nih.gov/pubmed/34341534},
   year = {2021},
   type = {Journal Article}
}

@article{RN340,
   author = {Choi, Hyejin and Sheng, Jianting and Gao, Dingcheng and Li, Fuhai and Durrans, Anna and Ryu, Seongho and Lee, Sharrell B and Narula, Navneet and Rafii, Shahin and Elemento, Olivier and Altorki, Nasser K and Wong, Stephen T C. and Mittal, Vivek},
   title = {Transcriptome Analysis of Individual Stromal Cell Populations Identifies Stroma-Tumor Crosstalk in Mouse Lung Cancer Model},
   journal = {Cell Reports},
   volume = {10},
   number = {7},
   pages = {1187-1201},
   ISSN = {2211-1247},
   DOI = {10.1016/j.celrep.2015.01.040},
   url = {https://doi.org/10.1016/j.celrep.2015.01.040},
   year = {2015},
   type = {Journal Article}
}

@article{RN418,
   author = {Chowdhury, Shrabanti and Ferri-Borgogno, Sammy and Yang, Peng and Wang, Wenyi and Peng, Jie and C Mok, Samuel and Wang, Pei},
   title = {Learning directed acyclic graphs for ligands and receptors based on spatially resolved transcriptomic data of ovarian cancer},
   journal = {Briefings in bioinformatics},
   volume = {26},
   number = {2},
   year = {2025},
   type = {Journal Article}
}

@article{RN332,
   author = {Cillo, Anthony R. and Kürten, Cornelius H. L. and Tabib, Tracy and Qi, Zengbiao and Onkar, Sayali and Wang, Ting and Liu, Angen and Duvvuri, Umamaheswar and Kim, Seungwon and Soose, Ryan J. and Oesterreich, Steffi and Chen, Wei and Lafyatis, Robert and Bruno, Tullia C. and Ferris, Robert L. and Vignali, Dario A. A.},
   title = {Immune Landscape of Viral- and Carcinogen-Driven Head and Neck Cancer},
   journal = {Immunity},
   volume = {52},
   number = {1},
   pages = {183-199.e9},
   ISSN = {1074-7613},
   DOI = {10.1016/j.immuni.2019.11.014},
   url = {https://doi.org/10.1016/j.immuni.2019.11.014},
   year = {2020},
   type = {Journal Article}
}

@article{RN411,
   author = {Ding, Qian and Yang, Wenyi and Xue, Guangfu and Liu, Hongxin and Cai, Yideng and Que, Jinhao and Jin, Xiyun and Luo, Meng and Pang, Fenglan and Yang, Yuexin},
   title = {Dimension reduction, cell clustering, and cell–cell communication inference for single-cell transcriptomics with {DcjComm}},
   journal = {Genome Biology},
   volume = {25},
   number = {1},
   pages = {241},
   ISSN = {1474-760X},
   year = {2024},
   type = {Journal Article}
}

@article{RN364Giotto,
   author = {Dries, Ruben and Zhu, Qian and Dong, Rui and Eng, Chee-Huat Linus and Li, Huipeng and Liu, Kan and Fu, Yuntian and Zhao, Tianxiao and Sarkar, Arpan and Bao, Feng and George, Rani E. and Pierson, Nico and Cai, Long and Yuan, Guo-Cheng},
   title = {{Giotto}: a toolbox for integrative analysis and visualization of spatial expression data},
   journal = {Genome Biology},
   volume = {22},
   number = {1},
   pages = {78},
   ISSN = {1474-760X},
   DOI = {10.1186/s13059-021-02286-2},
   url = {https://doi.org/10.1186/s13059-021-02286-2},
   year = {2021},
   type = {Journal Article}
}

@article{RN343,
   author = {Efremova, Mirjana and Vento-Tormo, Miquel and Teichmann, Sarah A. and Vento-Tormo, Roser},
   title = {{CellPhoneDB}: inferring cell–cell communication from combined expression of multi-subunit ligand–receptor complexes},
   journal = {Nature Protocols},
   volume = {15},
   number = {4},
   pages = {1484-1506},
   ISSN = {1750-2799},
   DOI = {10.1038/s41596-020-0292-x},
   url = {https://doi.org/10.1038/s41596-020-0292-x},
   year = {2020},
   type = {Journal Article}
}

@article{RN425PathFinder,
   author = {Feng, Jiarui and Song, Haoran and Province, Michael and Li, Guangfu and Payne, Philip RO and Chen, Yixin and Li, Fuhai},
   title = {{PathFinder}: a novel graph transformer model to infer multi-cell intra-and inter-cellular signaling pathways and communications},
   journal = {Frontiers in Cellular Neuroscience},
   volume = {18},
   pages = {1369242},
   ISSN = {1662-5102},
   year = {2024},
   type = {Journal Article}
}

@article{RN369,
   author = {Fischer, David S. and Schaar, Anna C. and Theis, Fabian J.},
   title = {Modeling intercellular communication in tissues using spatial graphs of cells},
   journal = {Nature Biotechnology},
   volume = {41},
   number = {3},
   pages = {332-336},
   ISSN = {1546-1696},
   DOI = {10.1038/s41587-022-01467-z},
   url = {https://doi.org/10.1038/s41587-022-01467-z},
   year = {2023},
   type = {Journal Article}
}

@article{RN434,
   author = {Fu, Yifeng and Qu, Hong and Qu, Dacheng and Zhao, Min},
   title = {Trajectory Inference with Cell–Cell Interactions ({TICCI}): intercellular communication improves the accuracy of trajectory inference methods},
   journal = {Bioinformatics},
   volume = {41},
   number = {2},
   pages = {btaf027},
   ISSN = {1367-4811},
   year = {2025},
   type = {Journal Article}
}

@article{RN345,
   author = {Garcia-Alonso, Luz and Lorenzi, Valentina and Mazzeo, Cecilia Icoresi and Alves-Lopes, João Pedro and Roberts, Kenny and Sancho-Serra, Carmen and Engelbert, Justin and Marečková, Magda and Gruhn, Wolfram H. and Botting, Rachel A. and Li, Tong and Crespo, Berta and van Dongen, Stijn and Kiselev, Vladimir Yu and Prigmore, Elena and Herbert, Mary and Moffett, Ashley and Chédotal, Alain and Bayraktar, Omer Ali and Surani, Azim and Haniffa, Muzlifah and Vento-Tormo, Roser},
   title = {Single-cell roadmap of human gonadal development},
   journal = {Nature},
   volume = {607},
   number = {7919},
   pages = {540-547},
   ISSN = {1476-4687},
   DOI = {10.1038/s41586-022-04918-4},
   url = {https://doi.org/10.1038/s41586-022-04918-4},
   year = {2022},
   type = {Journal Article}
}

@article{RN370,
   author = {Ghaddar, Bassel and De, Subhajyoti},
   title = {Reconstructing physical cell interaction networks from single-cell data using {Neighbor-seq}},
   journal = {Nucleic Acids Research},
   volume = {50},
   number = {14},
   pages = {e82-e82},
   ISSN = {0305-1048},
   DOI = {10.1093/nar/gkac333},
   url = {https://doi.org/10.1093/nar/gkac333},
   year = {2022},
   type = {Journal Article}
}

@article{RN281,
    author = {He, Changhan and Zhou, Peijie and Nie, Qing},
    title = {{exFINDER}: identify external communication signals using single-cell transcriptomics data},
    journal = {Nucleic Acids Research},
    volume = {51},
    number = {10},
    pages = {e58-e58},
    year = {2023},
    month = {04},
    issn = {0305-1048},
    doi = {10.1093/nar/gkad262},
    url = {https://doi.org/10.1093/nar/gkad262},
    eprint = {https://academic.oup.com/nar/article-pdf/51/10/e58/50527145/gkad262.pdf},
}

@article{RN273,
   author = {Hou, R. and Denisenko, E. and Ong, H. T. and Ramilowski, J. A. and Forrest, A. R. R.},
   title = {Predicting cell-to-cell communication networks using {NATMI}},
   journal = {Nat Commun},
   volume = {11},
   number = {1},
   pages = {5011},
   ISSN = {2041-1723 (Electronic)
2041-1723 (Linking)},
   DOI = {10.1038/s41467-020-18873-z},
   url = {https://www.ncbi.nlm.nih.gov/pubmed/33024107},
   year = {2020},
   type = {Journal Article}
}

@article{RN414,
	author = {Hou, Siyu and Ma, Wenjing and Zhou, Xiang},
	title = {{FastCCC}: a permutation-free framework for scalable, robust, and reference-based cell-cell communication analysis in single cell transcriptomics studies},
   journal = {Nature Communications},
   volume = {16},
   pages = {11428},
   ISSN = {2041-1723},
   year = {2025},
   type = {Journal Article}
}

@article{RN318,
   author = {Hu, Yuxuan and Peng, Tao and Gao, Lin and Tan, Kai},
   title = {{CytoTalk}: De novo construction of signal transduction networks using single-cell transcriptomic data},
   journal = {Science Advances},
   volume = {7},
   number = {16},
   pages = {eabf1356},
   ISSN = {2375-2548},
   DOI = {10.1101/2020.03.29.014464},
   year = {2021},
   type = {Journal Article}
}

@article{RN428scDCA,
   author = {Ji, Boya and Wang, Xiaoqi and Wang, Xiang and Xu, Liwen and Peng, Shaoliang},
   title = {{scDCA}: deciphering the dominant cell communication assembly of downstream functional events from single-cell {RNA-seq} data},
   journal = {Briefings in Bioinformatics},
   volume = {26},
   number = {1},
   year = {2024},
   type = {Journal Article}
}

@article{RN431SpaCCC,
   author={Ji, Boya and Wang, Xiaoqi and Qiao, Debin and Xu, Liwen and Peng, Shaoliang},
   journal={Big Data Mining and Analytics}, 
   title={{SpaCCC}: Large Language Model-Based Cell-Cell Communication Inference for Spatially Resolved Transcriptomic Data}, 
   year={2024},
   volume={7},
   number={4},
   pages={1129-1147},
   doi={10.26599/BDMA.2024.9020056}
}

@article{RN403,
   author = {Jiang, Junyao and Li, Jinlian and Huang, Sunan and Jiang, Fan and Liang, Yanran and Xu, Xueli and Wang, Jie},
   title = {{CACIMAR}: cross-species analysis of cell identities, markers, regulations, and interactions using single-cell {RNA} sequencing data},
   journal = {Briefings in Bioinformatics},
   volume = {25},
   number = {4},
   ISSN = {1477-4054},
   DOI = {10.1093/bib/bbae283},
   url = {https://doi.org/10.1093/bib/bbae283},
   year = {2024},
   type = {Journal Article}
}

@article{RN342,
   author = {Jin, Suoqin and Guerrero-Juarez, Christian F. and Zhang, Lihua and Chang, Ivan and Ramos, Raul and Kuan, Chen-Hsiang and Myung, Peggy and Plikus, Maksim V. and Nie, Qing},
   title = {Inference and analysis of cell-cell communication using {CellChat}},
   journal = {Nature Communications},
   volume = {12},
   number = {1},
   pages = {1088},
   ISSN = {2041-1723},
   DOI = {10.1038/s41467-021-21246-9},
   url = {https://doi.org/10.1038/s41467-021-21246-9},
   year = {2021},
   type = {Journal Article}
}

@article{jin2025cellchat,
  title={{CellChat} for systematic analysis of cell--cell communication from single-cell transcriptomics},
  author={Jin, Suoqin and Plikus, Maksim V and Nie, Qing},
  journal={Nature protocols},
  volume={20},
  number={1},
  pages={180--219},
  year={2025},
  publisher={Nature Publishing Group UK London}
}

@article{RN289,
   author = {Jin, Z. and Zhang, X. and Dai, X. and Huang, J. and Hu, X. and Zhang, J. and Shi, L.},
   title = {{InterCellDB}: A User-Defined Database for Inferring Intercellular Networks},
   journal = {Adv Sci (Weinh)},
   volume = {9},
   number = {22},
   pages = {e2200045},
   ISSN = {2198-3844 (Electronic)
2198-3844 (Linking)},
   DOI = {10.1002/advs.202200045},
   url = {https://www.ncbi.nlm.nih.gov/pubmed/35652265},
   year = {2022},
   type = {Journal Article}
}

@article{RN302,
   author = {Jung, S. and Singh, K. and Del Sol, A.},
   title = {{FunRes}: resolving tissue-specific functional cell states based on a cell-cell communication network model},
   journal = {Brief Bioinform},
   volume = {22},
   number = {4},
   ISSN = {1477-4054 (Electronic)
1467-5463 (Print)
1467-5463 (Linking)},
   DOI = {10.1093/bib/bbaa283},
   url = {https://www.ncbi.nlm.nih.gov/pubmed/33179736},
   year = {2021},
   type = {Journal Article}
}

@article{RN358CellNeighborEX,
   author = {Kim, Hyobin and Kumar, Amit and Lövkvist, Cecilia and Palma, António M. and Martin, Patrick and Kim, Junil and Bhoopathi, Praveen and Trevino, Jose and Fisher, Paul and Madan, Esha and Gogna, Rajan and Won, Kyoung Jae},
   title = {{CellNeighborEX}: deciphering neighbor-dependentgene expression from spatial transcriptomics data},
   journal = {Molecular Systems Biology},
   volume = {19},
   number = {11},
   pages = {e11670},
   ISSN = {1744-4292},
   DOI = {https://doi.org/10.15252/msb.202311670},
   url = {https://doi.org/10.15252/msb.202311670},
   year = {2023},
   type = {Journal Article}
}

@article{RN328DeepCOLOR,
   author = {Kojima, Yasuhiro and Mii, Shinji and Hayashi, Shuto and Hirose, Haruka and Ishikawa, Masato and Akiyama, Masashi and Enomoto, Atsushi and Shimamura, Teppei},
   title = {Single-cell colocalization analysis using a deep generative model},
   journal = {Cell Systems},
   volume = {15},
   number = {2},
   pages = {180-192. e7},
   ISSN = {2405-4712},
   DOI = {10.1101/2022.04.10.487815},
   year = {2024},
   type = {Journal Article}
}

@article{RN263,
   author = {Lagger, C. and Ursu, E. and Equey, A. and Avelar, R. A. and Pisco, A. O. and Tacutu, R. and de Magalhaes, J. P.},
   title = {{scDiffCom}: a tool for differential analysis of cell-cell interactions provides a mouse atlas of aging changes in intercellular communication},
   journal = {Nat Aging},
   volume = {3},
   number = {11},
   pages = {1446-1461},
   ISSN = {2662-8465 (Electronic)
2662-8465 (Linking)},
   DOI = {10.1038/s43587-023-00514-x},
   url = {https://www.ncbi.nlm.nih.gov/pubmed/37919434},
   year = {2023},
   type = {Journal Article}
}

@article{RN367MESSI,
   author = {Li, Dongshunyi and Ding, Jun and Bar-Joseph, Ziv},
   title = {Identifying signaling genes in spatial single-cell expression data},
   journal = {Bioinformatics},
   volume = {37},
   number = {7},
   pages = {968-975},
   ISSN = {1367-4803},
   DOI = {10.1093/bioinformatics/btaa769},
   url = {https://doi.org/10.1093/bioinformatics/btaa769},
   year = {2020},
   type = {Journal Article}
}

@article{RN292,
   author = {Li, D. and Velazquez, J. J. and Ding, J. and Hislop, J. and Ebrahimkhani, M. R. and Bar-Joseph, Z.},
   title = {{TraSig}: inferring cell-cell interactions from pseudotime ordering of {scRNA-Seq} data},
   journal = {Genome Biol},
   volume = {23},
   number = {1},
   pages = {73},
   ISSN = {1474-760X (Electronic)
1474-7596 (Print)
1474-7596 (Linking)},
   DOI = {10.1186/s13059-022-02629-7},
   url = {https://www.ncbi.nlm.nih.gov/pubmed/35255944},
   year = {2022},
   type = {Journal Article}
}

@article{RN366HoloNet,
   author = {Li, Haochen and Ma, Tianxing and Hao, Minsheng and Guo, Wenbo and Gu, Jin and Zhang, Xuegong and Wei, Lei},
   title = {Decoding functional cell–cell communication events by multi-view graph learning on spatial transcriptomics},
   journal = {Briefings in Bioinformatics},
   volume = {24},
   number = {6},
   ISSN = {1477-4054},
   DOI = {10.1093/bib/bbad359},
   url = {https://doi.org/10.1093/bib/bbad359},
   year = {2023},
   type = {Journal Article}
}

@article{RN362DeepLinc,
   author = {Li, Runze and Yang, Xuerui},
   title = {De novo reconstruction of cell interaction landscapes from single-cell spatial transcriptome data with {DeepLinc}},
   journal = {Genome Biology},
   volume = {23},
   number = {1},
   pages = {124},
   ISSN = {1474-760X},
   DOI = {10.1186/s13059-022-02692-0},
   url = {https://doi.org/10.1186/s13059-022-02692-0},
   year = {2022},
   type = {Journal Article}
}

@article{RN429scHyper,
   author = {Li, Wenying and Wang, Haiyun and Zhao, Jianping and Xia, Junfeng and Sun, Xiaoqiang},
   title = {{scHyper}: reconstructing cell–cell communication through hypergraph neural networks},
   journal = {Briefings in Bioinformatics},
   volume = {25},
   number = {5},
   pages = {bbae436},
   ISSN = {1467-5463},
   year = {2024},
   type = {Journal Article}
}

@article{RN378SpatialDM,
   author = {Li, Zhuoxuan and Wang, Tianjie and Liu, Pentao and Huang, Yuanhua},
   title = {{SpatialDM} for rapid identification of spatially co-expressed ligand–receptor and revealing cell–cell communication patterns},
   journal = {Nature Communications},
   volume = {14},
   number = {1},
   pages = {3995},
   ISSN = {2041-1723},
   DOI = {10.1038/s41467-023-39608-w},
   url = {https://doi.org/10.1038/s41467-023-39608-w},
   year = {2023},
   type = {Journal Article}
}

@article{RN410CytoSignal,
	author = {Liu, Jialin and Manabe, Hiroaki and Qian, Weizhou and Wang, Yichen and Gu, Yichen and Yan Chu, Angel Ka and Gadhvi, Gaurav and Song, Yuxuan and Ono, Noriaki and Welch, Joshua D.},
	title = {{CytoSignal} Detects Locations and Dynamics of Ligand-Receptor Signaling at Cellular Resolution from Spatial Transcriptomic Data},
	elocation-id = {2024.03.08.584153},
	year = {2024},
	doi = {10.1101/2024.03.08.584153},
	publisher = {Cold Spring Harbor Laboratory},
	URL = {https://www.biorxiv.org/content/early/2024/03/13/2024.03.08.584153},
	eprint = {https://www.biorxiv.org/content/early/2024/03/13/2024.03.08.584153.full.pdf},
	journal = {bioRxiv}
}

@article{RN338,
   author = {Liu, Qi and Hsu, Chih-Yuan and Li, Jia and Shyr, Yu},
   title = {Dysregulated ligand–receptor interactions from single-cell transcriptomics},
   journal = {Bioinformatics},
   volume = {38},
   number = {12},
   pages = {3216-3221},
   ISSN = {1367-4803},
   DOI = {10.1093/bioinformatics/btac294},
   url = {https://doi.org/10.1093/bioinformatics/btac294},
   year = {2022},
   type = {Journal Article}
}

@article{RN394HiVAE,
   author = {Liu, Shuhui and Zhang, Yupei and Peng, Jiajie and Shang, Xuequn},
   title = {An improved hierarchical variational autoencoder for cell–cell communication estimation using single-cell {RNA-seq} data},
   journal = {Briefings in Functional Genomics},
   volume = {23},
   number = {2},
   pages = {118-127},
   ISSN = {2041-2657},
   DOI = {10.1093/bfgp/elac056},
   url = {https://doi.org/10.1093/bfgp/elac056},
   year = {2023},
   type = {Journal Article}
}

@article{RN313,
   author = {Liu, Y. and Li, J. S. S. and Rodiger, J. and Comjean, A. and Attrill, H. and Antonazzo, G. and Brown, N. H. and Hu, Y. and Perrimon, N.},
   title = {{FlyPhoneDB}: an integrated web-based resource for cell-cell communication prediction in {Drosophila}},
   journal = {Genetics},
   volume = {220},
   number = {3},
   ISSN = {1943-2631 (Electronic)
0016-6731 (Print)
0016-6731 (Linking)},
   DOI = {10.1093/genetics/iyab235},
   url = {https://www.ncbi.nlm.nih.gov/pubmed/35100387},
   year = {2022},
   type = {Journal Article}
}

@article{RN419,
   author = {Liu, Yi and Zhang, Yuelei and Chang, Xiao and Liu, Xiaoping},
   title = {{MDIC3}: Matrix decomposition to infer cell-cell communication},
   journal = {Patterns},
   volume = {5},
   number = {2},
   ISSN = {2666-3899},
   year = {2024},
   type = {Journal Article}
}

@article{RN306,
   author = {Lummertz da Rocha, E. and Kubaczka, C. and Sugden, W. W. and Najia, M. A. and Jing, R. and Markel, A. and LeBlanc, Z. C. and Dos Santos Peixoto, R. and Falchetti, M. and Collins, J. J. and North, T. E. and Daley, G. Q.},
   title = {{CellComm} infers cellular crosstalk that drives haematopoietic stem and progenitor cell development},
   journal = {Nat Cell Biol},
   volume = {24},
   number = {4},
   pages = {579-589},
   ISSN = {1476-4679 (Electronic)
1465-7392 (Print)
1465-7392 (Linking)},
   DOI = {10.1038/s41556-022-00884-1},
   url = {https://www.ncbi.nlm.nih.gov/pubmed/35414020},
   year = {2022},
   type = {Journal Article}
}

@article{RN305,
   author = {Luo, J. and Deng, M. and Zhang, X. and Sun, X.},
   title = {{ESICCC} as a systematic computational framework for evaluation, selection, and integration of cell-cell communication inference methods},
   journal = {Genome Res},
   volume = {33},
   number = {10},
   pages = {1788-1805},
   DOI = {10.1101/gr.278001.123},
   url = {https://www.ncbi.nlm.nih.gov/pubmed/37827697},
   year = {2023},
   type = {Journal Article}
}

@article{RN423Niche-DE,
   author = {Mason, Kaishu and Sathe, Anuja and Hess, Paul R and Rong, Jiazhen and Wu, Chi-Yun and Furth, Emma and Susztak, Katalin and Levinsohn, Jonathan and Ji, Hanlee P and Zhang, Nancy},
   title = {{Niche-DE}: niche-differential gene expression analysis in spatial transcriptomics data identifies context-dependent cell-cell interactions},
   journal = {Genome biology},
   volume = {25},
   number = {1},
   pages = {14},
   ISSN = {1474-760X},
   year = {2024},
   type = {Journal Article}
}

@article{RN323,
   author = {Mitchel, Jonathan and Gordon, M. Grace and Perez, Richard K. and Biederstedt, Evan and Bueno, Raymund and Ye, Chun Jimmie and Kharchenko, Peter V.},
   title = {Coordinated, multicellular patterns of transcriptional variation that stratify patient cohorts are revealed by tensor decomposition},
   journal = {Nature Biotechnology},
   ISSN = {1546-1696},
   volume = {43},
   pages = {1192-1201},
   DOI = {10.1038/s41587-024-02411-z},
   url = {https://doi.org/10.1038/s41587-024-02411-z},
   year = {2024},
   type = {Journal Article}
}

@article{RN401,
   author = {Moratalla-Navarro, F. and Moreno, V. and Sanz-Pamplona, R.},
   title = {{TALKIEN}: crossTALK IntEraction Network. A web-based tool for deciphering molecular communication through ligand-receptor interactions},
   journal = {Mol Omics},
   volume = {19},
   number = {9},
   pages = {688-696},
   ISSN = {2515-4184 (Electronic)
2515-4184 (Linking)},
   DOI = {10.1039/d3mo00049d},
   url = {https://www.ncbi.nlm.nih.gov/pubmed/37403821},
   year = {2023},
   type = {Journal Article}
}

@article{RN319,
   author = {Nagai, James S and Leimkühler, Nils B and Schaub, Michael T and Schneider, Rebekka K and Costa, Ivan G},
   title = {{CrossTalkeR}: analysis and visualization of ligand–receptor networks},
   journal = {Bioinformatics},
   volume = {37},
   number = {22},
   pages = {4263-4265},
   ISSN = {1367-4803},
   DOI = {10.1101/2021.01.20.427390},
   year = {2021},
   type = {Journal Article}
}

@article{RN427,
   author = {Nagai, James S and Maié, Tiago and Schaub, Michael T and Costa, Ivan G},
   title = {{scACCorDiON}: a clustering approach for explainable patient level cell–cell communication graph analysis},
   journal = {Bioinformatics},
   volume = {41},
   number = {5},
   pages = {btaf288},
   ISSN = {1367-4811},
   year = {2025},
   type = {Journal Article}
}

@article{RN344,
   author = {Noël, Floriane and Massenet-Regad, Lucile and Carmi-Levy, Irit and Cappuccio, Antonio and Grandclaudon, Maximilien and Trichot, Coline and Kieffer, Yann and Mechta-Grigoriou, Fatima and Soumelis, Vassili},
   title = {Dissection of intercellular communication using the transcriptome-based framework {ICELLNET}},
   journal = {Nature Communications},
   volume = {12},
   number = {1},
   pages = {1089},
   ISSN = {2041-1723},
   DOI = {10.1038/s41467-021-21244-x},
   url = {https://doi.org/10.1038/s41467-021-21244-x},
   year = {2021},
   type = {Journal Article}
}

@article{RN380,
   author = {Palla, Giovanni and Spitzer, Hannah and Klein, Michal and Fischer, David and Schaar, Anna Christina and Kuemmerle, Louis Benedikt and Rybakov, Sergei and Ibarra, Ignacio L. and Holmberg, Olle and Virshup, Isaac and Lotfollahi, Mohammad and Richter, Sabrina and Theis, Fabian J.},
   title = {{Squidpy}: a scalable framework for spatial omics analysis},
   journal = {Nature Methods},
   volume = {19},
   number = {2},
   pages = {171-178},
   ISSN = {1548-7105},
   DOI = {10.1038/s41592-021-01358-2},
   url = {https://doi.org/10.1038/s41592-021-01358-2},
   year = {2022},
   type = {Journal Article}
}

@article{RN386,
   author = {Pancheva, Alexandrina and Wheadon, Helen and Rogers, Simon and Otto, Thomas D},
   title = {Using topic modeling to detect cellular crosstalk in {scRNA-seq}},
   journal = {PLOS Computational Biology},
   volume = {18},
   number = {4},
   pages = {e1009975},
   ISSN = {1553-734X},
   year = {2022},
   type = {Journal Article}
}

@article{RN391,
   author = {Peng, Lihong and Tan, Jingwei and Xiong, Wei and Zhang, Li and Wang, Zhao and Yuan, Ruya and Li, Zejun and Chen, Xing},
   title = {Deciphering ligand–receptor-mediated intercellular communication based on ensemble deep learning and the joint scoring strategy from single-cell transcriptomic data},
   journal = {Computers in Biology and Medicine},
   volume = {163},
   pages = {107137},
   ISSN = {0010-4825},
   DOI = {https://doi.org/10.1016/j.compbiomed.2023.107137},
   url = {https://www.sciencedirect.com/science/article/pii/S0010482523006029},
   year = {2023},
   type = {Journal Article}
}

@article{RN284,
   author = {Peng, L. and Yuan, R. and Han, C. and Han, G. and Tan, J. and Wang, Z. and Chen, M. and Chen, X.},
   title = {{CellEnBoost}: A Boosting-Based Ligand-Receptor Interaction Identification Model for Cell-to-Cell Communication Inference},
   journal = {IEEE Trans Nanobioscience},
   volume = {22},
   number = {4},
   pages = {705-715},
   ISSN = {1558-2639 (Electronic)
1536-1241 (Linking)},
   DOI = {10.1109/TNB.2023.3278685},
   url = {https://www.ncbi.nlm.nih.gov/pubmed/37216267},
   year = {2023},
   type = {Journal Article}
}

@article{RN382stLearn,
   author = {Pham, Duy and Tan, Xiao and Balderson, Brad and Xu, Jun and Grice, Laura F. and Yoon, Sohye and Willis, Emily F. and Tran, Minh and Lam, Pui Yeng and Raghubar, Arti and Kalita-de Croft, Priyakshi and Lakhani, Sunil and Vukovic, Jana and Ruitenberg, Marc J. and Nguyen, Quan H.},
   title = {Robust mapping of spatiotemporal trajectories and cell–cell interactions in healthy and diseased tissues},
   journal = {Nature Communications},
   volume = {14},
   number = {1},
   pages = {7739},
   ISSN = {2041-1723},
   DOI = {10.1038/s41467-023-43120-6},
   url = {https://doi.org/10.1038/s41467-023-43120-6},
   year = {2023},
   type = {Journal Article}
}

@article{RN324CellAgentChat,
	author = {Raghavan, Vishvak and Zheng, Yumin and Li, Yue and Ding, Jun},
	title = {Harnessing agent-based frameworks in {CellAgentChat} to unravel cell–cell interactions from single-cell and spatial transcriptomics},
	year = {2025},
   journal = {Genome Res},
   volume = {35},
   pages = {1646-1663},
	publisher = {Cold Spring Harbor Laboratory}
}

@article{RN371Renoir,
   author = {Rao, Narein and Kumar, Tanush and Pai, Rhea and Mishra, Archita and Ginhoux, Florent and Chan, Jerry and Sharma, Ankur and Zafar, Hamim},
   title = {Charting spatial ligand-target activity using {Renoir}},
   journal = {bioRxiv},
   pages = {2023.04.14.536833},
   DOI = {10.1101/2023.04.14.536833},
   url = {http://biorxiv.org/content/early/2024/11/01/2023.04.14.536833.abstract},
   year = {2024},
   type = {Journal Article}
}

@article{RN303,
   author = {Raredon, M. S. B. and Yang, J. and Kothapalli, N. and Lewis, W. and Kaminski, N. and Niklason, L. E. and Kluger, Y.},
   title = {Comprehensive visualization of cell-cell interactions in single-cell and spatial transcriptomics with {NICHES}},
   journal = {Bioinformatics},
   volume = {39},
   number = {1},
   ISSN = {1367-4811 (Electronic)
1367-4803 (Print)
1367-4803 (Linking)},
   DOI = {10.1093/bioinformatics/btac775},
   url = {https://www.ncbi.nlm.nih.gov/pubmed/36458905},
   year = {2023},
   type = {Journal Article}
}

@article{RN374SpaCET,
   author = {Ru, Beibei and Huang, Jinlin and Zhang, Yu and Aldape, Kenneth and Jiang, Peng},
   title = {Estimation of cell lineages in tumors from spatial transcriptomics data},
   journal = {Nature Communications},
   volume = {14},
   number = {1},
   pages = {568},
   ISSN = {2041-1723},
   DOI = {10.1038/s41467-023-36062-6},
   url = {https://doi.org/10.1038/s41467-023-36062-6},
   year = {2023},
   type = {Journal Article}
}

@inproceedings{RN373SpaCeNet,
   author = {Schrod, Stefan and Lück, Niklas and Lohmayer, Robert and Solbrig, Stefan and Völkl, Dennis and Wipfler, Tina and Shutta, Katherine H. and Guebila, Marouen Ben and Schäfer, Andreas and Beißbarth, Tim and Zacharias, Helena U. and Oefner, Peter J. and Quackenbush, John and Altenbuchinger, Michael},
   title = {{SpaCeNet}: Spatial Cellular Networks from omics data},
   booktitle={Research in Computational Molecular Biology},
   volume = {14758},
   pages={344-347},
   year={2024},
   organization={Springer}
}

@article{RN377SpaTalk,
   author = {Shao, Xin and Li, Chengyu and Yang, Haihong and Lu, Xiaoyan and Liao, Jie and Qian, Jingyang and Wang, Kai and Cheng, Junyun and Yang, Penghui and Chen, Huajun and Xu, Xiao and Fan, Xiaohui},
   title = {Knowledge-graph-based cell-cell communication inference for spatially resolved transcriptomic data with {SpaTalk}},
   journal = {Nature Communications},
   volume = {13},
   number = {1},
   pages = {4429},
   ISSN = {2041-1723},
   DOI = {10.1038/s41467-022-32111-8},
   url = {https://doi.org/10.1038/s41467-022-32111-8},
   year = {2022},
   type = {Journal Article}
}

@article{RN365GraphComm,
  title={{GraphComm} predicts cell cell communication using a graph based deep learning method in single cell {RNA} sequencing data},
  author={So, Emily and Hayat, Sikander and Nair, Sisira Kadambat and Wang, Bo and Haibe-Kains, Benjamin},
  journal={Scientific Reports},
  volume={15},
  number={1},
  pages={36914},
  year={2025},
  publisher={Nature Publishing Group UK London}
}

@article{RN407,
	author = {Solovey, Maria and Celik, Muhammet A. and Salcher, Felix R. and Abdalfattah, Mohmed and Ismail, Mostafa and Scialdone, Antonio and Ziemann, Frank and Colom{\'e}-Tatch{\'e}, Maria},
	title = {{Community} assesses differential cell communication using large multi-sample case-control {scRNAseq} datasets},
	elocation-id = {2024.03.01.582941},
	year = {2024},
	doi = {10.1101/2024.03.01.582941},
	publisher = {Cold Spring Harbor Laboratory},
	URL = {https://www.biorxiv.org/content/early/2024/03/04/2024.03.01.582941},
	eprint = {https://www.biorxiv.org/content/early/2024/03/04/2024.03.01.582941.full.pdf},
	journal = {bioRxiv}
}

@article{RN280,
   author = {Solovey, M. and Scialdone, A.},
   title = {{COMUNET}: a tool to explore and visualize intercellular communication},
   journal = {Bioinformatics},
   volume = {36},
   number = {15},
   pages = {4296-4300},
   ISSN = {1367-4811 (Electronic)
1367-4803 (Print)
1367-4803 (Linking)},
   DOI = {10.1093/bioinformatics/btaa482},
   url = {https://www.ncbi.nlm.nih.gov/pubmed/32399572},
   year = {2020},
   type = {Journal Article}
}

@article{RN272SPRUCE,
   author = {Subedi, S. and Park, Y. P.},
   title = {Single-cell pair-wise relationships untangled by composite embedding model},
   journal = {iScience},
   volume = {26},
   number = {2},
   pages = {106025},
   ISSN = {2589-0042 (Electronic)
2589-0042 (Linking)},
   DOI = {10.1016/j.isci.2023.106025},
   url = {https://www.ncbi.nlm.nih.gov/pubmed/36824286},
   year = {2023},
   type = {Journal Article}
}

@article{RN375spaCI,
   author = {Tang, Ziyang and Zhang, Tonglin and Yang, Baijian and Su, Jing and Song, Qianqian},
   title = {{spaCI}: deciphering spatial cellular communications through adaptive graph model},
   journal = {Briefings in Bioinformatics},
   volume = {24},
   number = {1},
   ISSN = {1477-4054},
   DOI = {10.1093/bib/bbac563},
   url = {https://doi.org/10.1093/bib/bbac563},
   year = {2022},
   type = {Journal Article}
}

@article{RN316CellPhoneDBv5,
   author = {Troulé, K. and Petryszak, R. and Cakir, B. and Cranley, J. and Harasty, A. and Prete, M. and Tuong, Z. K. and Teichmann, S. A. and Garcia-Alonso, L. and Vento-Tormo, R.},
   title = {{CellPhoneDB} v5: inferring cell-cell communication from single-cell multiomics data},
   journal = {Nat Protoc},
   DOI = {10.1038/s41596-024-01137-1},
   year = {2025},
   type = {Journal Article}
}

@article{RN355,
   author = {Tsuchiya, Takaho and Hori, Hiroki and Ozaki, Haruka},
   title = {{CCPLS} reveals cell-type-specific spatial dependence of transcriptomes in single cells},
   journal = {Bioinformatics},
   volume = {38},
   number = {21},
   pages = {4868-4877},
   ISSN = {1367-4811},
   DOI = {10.1093/bioinformatics/btac599},
   url = {https://doi.org/10.1093/bioinformatics/btac599},
   year = {2022},
   type = {Journal Article}
}

@inproceedings{kim2007nonnegative,
   author={Kim, Yong-Deok and Choi, Seungjin},
   booktitle={2007 IEEE Conference on Computer Vision and Pattern Recognition}, 
   title={Nonnegative Tucker Decomposition}, 
   year={2007},
   volume={},
   number={},
   pages={1-8},
   doi={10.1109/CVPR.2007.383405}
}

@article{RN399,
   author = {Tsuyuzaki, Koki and Ishii, Manabu and Nikaido, Itoshi},
   title = {{Sctensor} detects many-to-many cell–cell interactions from single cell {RNA-sequencing} data},
   journal = {BMC Bioinformatics},
   volume = {24},
   number = {1},
   pages = {420},
   ISSN = {1471-2105},
   DOI = {10.1186/s12859-023-05490-y},
   url = {https://doi.org/10.1186/s12859-023-05490-y},
   year = {2023},
   type = {Journal Article}
}

@article{RN266,
   author = {Vahid, M. R. and Kurlovs, A. H. and Andreani, T. and Auge, F. and Olfati-Saber, R. and de Rinaldis, E. and Rapaport, F. and Savova, V.},
   title = {{DiSiR}: fast and robust method to identify ligand-receptor interactions at subunit level from single-cell {RNA-sequencing} data},
   journal = {NAR Genom Bioinform},
   volume = {5},
   number = {1},
   pages = {lqad030},
   ISSN = {2631-9268 (Electronic)
2631-9268 (Linking)},
   DOI = {10.1093/nargab/lqad030},
   url = {https://www.ncbi.nlm.nih.gov/pubmed/36968431},
   year = {2023},
   type = {Journal Article}
}

@article{RN397,
   title={Mathematically mapping the network of cells in the tumor microenvironment},
   author={van Santvoort, Mike and Lapuente-Santana, {\'O}scar and Zopoglou, Maria and Zackl, Constantin and Finotello, Francesca and van der Hoorn, Pim and Eduati, Federica},
   journal={Cell Reports Methods},
   volume={5},
   number={2},
   year={2025},
   publisher={Elsevier}
}

@article{RN275BulkSignalR,
   author = {Villemin, J. P. and Bassaganyas, L. and Pourquier, D. and Boissiere, F. and Cabello-Aguilar, S. and Crapez, E. and Tanos, R. and Cornillot, E. and Turtoi, A. and Colinge, J.},
   title = {Inferring ligand-receptor cellular networks from bulk and spatial transcriptomic datasets with {BulkSignalR}},
   journal = {Nucleic Acids Res},
   volume = {51},
   number = {10},
   pages = {4726-44},
   ISSN = {1362-4962 (Electronic)
0305-1048 (Print)
0305-1048 (Linking)},
   DOI = {10.1093/nar/gkad352},
   url = {https://www.ncbi.nlm.nih.gov/pubmed/37144485},
   year = {2023},
   type = {Journal Article}
}

@article{RN406CLARA,
   author = {Wang, Honglin and Chung, Yeonsoo and Shin, Dong-guk},
   title = {Decipher cell communication with attention: {CLARA}},
   journal = {bioRxiv},
   pages = {2025.03. 09.642280},
   year = {2025},
   type = {Journal Article}
}

@article{RN426,
   author = {Wang, Haiyun and Zhao, Jianping and Nie, Qing and Zheng, Chunhou and Sun, Xiaoqiang},
   title = {Dissecting spatiotemporal structures in spatial Transcriptomics via diffusion-based adversarial learning},
   journal = {Research},
   volume = {7},
   pages = {0390},
   ISSN = {2639-5274},
   year = {2024},
   type = {Journal Article}
}

@article{RN379,
   author = {Wang, Jingwan and Li, Shiying and Chen, Lingxi and Li, Shuai Cheng},
   title = {{SPROUT}: spectral sparsification helps restore the spatial structure at single-cell resolution},
   journal = {NAR Genomics and Bioinformatics},
   volume = {4},
   number = {3},
   ISSN = {2631-9268},
   DOI = {10.1093/nargab/lqac069},
   url = {https://doi.org/10.1093/nargab/lqac069},
   year = {2022},
   type = {Journal Article}
}

@article{RN334,
   author = {Wang, K. and Patkar, S. and Lee, J. S. and Gertz, E. M. and Robinson, W. and Schischlik, F. and Crawford, D. R. and Schäffer, A. A. and Ruppin, E.},
   title = {Deconvolving Clinically Relevant Cellular Immune Cross-talk from Bulk Gene Expression Using {CODEFACS} and {LIRICS} Stratifies Patients with Melanoma to Anti-{PD}-1 Therapy},
   journal = {Cancer Discov},
   volume = {12},
   number = {4},
   pages = {1088-1105},
   ISSN = {2159-8274 (Print)
2159-8274},
   DOI = {10.1158/2159-8290.Cd-21-0887},
   year = {2022},
   type = {Journal Article}
}

@article{RN282,
   author = {Wang, L. and Zheng, Y. and Sun, Y. and Mao, S. and Li, H. and Bo, X. and Li, C. and Chen, H.},
   title = {{TimeTalk} uses single-cell {RNA-seq} datasets to decipher cell-cell communication during early embryo development},
   journal = {Commun Biol},
   volume = {6},
   number = {1},
   pages = {901},
   ISSN = {2399-3642 (Electronic)
2399-3642 (Linking)},
   DOI = {10.1038/s42003-023-05283-2},
   url = {https://www.ncbi.nlm.nih.gov/pubmed/37660148},
   year = {2023},
   type = {Journal Article}
}

@article{RN409,
   author = {Wang, Xinyi and Almet, Axel A and Nie, Qing},
   title = {Detecting global and local hierarchical structures in cell-cell communication using {CrossChat}},
   journal = {Nature Communications},
   volume = {15},
   number = {1},
   pages = {10542},
   ISSN = {2041-1723},
   year = {2024},
   type = {Journal Article}
}

@article{RN352,
   author = {Wilk, Aaron J. and Shalek, Alex K. and Holmes, Susan and Blish, Catherine A.},
   title = {Comparative analysis of cell–cell communication at single-cell resolution},
   journal = {Nature Biotechnology},
   volume = {42},
   number = {3},
   pages = {470-483},
   ISSN = {1546-1696},
   DOI = {10.1038/s41587-023-01782-z},
   url = {https://doi.org/10.1038/s41587-023-01782-z},
   year = {2024},
   type = {Journal Article}
}

@inproceedings{RN413,
   author = {Wu, Catherine J and Azizi, Elham},
   title = {{DIISCO}: A Bayesian Framework for Inferring Dynamic Intercellular Interactions from Time-Series Single-Cell Data},
   booktitle = {Research in Computational Molecular Biology},
   publisher = {Springer Nature},
   volume = {14758},
   pages = {390-395},
   ISBN = {1071639897},
   year = {2024},
   type = {Conference Proceedings}
}

@article{RN354,
   author = {Wu, Dongyuan and Gaskins, Jeremy T. and Sekula, Michael and Datta, Susmita},
   title = {Inferring Cell–Cell Communications from Spatially Resolved Transcriptomics Data Using a Bayesian Tweedie Model},
   journal = {Genes},
   volume = {14},
   number = {7},
   pages = {1368},
   ISSN = {2073-4425},
   url = {https://www.mdpi.com/2073-4425/14/7/1368},
   year = {2023},
   type = {Journal Article}
}

@article{RN405CellMsg,
   author = {Xia, Hong and Ji, Boya and Qiao, Debin and Peng, Shaoliang},
   title = {{CellMsg}: graph convolutional networks for ligand–receptor-mediated cell-cell communication analysis},
   journal = {Briefings in Bioinformatics},
   volume = {26},
   number = {1},
   ISSN = {1477-4054},
   DOI = {10.1093/bib/bbae716},
   url = {https://doi.org/10.1093/bib/bbae716},
   year = {2025},
   type = {Journal Article}
}

@article{RN277,
   author = {Ximerakis, M. and Lipnick, S. L. and Innes, B. T. and Simmons, S. K. and Adiconis, X. and Dionne, D. and Mayweather, B. A. and Nguyen, L. and Niziolek, Z. and Ozek, C. and Butty, V. L. and Isserlin, R. and Buchanan, S. M. and Levine, S. S. and Regev, A. and Bader, G. D. and Levin, J. Z. and Rubin, L. L.},
   title = {Single-cell transcriptomic profiling of the aging mouse brain},
   journal = {Nat Neurosci},
   volume = {22},
   number = {10},
   pages = {1696-1708},
   ISSN = {1546-1726 (Electronic)
1097-6256 (Linking)},
   DOI = {10.1038/s41593-019-0491-3},
   url = {https://www.ncbi.nlm.nih.gov/pubmed/31551601},
   year = {2019},
   type = {Journal Article}
}

@article{RN288,
   author = {Xin, Y. and Lyu, P. and Jiang, J. and Zhou, F. and Wang, J. and Blackshaw, S. and Qian, J.},
   title = {{LRLoop}: a method to predict feedback loops in cell-cell communication},
   journal = {Bioinformatics},
   volume = {38},
   number = {17},
   pages = {4117-4126},
   ISSN = {1367-4811 (Electronic)
1367-4803 (Print)
1367-4803 (Linking)},
   DOI = {10.1093/bioinformatics/btac447},
   url = {https://www.ncbi.nlm.nih.gov/pubmed/35788263},
   year = {2022},
   type = {Journal Article}
}

@article{RN312,
   author = {Xu, C. and Ma, D. and Ding, Q. and Zhou, Y. and Zheng, H. L.},
   title = {{PlantPhoneDB}: A manually curated pan-plant database of ligand-receptor pairs infers cell-cell communication},
   journal = {Plant Biotechnol J},
   volume = {20},
   number = {11},
   pages = {2123-2134},
   ISSN = {1467-7652 (Electronic)
1467-7644 (Print)
1467-7644 (Linking)},
   DOI = {10.1111/pbi.13893},
   url = {https://www.ncbi.nlm.nih.gov/pubmed/35842742},
   year = {2022},
   type = {Journal Article}
}

@article{RN325,
   author = {Yang, Wenyi and Wang, Pingping and Luo, Meng and Cai, Yideng and Xu, Chang and Xue, Guangfu and Jin, Xiyun and Cheng, Rui and Que, Jinhao and Pang, Fenglan},
   title = {{DeepCCI}: a deep learning framework for identifying cell–cell interactions from single-cell {RNA} sequencing data},
   journal = {Bioinformatics},
   volume = {39},
   number = {10},
   pages = {btad596},
   ISSN = {1367-4811},
   DOI = {10.1101/2022.11.11.516061},
   year = {2023},
   type = {Journal Article}
}

@article{RN412DeepTalk,
   author = {Yang, Wenyi and Wang, Pingping and Xu, Shouping and Wang, Tao and Luo, Meng and Cai, Yideng and Xu, Chang and Xue, Guangfu and Que, Jinhao and Ding, Qian},
   title = {Deciphering cell–cell communication at single-cell resolution for spatial transcriptomics with subgraph-based graph attention network},
   journal = {Nature Communications},
   volume = {15},
   number = {1},
   pages = {7101},
   ISSN = {2041-1723},
   year = {2024},
   type = {Journal Article}
}

@article{RN294scTenifoldXct,
   author = {Yang, Y. and Li, G. and Zhong, Y. and Xu, Q. and Lin, Y. T. and Roman-Vicharra, C. and Chapkin, R. S. and Cai, J. J.},
   title = {{scTenifoldXct}: A semi-supervised method for predicting cell-cell interactions and mapping cellular communication graphs},
   journal = {Cell Syst},
   volume = {14},
   number = {4},
   pages = {302-311 e4},
   ISSN = {2405-4720 (Electronic)
2405-4712 (Print)
2405-4712 (Linking)},
   DOI = {10.1016/j.cels.2023.01.004},
   url = {https://www.ncbi.nlm.nih.gov/pubmed/36787742},
   year = {2023},
   type = {Journal Article}
}

@article{RN363,
   author = {Yuan, Ye and Bar-Joseph, Ziv},
   title = {{GCNG}: graph convolutional networks for inferring gene interaction from spatial transcriptomics data},
   journal = {Genome Biology},
   volume = {21},
   number = {1},
   pages = {300},
   ISSN = {1474-760X},
   DOI = {10.1186/s13059-020-02214-w},
   url = {https://doi.org/10.1186/s13059-020-02214-w},
   year = {2020},
   type = {Journal Article}
}

@article{RN285,
   author = {Yuan, Y. and Cosme, C., Jr. and Adams, T. S. and Schupp, J. and Sakamoto, K. and Xylourgidis, N. and Ruffalo, M. and Li, J. and Kaminski, N. and Bar-Joseph, Z.},
   title = {{CINS}: Cell Interaction Network inference from Single cell expression data},
   journal = {PLoS Comput Biol},
   volume = {18},
   number = {9},
   pages = {e1010468},
   ISSN = {1553-7358 (Electronic)
1553-734X (Print)
1553-734X (Linking)},
   DOI = {10.1371/journal.pcbi.1010468},
   url = {https://www.ncbi.nlm.nih.gov/pubmed/36095011},
   year = {2022},
   type = {Journal Article}
}

@article{RN433,
   author = {Zhang, Han and Cui, Ting and Xu, Xiaoqiang and Sui, Guangyu and Fang, Qiaoli and Yang, Guanghao and Gong, Yizhen and Yang, Sanqiao and Lv, Yufei and Shang, Desi},
   title = {{SpaGraphCCI}: Spatial cell–cell communication inference through {GAT}‐based co‐convolutional feature integration},
   journal = {IET Systems Biology},
   volume = {19},
   number = {1},
   pages = {e70000},
   ISSN = {1751-8849},
   year = {2025},
   type = {Journal Article}
}

@article{RN404CellGAT,
   author = {Zhang, Tianjiao and Wu, Zhenao and Li, Liangyu and Ren, Jixiang and Zhang, Ziheng and Zhang, Jingyu and Wang, Guohua},
   title = {{CellGAT}: A {GAT}-Based Method for Constructing a Cell Communication Network Integrating Multiomics Information},
   journal = {Biomolecules},
   volume = {15},
   number = {3},
   pages = {342},
   ISSN = {2218-273X},
   url = {https://www.mdpi.com/2218-273X/15/3/342},
   year = {2025},
   type = {Journal Article}
}

@article{RN268,
   author = {Zhang, Y. and Liu, T. and Hu, X. and Wang, M. and Wang, J. and Zou, B. and Tan, P. and Cui, T. and Dou, Y. and Ning, L. and Huang, Y. and Rao, S. and Wang, D. and Zhao, X.},
   title = {{CellCall}: integrating paired ligand-receptor and transcription factor activities for cell-cell communication},
   journal = {Nucleic Acids Res},
   volume = {49},
   number = {15},
   pages = {8520-8534},
   ISSN = {1362-4962 (Electronic)
0305-1048 (Print)
0305-1048 (Linking)},
   DOI = {10.1093/nar/gkab638},
   url = {https://www.ncbi.nlm.nih.gov/pubmed/34331449},
   year = {2021},
   type = {Journal Article}
}

@article{RN430SEnSCA,
   author = {Zhou, Liqian and Wang, Xiwen and Peng, Lihong and Chen, Min and Wen, Hong},
   title = {{SEnSCA}: Identifying possible ligand‐receptor interactions and its application in cell–cell communication inference},
   journal = {Journal of Cellular and Molecular Medicine},
   volume = {28},
   number = {9},
   pages = {e18372},
   ISSN = {1582-1838},
   year = {2024},
   type = {Journal Article}
}

@article{RN417IGAN,
   author = {Zhu, Junchao and Dai, Hao and Chen, Luonan},
   title = {Revealing cell–cell communication pathways with their spatially coupled gene programs},
   journal = {Briefings in Bioinformatics},
   volume = {25},
   number = {3},
   pages = {bbae202},
   ISSN = {1467-5463},
   year = {2024},
   type = {Journal Article}
}

@article{RN432Spacia,
   author = {Zhu, James and Wang, Yunguan and Chang, Woo Yong and Malewska, Alicia and Napolitano, Fabiana and Gahan, Jeffrey C and Unni, Nisha and Zhao, Min and Yuan, Rongqing and Wu, Fangjiang},
   title = {Mapping cellular interactions from spatially resolved transcriptomics data},
   journal = {Nature methods},
   volume = {21},
   number = {10},
   pages = {1830-1842},
   ISSN = {1548-7091},
   year = {2024},
   type = {Journal Article}
}

@article{zohora2025cellnest,
  title={{CellNEST} reveals cell--cell relay networks using attention mechanisms on spatial transcriptomics},
  author={Zohora, Fatema Tuz and Paliwal, Deisha and Flores-Figueroa, Eugenia and Li, Joshua and Gao, Tingxiao and Notta, Faiyaz and Schwartz, Gregory W},
  journal={Nature Methods},
  volume = {22},
  pages={1505--1519},
  year={2025},
  publisher={Nature Publishing Group US New York}
}

@article{RN449,
   author = {Almet, Axel A. and Cang, Zixuan and Jin, Suoqin and Nie, Qing},
   title = {The landscape of cell–cell communication through single-cell transcriptomics},
   journal = {Current Opinion in Systems Biology},
   volume = {26},
   pages = {12-23},
   ISSN = {2452-3100},
   DOI = {https://doi.org/10.1016/j.coisb.2021.03.007},
   url = {https://www.sciencedirect.com/science/article/pii/S2452310021000081},
   year = {2021},
   type = {Journal Article}
}

@article{CCCreviewNie2023,
   author = {Wang, X. and Almet, A. A. and Nie, Q.},
   title = {The promising application of cell-cell interaction analysis in cancer from single-cell and spatial transcriptomics},
   journal = {Semin Cancer Biol},
   volume = {95},
   pages = {42-51},
   DOI = {10.1016/j.semcancer.2023.07.001},
   url = {https://www.ncbi.nlm.nih.gov/pubmed/37454878},
   year = {2023},
   type = {Journal Article}
}

@article{RN447,
   author = {Armingol, Erick and Baghdassarian, Hratch M and Lewis, Nathan E},
   title = {The diversification of methods for studying cell–cell interactions and communication},
   journal = {Nature Reviews Genetics},
   volume = {25},
   number = {6},
   pages = {381-400},
   ISSN = {1471-0056},
   year = {2024},
   type = {Journal Article}
}

@article{RN448,
   author = {Armingol, Erick and Officer, Adam and Harismendy, Olivier and Lewis, Nathan E},
   title = {Deciphering cell–cell interactions and communication from gene expression},
   journal = {Nature Reviews Genetics},
   volume = {22},
   number = {2},
   pages = {71-88},
   ISSN = {1471-0056},
   year = {2021},
   type = {Journal Article}
}

@incollection{RN439,
  title={An introduction to early developmental processes},
  author={Gilbert, Scott F},
  booktitle={Developmental Biology. 6th edition},
  year={2000},
  publisher={Sinauer Associates}
}

@article{RN445,
   author = {Hill, Steven M},
   title = {Receptor crosstalk: communication through cell signaling pathways},
   journal = {The Anatomical Record: An Official Publication of the American Association of Anatomists},
   volume = {253},
   number = {2},
   pages = {42-48},
   ISSN = {0003-276X},
   year = {1998},
   type = {Journal Article}
}

@article{RN444,
   author = {Meier-Schellersheim, Martin and Varma, Rajat and Angermann, Bastian R},
   title = {Mechanistic models of cellular signaling, cytokine crosstalk, and cell-cell communication in immunology},
   journal = {Frontiers in immunology},
   volume = {10},
   pages = {2268},
   ISSN = {1664-3224},
   year = {2019},
   type = {Journal Article}
}

@article{RN442,
   author = {Schmitz, M Lienhard and Weber, Axel and Roxlau, Thomas and Gaestel, Matthias and Kracht, Michael},
   title = {Signal integration, crosstalk mechanisms and networks in the function of inflammatory cytokines},
   journal = {Biochimica et Biophysica Acta (BBA)-Molecular Cell Research},
   volume = {1813},
   number = {12},
   pages = {2165-2175},
   ISSN = {0167-4889},
   year = {2011},
   type = {Journal Article}
}

@article{kanehisa2000kegg,
  title={{KEGG}: kyoto encyclopedia of genes and genomes},
  author={Kanehisa, Minoru and Goto, Susumu},
  journal={Nucleic acids research},
  volume={28},
  number={1},
  pages={27--30},
  year={2000},
  publisher={Oxford University Press}
}

@article{croft2010reactome,
  title={{Reactome}: a database of reactions, pathways and biological processes},
  author={Croft, David and O`kelly, Gavin and Wu, Guanming and Haw, Robin and Gillespie, Marc and Matthews, Lisa and Caudy, Michael and Garapati, Phani and Gopinath, Gopal and Jassal, Bijay and others},
  journal={Nucleic acids research},
  volume={39},
  number={suppl\_1},
  pages={D691--D697},
  year={2010},
  publisher={Oxford University Press}
}

@article{wang2019cell,
  title={Cell lineage and communication network inference via optimization for single-cell transcriptomics},
  author={Wang, Shuxiong and Karikomi, Matthew and MacLean, Adam L and Nie, Qing},
  journal={Nucleic acids research},
  volume={47},
  number={11},
  pages={e66--e66},
  year={2019},
  publisher={Oxford University Press}
}

@article{ProximID2018,
   author = {Boisset, J. C. and Vivie, J. and Grun, D. and Muraro, M. J. and Lyubimova, A. and van Oudenaarden, A.},
   title = {Mapping the physical network of cellular interactions},
   journal = {Nat Methods},
   volume = {15},
   number = {7},
   pages = {547-553},
   DOI = {10.1038/s41592-018-0009-z},
   url = {https://www.ncbi.nlm.nih.gov/pubmed/29786092},
   year = {2018},
   type = {Journal Article}
}

@article{RABID-seq2021,
   author = {Clark, I. C. and Gutierrez-Vazquez, C. and Wheeler, M. A. and Li, Z. and Rothhammer, V. and Linnerbauer, M. and Sanmarco, L. M. and Guo, L. and Blain, M. and Zandee, S. E. J. and Chao, C. C. and Batterman, K. V. and Schwabenland, M. and Lotfy, P. and Tejeda-Velarde, A. and Hewson, P. and Manganeli Polonio, C. and Shultis, M. W. and Salem, Y. and Tjon, E. C. and Fonseca-Castro, P. H. and Borucki, D. M. and Alves de Lima, K. and Plasencia, A. and Abate, A. R. and Rosene, D. L. and Hodgetts, K. J. and Prinz, M. and Antel, J. P. and Prat, A. and Quintana, F. J.},
   title = {Barcoded viral tracing of single-cell interactions in central nervous system inflammation},
   journal = {Science},
   volume = {372},
   number = {6540},
   DOI = {10.1126/science.abf1230},
   url = {https://www.ncbi.nlm.nih.gov/pubmed/33888612},
   year = {2021},
   type = {Journal Article}
}

@article{PIC-seq2020,
   author = {Giladi, A. and Cohen, M. and Medaglia, C. and Baran, Y. and Li, B. and Zada, M. and Bost, P. and Blecher-Gonen, R. and Salame, T. M. and Mayer, J. U. and David, E. and Ronchese, F. and Tanay, A. and Amit, I.},
   title = {Dissecting cellular crosstalk by sequencing physically interacting cells},
   journal = {Nat Biotechnol},
   volume = {38},
   number = {5},
   pages = {629-637},
   DOI = {10.1038/s41587-020-0442-2},
   url = {https://www.ncbi.nlm.nih.gov/pubmed/32152598},
   year = {2020},
   type = {Journal Article}
}

@article{benchmarkCCCWang,
   author = {Liu, Z. and Sun, D. and Wang, C.},
   title = {Evaluation of cell-cell interaction methods by integrating single-cell {RNA} sequencing data with spatial information},
   journal = {Genome Biol},
   volume = {23},
   number = {1},
   pages = {218},
   ISSN = {1474-760X (Electronic)
1474-7596 (Print)
1474-7596 (Linking)},
   DOI = {10.1186/s13059-022-02783-y},
   url = {https://www.ncbi.nlm.nih.gov/pubmed/36253792},
   year = {2022},
   type = {Journal Article}
}

@article{CCCreviewSu2024,
   author = {Su, J. and Song, Y. and Zhu, Z. and Huang, X. and Fan, J. and Qiao, J. and Mao, F.},
   title = {Cell-cell communication: new insights and clinical implications},
   journal = {Signal Transduct Target Ther},
   volume = {9},
   number = {1},
   pages = {196},
   ISSN = {2059-3635 (Electronic)
2095-9907 (Print)
2059-3635 (Linking)},
   DOI = {10.1038/s41392-024-01888-z},
   url = {https://www.ncbi.nlm.nih.gov/pubmed/39107318},
   year = {2024},
   type = {Journal Article}
}

@article{BRICseq2020,
   author = {Huang, L. and Kebschull, J. M. and Furth, D. and Musall, S. and Kaufman, M. T. and Churchland, A. K. and Zador, A. M.},
   title = {{BRICseq} Bridges Brain-wide Interregional Connectivity to Neural Activity and Gene Expression in Single Animals},
   journal = {Cell},
   volume = {183},
   number = {7},
   pages = {2040},
   DOI = {10.1016/j.cell.2020.12.009},
   url = {https://www.ncbi.nlm.nih.gov/pubmed/33357401},
   year = {2020},
   type = {Journal Article}
}

@article{CCCreviewFan2020,
   author = {Shao, X. and Lu, X. Y. and Liao, J. and Chen, H. J. and Fan, X. H.},
   title = {New avenues for systematically inferring cell-cell communication: through single-cell transcriptomics data},
   journal = {Protein \& Cell},
   volume = {11},
   number = {12},
   pages = {866-880},
   DOI = {10.1007/s13238-020-00727-5},
   url = {<Go to ISI>://WOS:000534433300001},
   year = {2020},
   type = {Journal Article}
}

@article{CCCreviewJin2022,
   author = {Jin, S. and Ramos, R.},
   title = {Computational exploration of cellular communication in skin from emerging single-cell and spatial transcriptomic data},
   journal = {Biochem Soc Trans},
   volume = {50},
   number = {1},
   ISSN = {1470-8752 (Electronic)
0300-5127 (Print)
0300-5127 (Linking)},
   DOI = {10.1042/BST20210863},
   url = {https://www.ncbi.nlm.nih.gov/pubmed/35191953},
   year = {2022},
   type = {Journal Article}
}

@article{cesaro2025advances,
    author = {Cesaro, Giulia and Nagai, James Shiniti and Gnoato, Nicolò and Chiodi, Alice and Tussardi, Gaia and Klöker, Vanessa and Musumarra, Carmelo Vittorio and Mosca, Ettore and Costa, Ivan G and Di Camillo, Barbara and Calura, Enrica and Baruzzo, Giacomo},
    title = {Advances and challenges in cell–cell communication inference: a comprehensive review of tools, resources, and future directions},
    journal = {Briefings in Bioinformatics},
    volume = {26},
    number = {3},
    pages = {bbaf280},
    year = {2025},
    month = {06},
    issn = {1477-4054},
    doi = {10.1093/bib/bbaf280},
    url = {https://doi.org/10.1093/bib/bbaf280},
    eprint = {https://academic.oup.com/bib/article-pdf/26/3/bbaf280/63527857/bbaf280.pdf},
}

@article{heldin2016signals,
  title={Signals and receptors},
  author={Heldin, Carl-Henrik and Lu, Benson and Evans, Ron and Gutkind, J Silvio},
  journal={Cold Spring Harbor perspectives in biology},
  volume={8},
  number={4},
  pages={a005900},
  year={2016},
  publisher={Cold Spring Harbor Lab}
}

@misc{clark2018biology,
  title={Biology},
  author={Clark, Mary Ann and Choi, Jung and Douglas, Matthew},
  year={2018},
  publisher={OpenStax}
}

@article{kornberg2014communicating,
  title={Communicating by touch--neurons are not alone},
  author={Kornberg, Thomas B and Roy, Sougata},
  journal={Trends in cell biology},
  volume={24},
  number={6},
  pages={370-376},
  year={2014},
  publisher={Elsevier}
}

@article{orzechowska2023sonic,
  title={Sonic hedgehog and {WNT} signaling regulate a positive feedback loop between intestinal epithelial and stromal cells to promote epithelial regeneration},
  author={Orzechowska-Licari, Emilia J and Bialkowska, Agnieszka B and Yang, Vincent W},
  journal={Cellular and Molecular Gastroenterology and Hepatology},
  volume={16},
  number={4},
  pages={607-642},
  year={2023},
  publisher={Elsevier}
}

@article{peng2025predicting,
  title={Predicting cell--cell communication by combining heterogeneous ensemble deep learning and weighted geometric mean},
  author={Peng, Lihong and Liu, Longlong and Huang, Liangliang and Bai, Zongzheng and Chen, Min and Chen, Xing},
  journal={Applied Soft Computing},
  volume={172},
  pages={112839},
  year={2025},
  publisher={Elsevier}
}

@article{feng2025orgaccc,
  title={{OrgaCCC}: Orthogonal graph autoencoders for constructing cell-cell communication networks on spatial transcriptomics data},
  author={Feng, Xixuan and Zhang, Shuqin and Li, Limin},
  journal={PLOS Computational Biology},
  volume={21},
  number={6},
  pages={e1013212},
  year={2025},
  publisher={Public Library of Science San Francisco, CA USA}
}

@article{qi2025interpretable,
  title={Interpretable niche-based cell--cell communication inference using multi-view graph neural networks},
  author={Qi, Juntian and Luo, Zhengchao and Li, Chuan-Yun and Wang, Jinzhuo and Ding, Wanqiu},
  journal={Nature Computational Science},
  pages={1-12},
  year={2025},
  publisher={Nature Publishing Group US New York}
}

@article{Xiao20240821608964GITIII,
  title={Inferring spatial single-cell-level interactions through interpreting cell state and niche correlations learned by self-supervised graph transformer},
  author={Xiao, Xiao and Zhang, Le and Zhao, Hongyu and Wang, Zuoheng},
  journal={Nature Machine Intelligence},
  pages={1--17},
  year={2025},
  publisher={Nature Publishing Group UK London}
}

@article{zhang2023defining,
  title={Defining and identifying cell sub-crosstalk pairs for characterizing cell--cell communication patterns},
  author={Zhang, Chenxing and Hu, Yuxuan and Gao, Lin},
  journal={Scientific Reports},
  volume={13},
  number={1},
  pages={15746},
  year={2023},
  publisher={Nature Publishing Group UK London}
}

@article{jerby2022dialogueDIALOGUE,
  title={{DIALOGUE} maps multicellular programs in tissue from single-cell or spatial transcriptomics data},
  author={Jerby-Arnon, Livnat and Regev, Aviv},
  journal={Nature biotechnology},
  volume={40},
  number={10},
  pages={1467-1477},
  year={2022},
  publisher={Nature Publishing Group US New York}
}

@article{flores2023multicellular,
  title={Multicellular factor analysis of single-cell data for a tissue-centric understanding of disease},
  author={Flores, Ricardo Omar Ramirez and Lanzer, Jan David and Dimitrov, Daniel and Velten, Britta and Saez-Rodriguez, Julio},
  journal={Elife},
  volume={12},
  pages={e93161},
  year={2023},
  publisher={eLife Sciences Publications Limited}
}

@article{argelaguet2018multi,
  title={Multi-Omics Factor Analysis—a framework for unsupervised integration of multi-omics data sets},
  author={Argelaguet, Ricard and Velten, Britta and Arnol, Damien and Dietrich, Sascha and Zenz, Thorsten and Marioni, John C and Buettner, Florian and Huber, Wolfgang and Stegle, Oliver},
  journal={Molecular systems biology},
  volume={14},
  number={6},
  pages={e8124},
  year={2018}
}

@article{argelaguet2020mofa,
  title={{MOFA+}: a statistical framework for comprehensive integration of multi-modal single-cell data},
  author={Argelaguet, Ricard and Arnol, Damien and Bredikhin, Danila and Deloro, Yonatan and Velten, Britta and Marioni, John C and Stegle, Oliver},
  journal={Genome biology},
  volume={21},
  number={1},
  pages={111},
  year={2020},
  publisher={Springer}
}

@article{Armingol20221102514917,
	author = {Armingol, Erick and Larsen, Reid O. and Gale, Lia and Cequeira, Martin and Baghdassarian, Hratch and Lewis, Nathan E.},
	title = {{Tensor-cell2cell} v2 unravels coordinated dynamics of protein- and metabolite-mediated cell-cell communication},
	elocation-id = {2022.11.02.514917},
	year = {2025},
	doi = {10.1101/2022.11.02.514917},
	publisher = {Cold Spring Harbor Laboratory},
	URL = {https://www.biorxiv.org/content/early/2025/09/01/2022.11.02.514917},
	eprint = {https://www.biorxiv.org/content/early/2025/09/01/2022.11.02.514917.full.pdf},
	journal = {bioRxiv}
}

@article{gao2024macc,
  title={{MACC}: a visual interactive knowledgebase of metabolite-associated cell communications},
  author={Gao, Jian and Mo, Saifeng and Wang, Jun and Zhang, Mou and Shi, Yao and Zhu, Chuhan and Shang, Yuxuan and Tang, Xinyue and Zhang, Shiyue and Wu, Xinwen and others},
  journal={Nucleic Acids Research},
  volume={52},
  number={D1},
  pages={D633-D639},
  year={2024},
  publisher={Oxford University Press}
}

@article{farr2024metalinksdb,
  title={{MetalinksDB}: a flexible and contextualizable resource of metabolite-protein interactions},
  author={Farr, Elias and Dimitrov, Daniel and Schmidt, Christina and Turei, Denes and Lobentanzer, Sebastian and Dugourd, Aurelien and Saez-Rodriguez, Julio},
  journal={Briefings in Bioinformatics},
  volume={25},
  number={4},
  pages={bbae347},
  year={2024},
  publisher={Oxford University Press}
}

@article{zhang2024predicting,
  title={Predicting intercellular communication based on metabolite-related ligand-receptor interactions with {MRCLinkdb}},
  author={Zhang, Yuncong and Yang, Yu and Ren, Liping and Zhan, Meixiao and Sun, Taoping and Zou, Quan and Zhang, Yang},
  journal={BMC biology},
  volume={22},
  number={1},
  pages={152},
  year={2024},
  publisher={Springer}
}

@article{chen2024ev,
  title={{EV-COMM}: A database of interspecies and intercellular interactions mediated by extracellular vesicles},
  author={Chen, Jingyu and Lin, Jing-Jing and Wang, Weiyi and Huang, Haining and Pan, Zhizhen and Ye, Guozhu and Dong, Sijun and Lin, Yi and Lin, Congtian and Huang, Qiansheng},
  journal={Journal of Extracellular Vesicles},
  volume={13},
  number={4},
  pages={e12442},
  year={2024},
  publisher={Wiley Online Library}
}

@article{wheeler2023droplet,
  title={Droplet-based forward genetic screening of astrocyte--microglia cross-talk},
  author={Wheeler, Michael A and Clark, Iain C and Lee, Hong-Gyun and Li, Zhaorong and Linnerbauer, Mathias and Rone, Joseph M and Blain, Manon and Akl, Camilo Faust and Piester, Gavin and Giovannoni, Federico and others},
  journal={Science},
  volume={379},
  number={6636},
  pages={1023-1030},
  year={2023},
  publisher={American Association for the Advancement of Science}
}

@article{Armingol20250509653038,
	author = {Armingol, Erick and Ashcroft, James and Mareckova, Magda and Prete, Martin and Lorenzi, Valentina and Mazzeo, Cecilia Icoresi and Lee, Jimmy Tsz Hang and Moullet, Marie and Bayraktar, Omer Ali and Becker, Christian and Zondervan, Krina and Garcia-Alonso, Luz and Lewis, Nathan E. and Vento-Tormo, Roser},
	title = {Atlas-scale metabolic activities inferred from single-cell and spatial transcriptomics},
	elocation-id = {2025.05.09.653038},
	year = {2025},
	doi = {10.1101/2025.05.09.653038},
	publisher = {Cold Spring Harbor Laboratory},
	URL = {https://www.biorxiv.org/content/early/2025/05/14/2025.05.09.653038},
	eprint = {https://www.biorxiv.org/content/early/2025/05/14/2025.05.09.653038.full.pdf},
	journal = {bioRxiv}
}

@article{miceli2024extracellular,
  title={Extracellular vesicles, {RNA} sequencing, and bioinformatic analyses: Challenges, solutions, and recommendations},
  author={Miceli, Rebecca T and Chen, Tzu-Yi and Nose, Yohei and Tichkule, Swapnil and Brown, Briana and Fullard, John F and Saulsbury, Marilyn D and Heyliger, Simon O and Gnjatic, Sacha and Kyprianou, Natasha and others},
  journal={Journal of Extracellular Vesicles},
  volume={13},
  number={12},
  pages={e70005},
  year={2024},
  publisher={Wiley Online Library}
}

@article{luo2022transcriptomic,
  title={Transcriptomic features in a single extracellular vesicle via single-cell {RNA} sequencing},
  author={Luo, Tao and Chen, Si-Yi and Qiu, Zhi-Xin and Miao, Ya-Ru and Ding, Yue and Pan, Xiang-Yu and Li, Yirong and Lei, Qian and Guo, An-Yuan},
  journal={Small methods},
  volume={6},
  number={11},
  pages={2200881},
  year={2022},
  publisher={Wiley Online Library}
}

@article{he2024sevtras,
  title={{SEVtras} delineates small extracellular vesicles at droplet resolution from single-cell transcriptomes},
  author={He, Ruiqiao and Zhu, Junjie and Ji, Peifeng and Zhao, Fangqing},
  journal={Nature Methods},
  volume={21},
  number={2},
  pages={259-266},
  year={2024},
  publisher={Nature Publishing Group US New York}
}

@article{Zhao20250624660984,
	author = {Zhao, Yiyong and Chintalapudi, Himanshu and Xu, Ziqian and Liu, Weiqiang and Hu, Yuxuan and Grassin, Ewa and Song, Minsun and Hong, SoonGweon and Lee, Luke P. and Dong, Xianjun},
	title = {{EVscope}: A Comprehensive Bioinformatics Pipeline for Accurate and Robust Analysis of Total {RNA} Sequencing from Extracellular Vesicles},
	elocation-id = {2025.06.24.660984},
	year = {2025},
	doi = {10.1101/2025.06.24.660984},
	publisher = {Cold Spring Harbor Laboratory},
	URL = {https://www.biorxiv.org/content/early/2025/06/27/2025.06.24.660984},
	eprint = {https://www.biorxiv.org/content/early/2025/06/27/2025.06.24.660984.full.pdf},
	journal = {bioRxiv}
}

@article{blair2025phospho,
  title={{Phospho-seq}: Integrated, multi-modal profiling of intracellular protein dynamics in single cells},
  author={Blair, John D and Hartman, Austin and Zenk, Fides and Wahle, Philipp and Brancati, Giovanna and Dalgarno, Carol and Treutlein, Barbara and Satija, Rahul},
  journal={Nature communications},
  volume={16},
  number={1},
  pages={1346},
  year={2025},
  publisher={Nature Publishing Group UK London}
}

@article{Opzoomer20240223581433,
	author = {Opzoomer, James W. and O{\textquoteright}Sullivan, Rhianna and Sufi, Jahangir and Madsen, Ralitsa and Qin, Xiao and Basiarz, Ewa and Tape, Christopher J.},
	title = {{SIGNAL-seq}: Multimodal Single-cell Inter- and Intra-cellular Signalling Analysis},
	elocation-id = {2024.02.23.581433},
	year = {2024},
	doi = {10.1101/2024.02.23.581433},
	publisher = {Cold Spring Harbor Laboratory},
	URL = {https://www.biorxiv.org/content/early/2024/03/11/2024.02.23.581433},
	eprint = {https://www.biorxiv.org/content/early/2024/03/11/2024.02.23.581433.full.pdf},
	journal = {bioRxiv}
}

@article{muller2011extracellular,
  title={Extracellular movement of signaling molecules},
  author={M{\"u}ller, Patrick and Schier, Alexander F},
  journal={Developmental cell},
  volume={21},
  number={1},
  pages={145-158},
  year={2011},
  publisher={Elsevier}
}

@book{purves2017neuroscience,
  title = {Neuroscience},
  author = {Purves, D and Augustine, GJ and Fitzpatrick, D and et al.},
  edition = {6th},
  isbn={9781605356372},
  year={2017},
  publisher={Oxford University Press, Incorporated}
}

@article{gruteser2022examination,
  title={Examination of intracellular {GPCR}-mediated signaling with high temporal resolution},
  author={Gruteser, Nadine and Baumann, Arnd},
  journal={International journal of molecular sciences},
  volume={23},
  number={15},
  pages={8516},
  year={2022},
  publisher={MDPI}
}

@book{alberts2022molecular,
  title={Molecular Biology of the Cell},
  author={Alberts, B. and Heald, R. and Johnson, A. and Morgan, D. and Raff, M. and Roberts, K. and Walter, P.},
  edition = {7th},
  isbn={9780393884647},
  year={2022},
  publisher={W. W. Norton, Incorporated}
}

@article{zhang2025modelingcelldynamicsinteractions,
   title={Modeling Cell Dynamics and Interactions with Unbalanced Mean Field Schr\"odinger Bridge}, 
   author={Zhenyi Zhang and Zihan Wang and Yuhao Sun and Tiejun Li and Peijie Zhou},
   year={2025},
   eprint={2505.11197},
   archivePrefix={arXiv},
   journal={arXiv preprint arXiv:2505.11197},
   primaryClass={cs.LG},
   url={https://arxiv.org/abs/2505.11197}, 
}

@article{zhang2025deciphering,
  title={Deciphering cell-fate trajectories using spatiotemporal single-cell transcriptomic data},
  author={Zhang, Zhenyi and Wang, Zihan and Sun, Yuhao and Shen, Jiantao and Peng, Qiangwei and Li, Tiejun and Zhou, Peijie},
  journal={Authorea Preprints, September 2025e. doi},
  volume={10},
  year={2025}
}

@article{hao2024large,
  title={Large-scale foundation model on single-cell transcriptomics},
  author={Hao, Minsheng and Gong, Jing and Zeng, Xin and Liu, Chiming and Guo, Yucheng and Cheng, Xingyi and Wang, Taifeng and Ma, Jianzhu and Zhang, Xuegong and Song, Le},
  journal={Nature methods},
  volume={21},
  number={8},
  pages={1481-1491},
  year={2024},
  publisher={Nature Publishing Group US New York}
}

@article{van2022challenges,
  title={Challenges and directions in studying cell--cell communication by extracellular vesicles},
  author={van Niel, Guillaume and Carter, David RF and Clayton, Aled and Lambert, Daniel W and Raposo, Gra{\c{c}}a and Vader, Pieter},
  journal={Nature reviews Molecular cell biology},
  volume={23},
  number={5},
  pages={369-382},
  year={2022},
  publisher={Nature Publishing Group UK London}
}

@article{gurung2021exosome,
  title={The exosome journey: from biogenesis to uptake and intracellular signalling},
  author={Gurung, Sonam and Perocheau, Dany and Touramanidou, Loukia and Baruteau, Julien},
  journal={Cell Communication and Signaling},
  volume={19},
  number={1},
  pages={47},
  year={2021},
  publisher={Springer}
}

@article{liu2023review,
  title={A review of the regulatory mechanisms of extracellular vesicles-mediated intercellular communication},
  author={Liu, Ya-Juan and Wang, Cheng},
  journal={Cell Communication and Signaling},
  volume={21},
  number={1},
  pages={77},
  year={2023},
  publisher={Springer}
}

@article{gao2022cellcallext,
  title={{CellCallEXT}: analysis of ligand--receptor and transcription factor activities in cell--cell communication of tumor immune microenvironment},
  author={Gao, Shouguo and Feng, Xingmin and Wu, Zhijie and Kajigaya, Sachiko and Young, Neal S},
  journal={Cancers},
  volume={14},
  number={19},
  pages={4957},
  year={2022},
  publisher={MDPI}
}

@article{witten2009penalized,
  title={A penalized matrix decomposition, with applications to sparse principal components and canonical correlation analysis},
  author={Witten, Daniela M and Tibshirani, Robert and Hastie, Trevor},
  journal={Biostatistics},
  volume={10},
  number={3},
  pages={515--534},
  year={2009},
  publisher={Oxford University Press}
}

@article{tucker1966some,
  title={Some mathematical notes on three-mode factor analysis},
  author={Tucker, Ledyard R},
  journal={Psychometrika},
  volume={31},
  number={3},
  pages={279--311},
  year={1966},
  publisher={Springer}
}

@article{carroll1970analysis,
  title={Analysis of individual differences in multidimensional scaling via an {N}-way generalization of {“Eckart-Young”} decomposition},
  author={Carroll, J Douglas and Chang, Jih-Jie},
  journal={Psychometrika},
  volume={35},
  number={3},
  pages={283--319},
  year={1970},
  publisher={Springer-Verlag}
}

@article{harshman1970foundations,
  title={Foundations of the {PARAFAC} procedure: Models and conditions for an “explanatory” multi-modal factor analysis},
  author={Harshman, Richard A and others},
  journal={UCLA working papers in phonetics},
  volume={16},
  number={1},
  pages={84},
  year={1970},
  publisher={Los Angeles, CA}
}

@article{Dai20231215571918,
	author = {Dai, Qile and Yang, Jingjing and Epstein, Michael P.},
	title = {Identifying condition-related cell-cell communication events using supervised tensor analysis},
	elocation-id = {2023.12.15.571918},
	year = {2025},
	doi = {10.1101/2023.12.15.571918},
	publisher = {Cold Spring Harbor Laboratory},
	URL = {https://www.biorxiv.org/content/early/2025/06/19/2023.12.15.571918},
	eprint = {https://www.biorxiv.org/content/early/2025/06/19/2023.12.15.571918.full.pdf},
	journal = {bioRxiv}
}

@article{burdziak2023epigenetic,
  title={Epigenetic plasticity cooperates with cell-cell interactions to direct pancreatic tumorigenesis},
  author={Burdziak, Cassandra and Alonso-Curbelo, Direna and Walle, Thomas and Reyes, Jos{\'e} and Barriga, Francisco M and Haviv, Doron and Xie, Yubin and Zhao, Zhen and Zhao, Chujun Julia and Chen, Hsuan-An and others},
  journal={Science},
  volume={380},
  number={6645},
  pages={eadd5327},
  year={2023},
  publisher={American Association for the Advancement of Science}
}

@book{LimMayer2024,
  title = {Cell Signaling, 2nd edition: Principles and Mechanisms},
  author = {Lim, W. A. and Mayer, B. J.},
  year = {2024},
  edition = {2nd},
  publisher = {CRC Press},
  doi = {10.1201/9780429298844},
  url = {https://doi.org/10.1201/9780429298844}
}

@article{atakan2014molecular,
  title={Molecular Communications and Nanonetworks},
  author={Atakan, Baris},
  journal={PSpringer-Verlag New York},
  volume={1},
  pages={1--103},
  year={2014},
  publisher={Springer}
}

@article{fagotto1996cell,
  title={Cell contact-dependent signaling},
  author={Fagotto, Fran{\c{c}}ois and Gumbiner, Barry M},
  journal={Developmental biology},
  volume={180},
  number={2},
  pages={445--454},
  year={1996},
  publisher={Elsevier}
}

@article{desjardins1981endocrine,
  title={Endocrine signaling and male reproduction},
  author={Desjardins, Claude},
  journal={Biology of reproduction},
  volume={24},
  number={1},
  pages={1--21},
  year={1981},
  publisher={Oxford University Press}
}

@article{newton2016second,
  title={Second messengers},
  author={Newton, Alexandra C and Bootman, Martin D and Scott, John D},
  journal={Cold Spring Harbor perspectives in biology},
  volume={8},
  number={8},
  pages={a005926},
  year={2016},
  publisher={Cold Spring Harbor Lab}
}

@article{nishi2015crosstalk,
  title={Crosstalk between signaling pathways provided by single and multiple protein phosphorylation sites},
  author={Nishi, Hafumi and Demir, Emek and Panchenko, Anna R},
  journal={Journal of molecular biology},
  volume={427},
  number={2},
  pages={511--520},
  year={2015},
  publisher={Elsevier}
}

@article{RN454stMLnet,
  title={Dissecting multilayer cell--cell communications with signaling feedback loops from spatial transcriptomics data},
  author={Yan, Lulu and Cheng, Jinyu and Nie, Qing and Sun, Xiaoqiang},
  journal={Genome Research},
  volume={35},
  number={6},
  pages={1400--1414},
  year={2025},
  publisher={Cold Spring Harbor Lab}
}

@article {LiangSteamboat,
	author = {Liang, Shaoheng and Tang, Junjie and Wang, Guanghan and Ma, Jian},
	title = {{Steamboat}: Attention-based multiscale delineation of cellular interactions in tissues},
	elocation-id = {2025.04.06.647437},
	year = {2025},
	doi = {10.1101/2025.04.06.647437},
	publisher = {Cold Spring Harbor Laboratory},
	URL = {https://www.biorxiv.org/content/early/2025/04/10/2025.04.06.647437},
	eprint = {https://www.biorxiv.org/content/early/2025/04/10/2025.04.06.647437.full.pdf},
	journal = {bioRxiv}
}

@article {Avior2013,
	author = {Avior, Yishai and Bomze, David and Ramon, Ory and Nahmias, Yaakov},
	title = {Flavonoids as dietary regulators of nuclear receptor activity},
	elocation-id = {2025.04.06.647437},
	year = {2013},
	doi = {10.1039/c3fo60063g},
	publisher = {The Royal Society of Chemistry},
	URL = {http://dx.doi.org/10.1039/C3FO60063G},
	journal = {Food Funct},
   volume = {4},
   issue = {6},
   pages = {831-844}
}

@article{RN456AMICI,
	author = {Hong, Justin and Desai, Khushi and Nguyen, Tu Duyen and Nazaret, Achille and Levy, Nathan and Ergen, Can and Plitas, George and Azizi, Elham},
	title = {{AMICI}: Attention Mechanism Interpretation of Cell-cell Interactions},
	elocation-id = {2025.09.22.677860},
	year = {2025},
	doi = {10.1101/2025.09.22.677860},
	publisher = {Cold Spring Harbor Laboratory},
	URL = {https://www.biorxiv.org/content/early/2025/09/24/2025.09.22.677860},
	eprint = {https://www.biorxiv.org/content/early/2025/09/24/2025.09.22.677860.full.pdf},
	journal = {bioRxiv}
}

@article{RN339LLM,
   author = {Cui, H. and Tejada-Lapuerta, A. and Brbic, M. and Saez-Rodriguez, J. and Cristea, S. and Goodarzi, H. and Lotfollahi, M. and Theis, F. J. and Wang, B.},
   title = {Towards multimodal foundation models in molecular cell biology},
   journal = {Nature},
   volume = {640},
   number = {8059},
   pages = {623-633},
   DOI = {10.1038/s41586-025-08710-y},
   url = {https://www.ncbi.nlm.nih.gov/pubmed/40240854},
   year = {2025},
   type = {Journal Article}
}

\end{document}